\documentclass[twocolumn]{aastex631}
\hypersetup{linkcolor=blue,citecolor=blue,filecolor=cyan,urlcolor=blue}

\graphicspath{{figures/}{../figures/}}
\usepackage{newtxtext,newtxmath}

\usepackage[T1]{fontenc}

\usepackage{graphicx}	
\usepackage{amsmath}	
\usepackage{amssymb}	
\usepackage{hyperref} 
\usepackage{natbib}
\usepackage{multirow}
\usepackage{enumitem}
\setlist[enumerate]{label=(\roman*),itemsep=0pt,topsep=0pt,parsep=0pt,partopsep=0pt}

\usepackage{graphicx}

\begin{document}

\title{Preliminary title: "Our Milky Way as a laboratory: uncovering early metal enrichment"}
\title{{\sc NEFERTITI:} Linking early galaxy formation to the assembly of the Milky Way}

\correspondingauthor{Ioanna Koutsouridou}
\email{ioanna.koutsouridou@unifi.it}

\author[0000-0002-3524-7172]{Ioanna Koutsouridou}
\affiliation{Dipartimento di Fisica e Astrofisica
Univerisitá degli Studi di Firenze,
via G. Sansone 1, Sesto Fiorentino, Italy}
\affiliation{INAF/Osservatorio Astrofisico di Arcetri, Largo E. Fermi 5, I-50125 Firenze, Italy}

\author[0000-0001-7298-2478]{Stefania Salvadori}
\affiliation{Dipartimento di Fisica e Astrofisica
Univerisitá degli Studi di Firenze,
via G. Sansone 1, Sesto Fiorentino, Italy}
\affiliation{INAF/Osservatorio Astrofisico di Arcetri, Largo E. Fermi 5, I-50125 Firenze, Italy}

\author[0000-0001-9155-9018]{{\'A}sa Sk{\'u}lad{\'o}ttir}
\affiliation{Dipartimento di Fisica e Astrofisica
Univerisitá degli Studi di Firenze,
via G. Sansone 1, Sesto Fiorentino, Italy}
\affiliation{INAF/Osservatorio Astrofisico di Arcetri, Largo E. Fermi 5, I-50125 Firenze, Italy}

\author[0000-0001-5487-0392]{Viola Gelli}
\affiliation{Cosmic Dawn Center (DAWN), Denmark}
\affiliation{Niels Bohr Institute, University of Copenhagen, Jagtvej 128, 2200 Copenhagen N, Denmark}

\author[0009-0006-4326-6097]{Elka Rusta}
\affiliation{Dipartimento di Fisica e Astrofisica
Univerisitá degli Studi di Firenze,
via G. Sansone 1, Sesto Fiorentino, Italy}
\affiliation{INAF/Osservatorio Astrofisico di Arcetri, Largo E. Fermi 5, I-50125 Firenze, Italy}

\author[0009-0006-7675-2614]{Lapo Querci}
\affiliation{Dipartimento di Fisica e Astrofisica
Univerisitá degli Studi di Firenze,
via G. Sansone 1, Sesto Fiorentino, Italy}
\affiliation{INAF/Osservatorio Astrofisico di Arcetri, Largo E. Fermi 5, I-50125 Firenze, Italy}

\author[0000-0001-5200-3973]{David S. Aguado}
\affiliation{Instituto de Astrofísica de Canarias, Vía Láctea, 38205, La Laguna, Tenerife, Spain}
\affiliation{Universidad de La Laguna, Departamento de Astrofísica, 38206, La Laguna, Tenerife, Spain}

\author[0009-0003-0816-2880]{Alice Mori}
\affiliation{Dipartimento di Fisica e Astrofisica
Univerisitá degli Studi di Firenze,
via G. Sansone 1, Sesto Fiorentino, Italy}
\affiliation{INAF/Osservatorio Astrofisico di Arcetri, Largo E. Fermi 5, I-50125 Firenze, Italy}


\begin{abstract}
We use a new implementation of the {\sc NEFERTITI} galaxy formation model, coupled to $\sim 30$ high-resolution Caterpillar dark-matter simulations of Milky Way (MW) analogues, to connect early galaxy formation with the MW's assembly down to $z=0$. Our locally-constrained model resolves minihaloes hosting the first Pop~III stars and self-consistently tracks inhomogeneous ionization and chemical enrichment. Pop~III star formation begins at $z\simeq27$, peaks at $z\simeq10-15$, and persists down to $z\lesssim5$, producing Pop~III systems with $M_*\sim10-5\times10^5\:{\rm M_\odot}$. The present-day descendants of Pop~III stars span ${\rm [Fe/H]<-9}$ to ${\rm [Fe/H]\approx-1}$, with the most metal-poor stars typically enriched by a few (1-4) low-energy supernova progenitors. Pair-instability supernova descendants more commonly form in massive haloes ($M_{\rm vir}>10^8\:{\rm M_\odot}$), often externally enriched, reflecting the strong feedback and delayed recovery following energetic explosions. These early systems serve as building blocks for the present-day Galaxy's metal-poor component: although 90$\%$ of the total stellar mass formed \textit{in situ}, the accreted component dominates at $[{\rm Fe/H}]<-1$ and accounts for nearly all stars with $[{\rm Fe/H}]<-3$. This accreted population is largely built by a few ($\sim5$) massive ($M_*>10^8\:{\rm M_\odot}$) destroyed dwarfs, but lower-mass systems become increasingly important at low metallicities, with ultra-faint and classical dSph analogues contributing $\sim25\%$ at $[{\rm Fe/H}]<-3$. Our model simultaneously reproduces the properties of metal-poor MW stars and the JWST "Hebe" galaxy at $z\sim11$, supporting its identification as a pure Pop~III system. Ultimately, {\sc NEFERTITI} is a key tool to interpret upcoming local and high-$z$ observations linking the near- and far-field cosmology.

\end{abstract}

\keywords{
Milky Way formation ---
Galactic archaeology  ---
Population III stars ---
Chemical enrichment ---
High-redshift galaxies
}

\section{Introduction} \label{sec:intro}


According to the currently favoured $\Lambda$CDM cosmological model, the first metal-free Population~III (Pop~III) stars formed about 100-200 Myr after the Big Bang (at redshift $z\sim30-20$) in low-mass dark matter (DM) structures called minihalos (i.e., halos with virial temperatures $T_{\rm vir}<2 \times 10^4\:$K; \citealt{Bromm2013, Klessen2023}). Their formation marked a fundamental shift in cosmic history, ending the cosmic dark ages and initiating the first phases of galaxy evolution. 
Born out of pristine gas left over from primordial nucleosynthesis, Pop~III stars produced the first metals (i.e., elements heavier than lithium) and dust grains, and released them into the surrounding medium via supernova (SN) explosions and stellar winds. In addition, they were the sources of the first ionizing and photodissociating photons, contributing to the reionization of the Universe and heating the surrounding gas, thereby suppressing star formation in the lowest-mass halos.

The nature and impact of this combined radiative, mechanical, and chemical feedback remain a fundamental unknown in cosmology, as it depends critically on the poorly constrained properties of Pop~III stars, primarily their initial mass function (IMF). 
While the James Webb Space Telescope (JWST) has begun to provide the first candidates for Pop~III-host systems at high redshift \citep[e.g.,][]{Vanzella2023, Nakajima2025, Maiolino2026}, direct observation of individual Pop~III stars is still heavily restricted by current sensitivity limits \citep{Zackrisson2024}. Consequently, Galactic Archaeology remains a powerful complementary way to probe the early Universe, through the study of the oldest, most metal-poor stars in the Milky Way (MW) and its satellite galaxies.  Long-lived "second-generation" stars  are predicted to span a wide range of iron abundance, from below ${\rm [Fe/H]} = -7$ up to $-1$, and to constitute a significant fraction of the extremely metal-poor (EMP; ${\rm [Fe/H]} < -3$) population \citep{Hartwig+18, Vanni23, Koutsouridou2023}. By preserving the chemical fingerprints of their progenitors within their photospheres, they allow us to indirectly decode the properties of the very first stars  \citep[e.g.,][]{Iwamoto2005, Salvadori2007, Komiya2010, Ishigaki2018, Koutsouridou2024, Hartwig2024, Rossi2025}.

Within the $\Lambda$CDM framework, galaxies such as our own form through the combination of \textit{in situ} star formation and the hierarchical accretion of lower-mass systems. For MW-mass galaxies, the vast majority of stellar mass ($\sim 80-90\%$) is formed internally \citep{Rodriguez-Gomez2016, Davison2020, Fu2024}. Yet, the "accreted" component becomes increasingly important  when examining the oldest, most metal-poor populations (${\rm [Fe/H]} < -1$; e.g.,~\citealt{DiMatteo2019, Monachesi2019, Sestito2021}).

Over the past decade, the Gaia mission, combined with large spectroscopic surveys, has provided 6D phase-space and chemical data for a vast number of MW stars, providing a window into early galaxy formation and the MW assembly history  \citep[e.g.,][]{Deason2024}. Most notably, Gaia revealed that the inner stellar halo is dominated by a single, massive accretion event known as Gaia-Sausage-Enceladus (GSE; e.g., \citealt{Belokurov2018,Helmi2018, Myeong2018, Haywood2018, Fattahi2019,Mackereth2019}). This progenitor, thought to have merged with the MW 8–11 Gyr ago, is joined by an increasing census of other substructures, including evidence for an earlier massive merger \citep{Kruijssen2020,Horta2021}, as well as lower-mass streams and overdensities 
\citep[e.g.,][]{Majewski2003,Ibata2019,Ibata2021,Myeong2019, Koppelman2019, Naidu2020, Horta2023b}.

Yet reconstructing the MW’s accretion history remains fundamentally challenging. Over time, orbital mixing, dynamical heating and overlapping chemical signatures can blur the distinct signatures of past mergers; a single massive merger can manifest as multiple chemo-dynamical features, while stars from different progenitors, or even in situ populations, may appear indistinguishable in phase space \citep{Jean-Baptiste2017, Koppelman2020,Khoperskov2023, Mori2024, Thomas2025,Buder2025}. The problem is compounded for the lowest-mass progenitors ($M_*<10^6\:{\rm M_\odot}$), whose debris is traced by too few stars in current surveys to be clearly identified as separate halo components \citep{Naidu2020, Deason2023}.

Nevertheless, there are indications that this low-mass regime may be an  important contributor to the MW metal-poor tail. In particular, \cite{Bonifacio2021} concluded that the GSE remnant is strongly deficient in very metal-poor (VMP; ${\rm [Fe/H]}<-2$) stars compared to the broader halo population, and suggested that the lowest-metallicity populations were assembled, at least in part, from smaller galaxies. Ultra-faint dwarfs (UFDs; $M_*<10^5\:{\rm M_\odot}$) are natural candidates for such contributors, since both their metallicity distribution functions (MDFs) and abundance patterns for elements through the iron peak closely resemble those of halo stars at ${\rm [Fe/H]}<-2$ \citep{Kirby2008, Frebel2010, Simon2019}. For example, the halo hosts a large fraction of carbon-enhanced metal-poor (CEMP) stars—common in surviving UFDs  but largely absent in present-day massive dwarfs \citep[e.g.,][]{Lucchesi2024}. However, the heaviest elements point to a more complex picture; neutron-capture abundances in VMP halo stars are more readily reproduced by more massive classical dwarf spheroidals (dSphs) than by UFDs \citep[e.g.,][]{Tafelmeyer2010, Mashonkina2017,Ji2019,Skuladottir2024b}.


Theoretical models offer an important route to reconstructing the MW’s accretion history, especially where observational identification of individual progenitors becomes difficult. In broad agreement with current observations, these studies find that the accreted stellar component of MW-like galaxies is mainly built by a few massive, GSE-like progenitors, while the contribution of low-mass and ultra-faint dwarfs is expected to be small \citep[e.g.,][]{Bullock2005, Cooper2010, Deason2016, DSouza2018, Monachesi2019, Fattahi2020, Cunningham2022, Horta2023}. However, most of this work has focused on the global halo mass budget and chemo-dynamical structure, rather than on the origin of its most metal-poor stars. As a result, it remains much less clear which progenitors dominate the very metal-poor regime, particularly at ${\rm [Fe/H]}<-3$, especially since models that both resolve ultra-faint systems and include Pop~III star formation remain scarce.

Here, we address this question using {\sc NEFERTITI} \citep{Koutsouridou2023, Koutsouridou2025}, a state-of-the-art galaxy formation and chemical evolution model coupled with the \textit{Caterpillar} suite of 31 high-resolution $N$-body simulations of MW-analogues \citep{Griffen2016,Griffen2018}.
Our framework follows the origin and assembly of the MW down to ${\rm [Fe/H]}<-7$ in a fully cosmological context, resolving the minihalos that hosted the first episodes of star formation and tracking the formation of all individual Pop~III and metal-poor stars. It self-consistently accounts for the inhomogeneous chemical enrichment of the intergalactic medium (IGM) and for the patchy nature of reionization, and is locally calibrated against observations.
This approach allows us to study the environments that hosted the first stars and their descendants, trace their hierarchical assembly into the Galaxy, and link present-day observables to the properties of the earliest star-forming systems, now observed at high redshift thanks to the JWST.

\section{The {\sc NEFERTITI} framework}

{\sc NEFERTITI} (NEar FiEld cosmology: Re-Tracing Invisible TImes), first introduced in \citet{Koutsouridou2023}, is a cosmological semi-analytic model (SAM) of galaxy formation and chemical enrichment designed to study the unknown properties of the first stars and trace the earliest phases of galaxy formation. {\sc NEFERTITI} can run on halo merger trees extracted from $N$-body simulations or Monte Carlo techniques, and grounds on our previous experience with SAMs for the Local Group formation (e.g.,~\citealp{Salvadori2007, Salvadori2010, Salvadori2015, pagnini+23}).

Here, we combine {\sc NEFERTITI} with a suite of high-resolution cosmological $N-$body simulations of MW analogues (Sec.\ref{DM sims}), allowing us to sample a wide diversity of assembly histories. Along each DM merger tree, {\sc NEFERTITI} follows the evolution of the baryonic component, assuming that gas is initially pristine and set by the universal baryon fraction. Once star formation begins, the model self-consistently follows the different feedback processes (radiative, chemical, and mechanical), which regulate the gas content and chemical composition of galaxies through SN-driven outflows and the suppression of gas accretion.

Compared to previous implementations \citep{Koutsouridou2023, Koutsouridou2025} the present version of {\sc NEFERTITI} introduces several important advances. The most significant are a new algorithm to track the spatial and temporal propagation of ionized regions (Sec.~\ref{Ionization}) and SN-driven metal bubbles  (Sec.~\ref{Inhomogeneous_mixing}). This enables a self-consistent treatment of reionization and inhomogeneous metal enrichment, allowing us to capture how the location and environment of a halo influence both its enrichment history and ability to form stars. Additional developments are the inclusion of Type~Ia SNe (SNe~Ia; Sec.~\ref{SF-SSPs}) and stellar rotation for Pop~II/I stars (Sec.~\ref{sec: metals}). A full overview of the {\sc NEFERTITI} framework is detailed in the following Secs.~\ref{DM sims}-\ref{Inhomogeneous_mixing}.

\subsection{\textit{N}-body simulations of MW-analogues}
\label{DM sims}
In this work, we adopt a suite of 31 DM-only zoom-in cosmological simulations of MW-sized halos from the \textit{Caterpillar} project \citep{Griffen2016,Griffen2018}, 
The simulations follow the \citet{Planck2014} cosmology and were selected from a parent volume of $(100 h^{-1} {\rm Mpc})^3$, requiring  host halos at $z=0$ to have: virial masses in the range $0.7-3\times10^{12}\:{\rm M_\odot}$; no neighbouring halos more massive that $7\times10^{13}\:{\rm M_\odot}$ within 7 Mpc; and no companions with $M_{\mathrm{vir}} \ge 0.5\,M_{\mathrm{host}}$ within 2.8 Mpc. The latter two criteria were adopted to exclude systems near massive clusters or in close pairs, as such environments would greatly enlarge the Lagrangian volume and make the simulations prohibitively expensive at the desired resolution. A weak constrain on the assembly history was also imposed by excluding 
systems that experienced a major merger (mass ratio $\gtrsim 1:3$) since $z\simeq 0.05$. This selection results into a representative sample of $\sim 10^{12}\:{\rm M_\odot}$ halos, including a diversity of merger histories, some differing from the Milky Way, which likely had its last major merger at $z\sim2$ \citep{Belokurov2018, Helmi2018, Naidu2021}. 

The simulations are evolved from $z=30$ to $z=0$, with  snapshots every $\sim 6$--12 Myr at $z>6$, and  $\sim 40$--60 Myr at $z=6$--0. Dark matter halos are identified with ROCKSTAR \citep{Behroozi2013}
and assigned virial masses, $M_{\rm vir}$, using
the evolution of the virial relation from \citet{Bryan1998}. With a high-resolution dark-matter particle mass of $m_{\rm{DM}} = 3\times10^{4}\:{\rm M_\odot}$, these simulations rank among the  highest DM mass-resolution simulations of MW-sized halos currently available (e.g., Torrigiani Malaspina et al., in prep.), and resolve minihalos down to a few times $10^{5}\:{\rm M_\odot}$, making them ideal to track the first star-forming sites and model Pop~III star formation.



\subsection{Gas accretion}
\label{Gas accretion}

We assume that gas from the intergalactic medium is continuously accreted onto DM halos at a rate proportional to their DM growth:
\begin{equation}
    \dot{M}_{\rm halo, accr} = f_{\rm b}\dot{M}_{\rm vir},
\label{e:Mhalo_accr}
\end{equation}
where $f_{\rm b}=\Omega_{\rm b}/\Omega_{\rm m}$ is the universal baryon fraction. We treat this incoming material as cold gas associated with filamentary flows.
From the filaments, gas subsequently streams onto the central galaxy, replenishing the reservoir available for star formation, on a free-fall timescale: 
\begin{equation}
    t_{\rm ff} = \Big( \frac{3\pi}{32 G \rho}\Big)^{1/2},
\end{equation}
where $G$ is the gravitational constant and $\rho = \Delta_c \rho_{\rm crit}$ is the total (dark+baryonic) density of the halo at redshift $z$, with $\Delta_c$ the overdensity from \citet{Bryan1998}.

{\sc NERFERTITI} currently does not model shock heating of the infalling cold gas; to account for this mechanism, responsible for star formation quenching in massive galaxies \cite{Birnboim2003,Dekel2006}, we impose a complete shutdown of accretion onto halos above the critical mass $M_{\rm vir, shock}= 2 \times 10^{12}\:{\rm M_\odot}$, a prescription shown to reproduce the galaxy colour–magnitude distributions and stellar mass functions at $0 < z < 2.5$ \citep{Cattaneo2006, Koutsouridou2022}. In practice, only a subset of MW-like halos in our merger trees cross this threshold, and only at late times ($z<0.8$). It therefore has no impact on our results for stars formed at earlier phases.

Moreover, we assume that subhalos stop accreting gas after infall into a more massive host halo. This environmental starvation mechanism is thought to dominate star formation quenching in low-mass ($M_*<10^{10}\:{\rm M_\odot}$) satellite galaxies \cite[e.g.,][]{ vandenBosch2008}. Additional enviromental mechanisms, that operate in dense environments, such as stripping of the satellites' interstellar medium (ISM) and pressure induced star formation bursts \cite[e.g.,][]{Kapferer2009, Bekki2014,Koutsouridou2019} are currently not included.

Finally, we account for the suppression of gas accretion onto minihalos caused by reionization (see Sec.~\ref{Ionization}).

\subsection{Star formation and stellar populations}
\label{SF-SSPs}

At each sub-timestep of the SAM, $\delta t_{\rm s}=5\:$Myr, we compute the star formation rate in each galaxy as: 
\begin{equation}
    {\rm SFR} = \epsilon_{\rm SF} \frac{M_{\rm gas}}{t_{\rm ff}},
\label{e:sfr}
\end{equation}
where the star formation efficiency $\epsilon_{\rm SF}$ is a free parameter of our model and $M_{\rm gas}$ is the ISM gas mass. In minihalos, the star formation is reduced by a factor of $2/[1 + (2 \times 10^4 \: {\rm K}/T_{\rm vir})^3]$ to account for the ineffective cooling by molecular hydrogen \citep{Salvadori2009}. 

We assume that Pop~III stars form if the ISM metallicity is below a critical value, $Z_{\rm ISM} < Z_{\rm crit} = 5.15 \times 10^{-5}\:{\rm Z_\odot}$ (\citealt{deBen2017}; assuming $Z_\odot = 0.0134$ from \citealt{Asplund2009}). Otherwise, Pop~II/I stars form. In both cases, stars are formed following a \cite{Larson1980} type IMF:

\begin{equation}
\phi(m_\star) = \frac{d N}{d m_\star} \propto m_\star^{-x} \exp \bigg( - \frac{m_{\rm ch}}{m_\star} \bigg).
\label{eq:imf}
\end{equation}

\noindent For Pop~III stars we adopt $x=2.35$, $m_{\rm ch}=10\:{\rm M_\odot}$, and a mass range $m_\star = (0.8-1000)\:{\rm M_\odot}$, consistent with observational constraints \citep{Hartwig15,Rossi+21,pagnini+23,Koutsouridou2023,Koutsouridou2024,Rusta2026} and hydrodynamic simulations of Pop~III star formation \citep{Hirano14,Hirano15}. 
The impact of varying $x$ and $m_{\rm ch}$ on the properties of Galactic halo stars at $z=0$ is discussed in \cite{Koutsouridou2023,Koutsouridou2024}. 
For Pop~II/I stars, we assume a Salpeter slope of $x=2.35$ \citep{Salpeter1955}, $m_{\rm ch} = 0.35\:{\rm M_\odot}$, and $m_\star = (0.08-100)\:{\rm M_\odot}$. 

A key feature of {\sc NEFERTITI} is that all Pop~III and metal-poor (MP; [Fe/H]$\leq -1$) stars are followed individually. For each star-formation burst with total mass computed through Eq.~\ref{e:sfr}, we construct a simple stellar population (SSP) by randomly sampling stellar masses according to the adopted IMF \citep{Rossi+21}. If a sampled mass exceeds the star-forming gas available at that time, we do not truncate the IMF. Instead, we postpone the burst in that halo until sufficient gas accumulates to form the star \citep{Koutsouridou2024}. This prevents low-mass, weakly star-forming halos from systematically missing massive stars and ensures that the IMF realized across the full halo population matches the assumed one on average. The evolution of each star is then tracked individually using the stellar lifetimes of \citet[][strong mass loss set]{Schaerer2002} for Pop~III stars and  \citet{raiteri1996simulations} for Pop~II/I stars.\\

In addition, we account for the delayed occurrence of SNe~Ia associated with each SSP. We model their rate using the delay-time distribution (DTD) of \cite{Maoz2012}, following the parametrization by \cite{Vogelsberger2013}:
\begin{equation}
    g(t) = 
\begin{cases}
\text{0} & \text{if } t<\tau_{\rm 8\:M_\odot}\\
\text{$A_{\rm SNeIa} \Big(\frac{t}{\tau_{\rm 8\:M_\odot}} \Big)^{-s} \frac{s-1}{\tau_{\rm 8\:M_\odot}}$} & \text{if } t\geq\tau_{\rm 8\:M_\odot},
\end{cases}
\end{equation}
where $\tau_{\rm 8\:M_\odot} = 0.04\:$Gyr is the offset time between the birth of the SSP and the first expected SNIa event. In the above,  $s=1.12$ and $A_{\rm SNeIa} = 2.5792 \times 10^{-3}\:{\rm SN/M_\odot}$ is a normalization constant such that the Hubble-time-integrated number of SNIa per unit stellar mass formed is equal to $\int_{0.040}^{13.8} g(t) dt=1.3 \times 10^{-3}\:$SN/${\rm M_\odot}$ \citep{Maoz2012}.

For each SSP, the expected number of SNe Ia occurring within a timestep $[t_i,t_{i+1}]$ is then:
\begin{equation}
   \frac{N_{3-8\:{\rm M_\odot}}^{\rm SSP}}{{N_{3-8\:{\rm M_\odot}}^{\rm IMF}}} \int_{t_i}^{t_{i+1}} g(t')dt' ,
\end{equation}
where $N_{3-8\:{\rm M_\odot}}^{\rm SSP}$ is the actual number of stars formed in the SSP with initial masses $3-8\:{\rm M_\odot}$, i.e., within the commonly assumed mass range of SN Ia progenitors, and $N_{3-8\:{\rm M_\odot}}^{\rm IMF}\approx 0.0366\:{\rm M_\odot^{-1}}$ is their expected number per unit stellar mass formed according to our assumed Pop~II/I IMF. This scaling captures deviations from the theoretical IMF introduced by our stochastic IMF sampling.

\subsection{Radiative Feedback}


Radiation from young massive stars shapes the evolution of the least massive halos in two ways. Ionizing UV photons ($E>13.6\:$eV) ionize and heat the surrounding gas to $\sim2\times 10^4\:$K, impeding its accretion onto halos with lower virial temperatures, and can also overheat (photoevaporate) the gas already present in these systems, preventing it from forming stars \citep{Bark01}. Lyman-Werner (LW) photons (12.2 eV < E < 13.6 eV), instead, do not directly halt baryonic infall but photo-dissociate molecular hydrogen, thereby eliminating the only available coolant in pristine minihalos. As a result, star formation is postponed until the halo grows massive enough to reach the atomic-cooling regime at $T_{\rm vir}\approx 2 \times 10^{4}\,{\rm K}$ \citep{Bromm2004}. 

\label{LW-Ionization}

\newpage

\subsubsection{Photo-dissociation of H$_2$}
We approximate LW feedback as a spatially uniform background that sets a minimum halo mass for Pop~III star formation. This choice is motivated by the fact that LW photons propagate over very large distances, with a mean free path of $\sim 100 \:$cMpc \citep{Ahn2009}, larger that the effective comoving volume probed by our model. Although the LW field fluctuates, the vast majority of halos experience fluxes close to the mean, while strong local enhancements affect only rare close-pair configurations \citep{Dijkstra2008, Agarwal2012, Holzbauer2012}.

We adopt the evolution of the minimum mass threshold from \citet[see their Figure 2 and Section 2]{Salvadori2012}, which starts from the mass corresponding to $T_{\rm vir}=2\times 10^3\:{\rm K}$ at $z=30$ \citep{Tegmark1997, Dijkstra2008}, and increases with redshift as the LW background builds up \citep{Ahn2009}. Below this threshold, pristine minihalos are unable to form Pop~III stars. 
However, if the gas is enriched then metal-line cooling and, at high densities, dust cooling, can compensate for the reduced H$_2$ and enable fragmentation. Estimates for this transition metallicity range from $10^{-4} - 10^{-3.5}\,Z_\odot$ at low densities down to $10^{-6} - 10^{-5}\,Z_\odot$ when dust cooling becomes efficient \citep{bromm02,Schneider2003,Omukai2008,Safranek2010,Dopcke2013,Smith2015, Smith2024}. Motivated by these results, we assume that halos below the LW-regulated mass threshold can form stars if their gas metallicity exceeds $Z_{\rm crit}$. This treatment captures the transition from LW-quenched Pop~III formation to Pop~II star formation triggered by external or prior enrichment.\\

\begin{figure}
\begin{center}
\includegraphics[width=0.99\hsize]{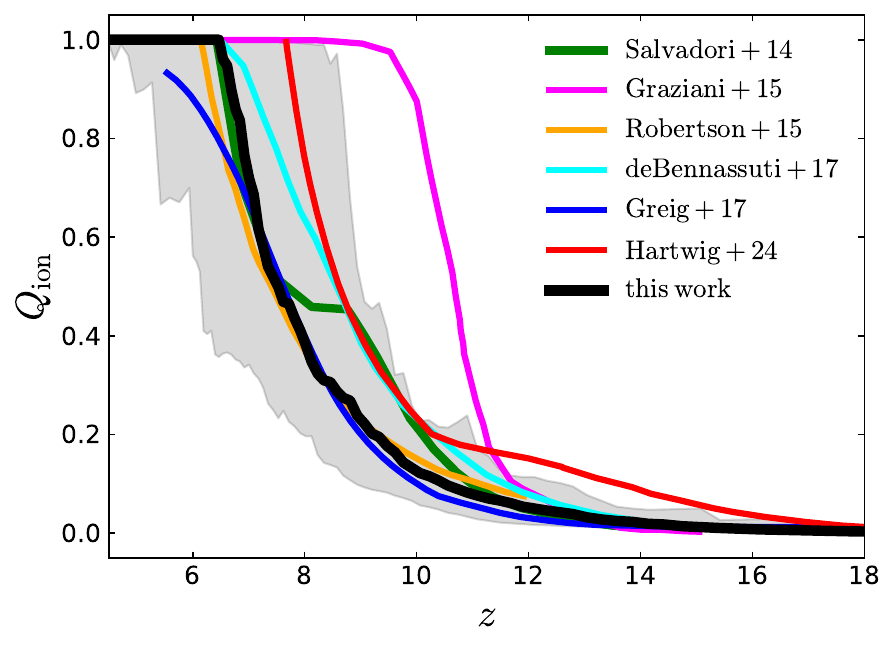}\\
\caption{The filling factor of ionized regions in the simulated MW volume, as a function of redshift. The black line shows the median of the 31 DM simulations, while the shaded area marks the range between the earliest and latest reionization histories. Our predictions agree well with other models (coloured lines).}
\label{fig: filling}
\end{center}
\end{figure}

\begin{figure*}
\begin{center}
\includegraphics[width=0.49\textwidth]{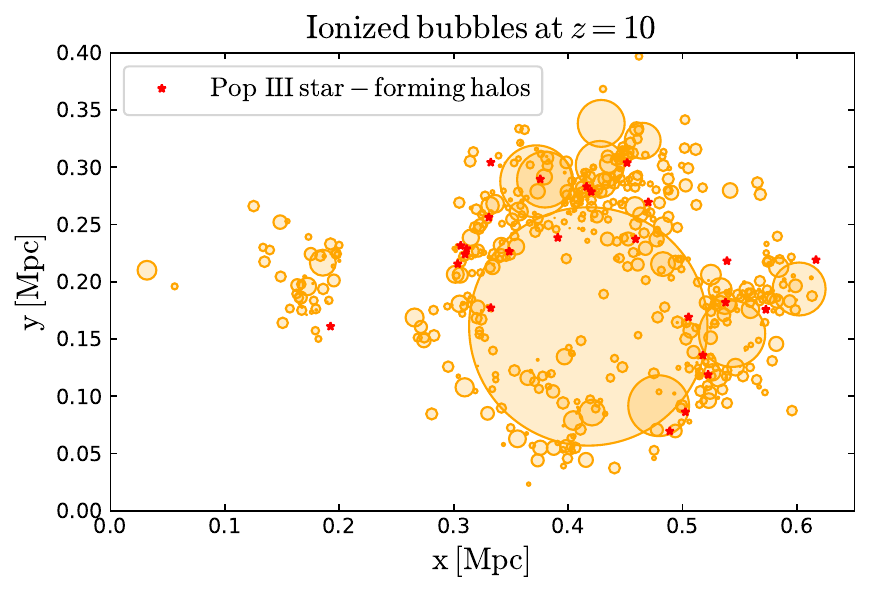}
\hfill
\includegraphics[width=0.49\textwidth]{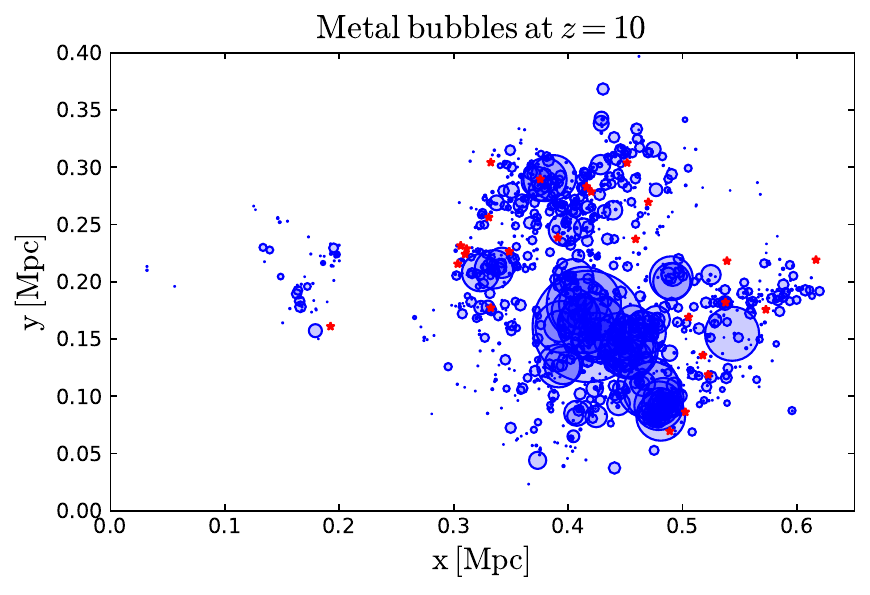}
\caption{Projection of ionized bubbles (left) and SN-driven metal-enriched bubbles (right) at $z=10$ for the Caterpillar merger tree Cat-10. Cat-10 has a MW–like assembly history, with the last major merger (mass ratio >1:3) at $z>1.5$. Note that SN bubbles may overlap, while ionized bubbles do not (see Sec.~\ref{LW-Ionization}). Pop~III star-forming halos at this snapshot, shown with red stars, lie outside the metal-enriched volumes.}
\label{fig: Bubbles}
\end{center}
\end{figure*}

\subsubsection{Ionization}
\label{Ionization}

In contrast to LW photons, ionizing photons have a much shorter mean free path and remain in the vicinity of the halo producing a strongly inhomogenous ionizing background, especially at early times \citep[e.g.,][]{Mesinger2009,Rahmati2018}. We, therefore, opt to follow the spatial and temporal evolution of ionized regions explicitly. As commonly adopted in SAMs \citep{Salvadori2014, Magg2018, Hartwig2022}, we assume that each star-forming halo is surrounded by an ionized region with comoving volume $V_{\rm ion,c}$ whose evolution is set by the balance between newly produced ionizations and recombinations:
\begin{equation}
    \dot{V}_{\rm ion,c} = \frac{\dot{N}_{\rm ion}}{n_{\rm H}^{\rm 0}} - \frac{V_{\rm ion,c}}{t_{\rm rec}}.
\end{equation}
Here $n_{\rm H}^0 = 0.75*\rho_{\rm crit}*\Omega_b/m_{\rm proton}$ is the mean comoving hydrogen number density in the IGM, and 
\begin{equation}
    t_{\rm rec}=[C \, \alpha_B \,n_{\rm H}^{\rm 0} (1+z)^3]^{-1}
\end{equation}
is the hydrogen recombination time, where $C$ is the clumping factor and $\alpha_B= 2.6 \times 10^{-13}\:{\rm cm^3 s^{-1}}$ is the case-B recombination coefficient at $10^4\:$K \citep{Draine2011}. The ionizing photon production rate is:
\begin{equation}
    \dot{N}_{\rm ion} = \frac{\dot{M}_\star}{m_{\rm proton}}  a_i f_{\rm esc},
\end{equation}
where $\dot{M}_\star$ is the SFR at the time, $a_i$ is the number of ionizing photons per stellar baryon and $f_{\rm esc}$ is their escape fraction. This approximation assumes that all ionizing photons are emitted instantaneously at formation, which is reasonable given that the ionizing budget is dominated by short-lived massive stars.

Following \cite{Salvadori2014}, we set $C=3$ and $f_{\rm esc}= 0.1$ for Pop~II/I stars, which provide a reionization history that is complete by $z \sim 6.5$ (see Fig.~\ref{fig: filling}). For Pop~III we adopt a higher $f_{\rm esc}= 0.5$ \citep{Visbal2020}.  For the photon yields we adopt $a_i=5,000$ for Pop~II/I stars \citep{Ferrara2016} and $a_i=94,600$ for Pop~III. The latter is expected for a metal-free $\sim20\:{\rm M_\odot}$ star (with moderate initial rotation $u_{\rm init}/u_{\rm crit}=0.3$) over its lifetime \citep{Yoon2012}\footnote{The mean stellar mass formed with our assumed Pop~III IMF is $22\:{\rm M_\odot}$.}. 
In future work, we will compute $a_i$ self-consistently from the effective IMF and metallicity \citep[e.g., as done in post-processing by][]{rusta2024}, enabling direct tests of how Pop~III IMF variations affect the ionization of the Milky Way environment.

To compute the evolution of $V_{\rm ion,c}$ we use the numerically stable implicit Euler scheme. Over a timestep $\delta t_{\rm s}$ from $t_i$ to $t_{i+1}$ this gives:
\begin{equation}
    V_{\rm ion,c}(t_{\rm i+1}) = \bigg(V_{\rm ion,c}(t_{\rm i}) + \frac{\dot{N}_{\rm ion}\delta t_{\rm s}}{n_{\rm H}^{0}}\bigg)(1 + \delta t_{\rm s}/t_{\rm rec}  )^{-1},
\end{equation}
which accounts for the effect of the cosmological expansion of the ionized regions. 
Assuming that ionized regions are spherical we can compute their physical radius at each redshift: 
\begin{equation}
R_{\rm ion, p}=(3 V_{\rm ion, c}/4\pi)^{1/3}/(1+z) .
\end{equation}
When two (or more) spheres overlap, we adjust their radii, while keeping their centres fixed, such that the total ionized volume is conserved and the updated spheres are just tangent. If no such solution exists, we add the volume of the smaller sphere to the largest one. This avoids ionizing the same volume multiple times in clustered environments (see Fig.~\ref{fig: Bubbles}). When halos merge, we sum their ionized volumes.

Minihalos whose centres fall inside an ionized region are assumed to: $i)$ stop accreting gas, and $ii)$ lose all previously accumulated ISM gas through photoevaporation, 
unless the gas metallicity exceeds $Z_{\rm crit}$, in which case the gas is retained and Pop~II/I star formation can proceed.\\

Fig.~\ref{fig: filling} shows the resulting median volume filling factor of ionized regions across the 31 Caterpillar merger trees:
\begin{equation}
    Q_{\rm ion}(z) = \frac{\sum_i V_{\rm ion,p}(z)}{V_{\rm MW}(z)},
\end{equation}
where we approximate the proper MW progenitor volume, $V_{MW}(z)$, as the virial mass of the MW at $z=0$ divided by the mean cosmic density, $V_{\rm MW}(z)= M_{\rm vir, peak}(z=0)/(\Omega_{\rm M} \, \rho_{\rm crit})/(1+z)^3$, yielding volumes of $20-70\:{\rm Mpc}^3$. 


We find that ionization within the MW volume is typically completed by $z=6.3$ (Fig.~\ref{fig: filling}), in agreement with the predictions for the MW environment by \cite{Salvadori2014} and with global constraints from \citet{Robertson2015} and \citet{Greig2017}. Our earlier reionization histories within the shaded region are also compatible with the models of \cite{deBen2017} and \cite{Hartwig2024} which are complete by $z=6.5$ and $z=7.6$, respectively. 
Earlier completion at $z\simeq9$, reported by \citet{Graziani2015}, who coupled a SAM to a radiative transfer code, lies outside our predicted range.

\subsection{Metals injected and supernova feedback}

\label{sec: metals}
{\sc NEFERTITI} follows metals from C to Zn as yielded by AGB stars, stellar winds from massive stars, SNe, SNe~Ia, and in the case of Pop~III stars also by PISNe \citep{Koutsouridou2025}. Furthermore, we model mechanical feedback from all SN types. 

For Pop~III stars we adopt the yields from \citet{Meynet2002} for AGB stars, $m_\star=[2-7]\:{\rm M_\odot}$, from \citet{Heger2002} for massive PISNe, $m_\star=[140-260]\:{\rm M_\odot}$, and those from \cite{Heger+woosley10} for SNe of $[10-100]\:{\rm M_\odot}$. The latter are given for 10 explosion energies and 14 mixing efficiencies, both of which strongly influence the final ejecta \citep[e.g., see discussion][]{Koutsouridou2023}. We refer to models with explosion energy $E_\star=0.3$ and $0.6 \times 10^{51}\:$erg as faint SNe, with $E_\star=0.9, 1.2$, and $1.5 \times 10^{51}\:$erg as core-collapse SNe (ccSNe), with $E_\star=1.8,\: 2.4$, and $3 \times 10^{51}\:$erg as high-energy SNe, and with 
$E_\star=5$ and $10 \times 10^{51}\:$erg as hypernovae.
For each star, we assign these parameters stochastically, drawing $E_\star$ from an energy distribution function (EDF):
\begin{equation}
    {\rm d}N/{\rm d}E\propto E_\star^{-\alpha_e},
\label{eq: edf}
\end{equation}
where $\alpha_e=2$ is calibrated to reproduce the halo CEMP fraction (see Sec.~\ref{Calibration_earliest}), and by adopting a uniform distribution over the available mixing efficiencies. 

For Pop~II/I stars, we adopt the AGB yields from \citet{Karakas2010} and massive-star yields from \citet{Limongi2018}, the latter given for three initial rotation velocities (0, 150 and 300 km/s).
Motivated by chemical-evolution studies suggesting that the average rotation decreases with metallicity \citep{Prantzos2018, Rizzuti2021}, we implement a metallicity-dependent rotation prescription. We follow the mean relation of \citeauthor{Rizzuti2021} (\citeyear{Rizzuti2021}; their Eq. 7) and assign individual velocities to each Pop~II/I star from a Gaussian with 
$\sigma=100\:$km/s. We note that using the \citet{Prantzos2018} relation which predicts on average higher velocities (by $\sim50-60\:$km/s), yields similar enrichment histories and global abundance trends.

Finally, for SNe~Ia, we adopt the explosion energy ($E_{\rm SNeIa} = 1.3 \times 10^{51}\:$erg) and stellar yields from \citet[models W7 and W70 for solar-metallicity and zero--metallicity stars, respectively]{Iwamoto99}, and interpolate for intermediate metallicities.\\

We assume that AGB stars and massive-star winds return their ejecta directly to the ISM of the host galaxy, without driving gas out of the halo; mechanical feedback is instead associated with SN explosions.
For the different SN types, i.e., Pop~III PISNe, and all ccSNe and SNe~Ia, we assume that SN-driven outflows eject gas from each halo at a rate: 
\begin{equation}
    \dot{M}_{\rm gas,ej} = \epsilon_{\rm wind} \frac{2 \sum_{i} \dot{N}^i_{\rm SN} E^i_{\rm SN}}{u_{\rm esc}^2}.
\label{e:Mgas_eje}
\end{equation}
Here the wind efficiency $\epsilon_{\rm wind}$ is a free parameter of our model, $\dot{N^i_{\rm SN}}$ is the explosion rate of SN of type $i$ with explosion energy $E^i_{\rm SN}$, and $u_{\rm esc} = \sqrt{G M_{\rm vir}/R_{\rm vir}}$ is the escape velocity of the halo. If the predicted ejected mass over a sub-timestep exceeds the available gas mass within the galaxy, we assume that the leftover SN energy is used to unbind the cold galactic filaments, removing from them a mass equal to the excess. The outflowing gas and metals enrich the surrounding medium and may later re-accrete onto the same halo, or accrete onto and enrich neighboring halos as described in the following Section.

\subsection{Inhomogeneous metal mixing in the IGM}
\label{Inhomogeneous_mixing}

Similar to our treatment of ionizing radiation, we model the enrichment of the IGM using a simple spherical bubble solution. We assume that each star-forming halo is surrounded by a metal-enriched bubble whose expansion has been driven by the cumulative energy injected by SNe since the halo formed (see Fig.~\ref{fig: Bubbles}). 
We can estimate the bubble radius using the  
classical superbubble solution of \citet{Weaver1977} for continuous energy injection:
\begin{equation}
    R_s(t) = \bigg(\frac{250}{308\pi} \bigg)^{1/5} L_w^{1/5} \rho^{-1/5} t^{3/5},
\end{equation}
 where $\rho$ is the mean surrounding baryonic density, $t$ is the time elapsed since the first SN explosion in the halo and $L_W$ is the mechanical luminosity of the wind, assumed to be constant in the \citet{Weaver1977} solution. We approximate $L_w$ as the total SN energy released within the halo divided by the time since the first explosion, $L_w = E_{\rm SN,tot}/t$. 
 
 We assume that the shell expansion stalls once its speed $dR_s/dt$ drops to the local sound speed $c_s$, and we distinguish two regimes for the ambient medium. Initially, the superbubble expands within the halo virial radius, where $\rho = \Delta_c \,\rho_{\rm crit} \,\Omega_b \,(1+z)^3$ and $c_s = \sqrt{\gamma/2} \times V_{\rm vir}$ \citep{Bark01}. Once the shell reaches the halo boundary, $R_s > R_{\rm vir}$, it propagates into the IGM, where $\rho = \rho_{\rm crit} \, \Omega_b \,(1+z)^3$ and $c_s = 15 \:$km/s, appropriate for an ionized medium \citep{Shapiro2004}, since ionization fronts typically propagate faster than metal bubbles.

Unlike our treatment of the ionized bubbles, the above formulation neglects the effects of cosmological expansion. Moreover, the Weaver solution is evaluated using the gas density at the redshift of interest, even though the cumulative SN energy includes contributions from earlier explosions that occurred when the surrounding medium was denser. This introduces two competing effects of cosmic expansion: while the Hubble flow would stretch the bubbles, the higher ambient densities at earlier times would have opposed their growth, leaving the net impact uncertain.\\

For each expanding bubble, we follow its chemical content: the mass of each element, total metal mass, $m_{\rm Z, bubble}$, and the contribution from different SN types, Pop III (faint, ccSNe, high energy, hypernovae, PISNe and SNe Ia) and Pop II (ccSNe and SNe Ia). When two halos that host enriched bubbles merge, we sum their volumes, matter and metal contents.

We assume that metals are uniformly mixed within each bubble, so its metallicity is given by 
\begin{equation}
    Z_{\rm bubble} = \frac{m_{\rm Z, bubble}}{m_{\rm bubble}},
\end{equation}
where $m_{\rm Z, bubble}$ (as well as the mass of each individual element) is updated at every timestep by adding the newly ejected metals from the halo and subtracting the metals that (re-)accrete onto the central halo and onto neighbouring halos whose centers lie within the bubble. The total gas mass contained within the bubble is computed as
\begin{equation}
   m_{\rm bubble}  = {\rho_{\rm bubble}(4\pi R_s^3/3) + M_{\rm gas, eje}}
\end{equation}
where $\rho_{\rm bubble}$ is the average baryonic density within the bubble (changing abruptly from $R<R_{\rm vir}$ to $R>R_{\rm vir}$ as described above) and $M_{\rm gas, eje}$ is the total gas mass ejected from the halo (Eq.~\ref{e:Mgas_eje}). Metallicities are summed where bubbles overlap (see Fig.~\ref{fig: Bubbles}).

The return of mass and metals from the enriched bubbles to their enclosed halos proceeds as follows. 
Consider a halo $i$ that accretes mass $M_{\rm halo, accr, i}$ during a timestep (given by Eq.~\ref{e:Mhalo_accr}). If its center lies within $n$ enriched bubbles (e.g., its own plus $n-1$ bubbles belonging to neighboring halos) we split this accretion across these bubbles based on their geometric overlap with the halo. For each bubble $j$, we compute the overlap fraction $f_{\rm overlap,i,j}$ between the halo volume (approximated as a sphere of radius $R_{\rm vir}$) and the bubble (for the halo’s own bubble, $f_{\rm overlap,i,j}=1$). We then define the normalized overlap weights
\begin{equation}
    w_{i,j} \equiv \frac{f_{\rm overlap,{i,j}}}{\sum_{j} f_{\rm overlap,{i,j}} },
\end{equation}
so that $\sum_j w_{i,j} = 1$ and $w_{i,j}M_{\rm halo, accr, i}$ is halo $i$'s \textit{requested} mass from bubble $j$.

Accretion is applied simultaneously for all halos during each timestep. Therefore, if bubble 
$j$ is shared by multiple halos, its available gas mass 
$m_{\rm bubble,j}$ is divided among them in proportion to these requests, so that the bubble is not depleted by one halo before the others accrete. The mass accreted by halo $i$ from bubble $j$ is thus given by:
\begin{equation}
M_{i\leftarrow j} = w_{i,j}\,M_{\rm halo,accr,i}\,
\frac{m_{\rm bubble,j}}{\sum_i w_{i,j}M_{\rm halo,accr,i}}.
\end{equation}
If the total available gas mass in the contributing bubbles is not enough to supply the full $M_{\rm halo,accr,i}$, the remaining fraction is drawn from the pristine surrounding medium.



\section{Validating {\sc NEFERTITI} against data}
\label{Calibration}

\subsection{Global Properties}
\label{global properties}

Our model contains two fundamental free parameters, the star formation efficiency $\epsilon_{\rm SF}$ and the wind efficiency $\epsilon_{\rm wind}$ (Eqs.~\ref{e:sfr} and \ref{e:Mgas_eje}), which strongly affect the present-day global properties of the MW and are calibrated to reproduce them as done in 
\cite{Koutsouridou2023}. The adopted values,  $\epsilon_{\rm SF} = 0.85$ and $\epsilon_{\rm wind}=0.008$, yield MW-analogues across the 31 Caterpillar DM simulations with stellar masses, gas-to-stellar-mass ratios (where gas refers to the star-forming ISM excluding the hot ionized circumgalactic medium) and SFRs consistent with observational estimates (as shown in Table~\ref{table: global_properties}). 

While our star formation efficiency is close to that found in our previous work \citep[$\epsilon_{\rm SF}=0.8$;][]{Koutsouridou2023}, the wind efficiency and consequently the mass loading factor $\eta \equiv \dot{M}_{\rm gas,ej}/{\rm SFR}$, is four times higher. This reflects the fact that the MW analogues in the Caterpillar suite span virial masses
$M_{\rm vir} = (0.77$--$2.78)\times10^{12}\:{\rm M_\odot}$\footnote{Observational estimates for the MW virial mass, $M_{\rm vir}\simeq (0.6$--$1.5)\times10^{12}\:{\rm M_\odot}$, are broadly consistent with this range \citep{McMillan2017, Callingham2019, Posti2019, Eadie2019, Watkins2019, Cautun2020, Labini2023}.}, with a median
$M_{\rm vir}=1.64\times10^{12}\:{\rm M_\odot}$, larger
than in our previous simulation, $M_{\rm vir}=0.78\times10^{12}\:{\rm M_\odot}$. 
More massive halos require stronger winds to overcome their deeper
potential wells if one wants to reproduce the same stellar mass.

\begin{table}[t]
\centering
\footnotesize
\caption{Global properties of the MW at $z=0$. The predicted properties correspond to the median values (with 16th--84th percentiles) for the 31 Caterpillar DM
simulations, assuming $\epsilon_{\rm SF}=0.85$ and $\epsilon_{\rm wind}=0.008$. For details and references see Sec.~\ref{global properties} and Sec~\ref{sec: MWobs}.}
\tabcolsep=0.25cm
\begin{tabular}{l c c}
\hline
\hline
Description & Model median & Observed value \\
\hline
Stellar mass & $5  \pm 1.7 \times 10^{10}\:{\rm M_\odot}$ & $(5$--$6.4)\times10^{10}\,{\rm M_\odot}$\\
Gas-to-stellar-mass ratio & $0.13^{+0.2}_{-0.08}$ & $0.1$--$0.2$\\
Star formation rate & $2^{+4.7}_{-1.1}\:{\rm M_\odot/yr}$ & $1$--$3\:{\rm M_\odot/yr}$\\
Mass loading factor & $0.39 \pm 0.06$ & $>0.1$--$0.38$ \\
Stellar metallicity & $1.3 \pm 0.1\:{\rm Z_\odot}$ & $1.58 \pm 0.8\:{\rm Z_\odot}$ \\
Gas-phase metallicity & $2.4^{+0.4}_{-1.2}\:{\rm Z_\odot}$ & $2.45 \pm 0.4\:{\rm Z_\odot}$ \\
SNIa rate & $0.12^{+0.08}_{-0.04}\:$SNuM & $0.11$--$0.14\:$SNuM \\
ccSN rate & $2.7^{+6.4}_{-1.4}\:(100\,{\rm yr})^{-1}$ & $1.6$--$3.9\:(100\,{\rm yr})^{-1}$  \\
MDF & \multicolumn{2}{c}{see Fig.~\ref{fig: MDF-CEMP}} \\
CEMP fraction & \multicolumn{2}{c}{see Fig.~\ref{fig: MDF-CEMP}} \\
\hline
\end{tabular}
\label{table: global_properties}
\end{table}

Once $\epsilon_\star$ and $\epsilon_w$ are calibrated to reproduce the stellar and gas mass of the MW, our model naturally reproduces its chemical properties. In particular, the predicted stellar and gas-phase metallicities are in good agreement with the observed mass--metallicity relations for local star-forming galaxies at $M_*=5 \times 10^{10}\:{\rm M_\odot}$: $Z_* = 1.58 \pm 0.8\:{\rm Z_\odot}$ \citep{Gallazzi2005} and $Z_{\rm gas} = 2.45 \pm 0.4\:{\rm Z_\odot}$\footnote{The oxygen abundance given in \cite{Tremonti2004} has been transformed to total metallicity assuming a solar-scaled composition and $Z_\odot = 0.0134$.} \citeauthor{Tremonti2004} (\citeyear{Tremonti2004}; see also Fig. 10 from \cite{Andrews2013} and references therein). Finally, the predicted SN rates are consistent with the observed values (see also Sec.~\ref{sec: metals}). Table~\ref{table: global_properties} lists the overall comparison to the observed values of the MW, which are referenced and discussed in more detail in Appendix~\ref{sec: MWobs}.


\subsection{Earliest chemical enrichment}
\label{Calibration_earliest}


The properties of the very metal-poor MW stellar population depend sensitively on the still unknown nature of the first Pop~III stars, in particular on their IMF and the EDF of the first SNe (Eqs.~\ref{eq:imf} and \ref{eq: edf}). In this work, we do not vary these parameters but adopt the values calibrated in \citet{Koutsouridou2023}, where they were shown to reproduce the observed metallicity distribution function (MDF) and the fraction of CEMP stars ($F_{\rm CEMP}$) in the Galactic halo. However, since {\sc NEFERTTITI} is now coupled to a different suite of DM simulations, we should first verify that this calibration remains valid.

\begin{figure}
\begin{center}
\includegraphics[width=0.99\hsize]{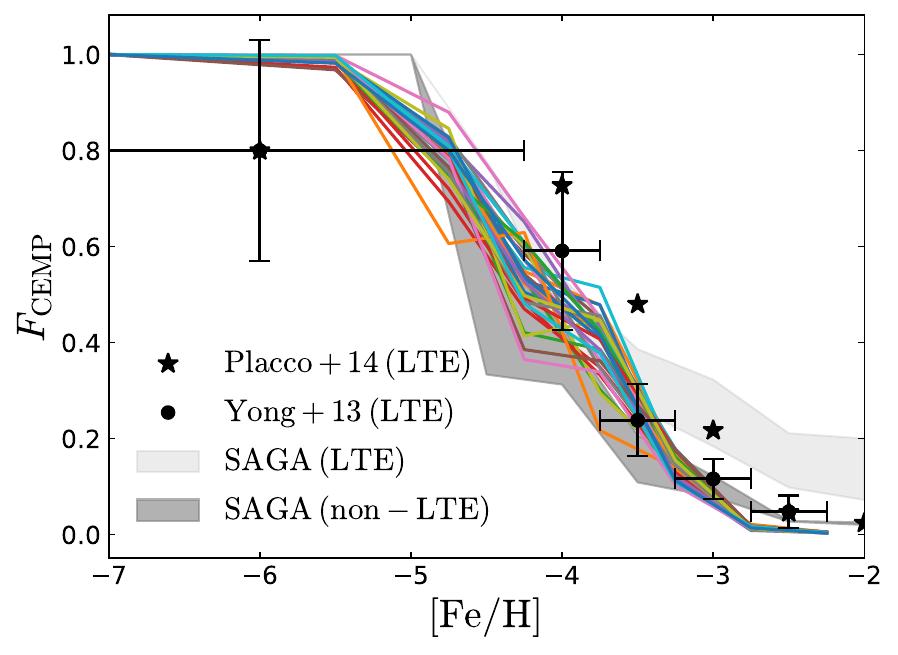}\\
\includegraphics[width=0.99\hsize]{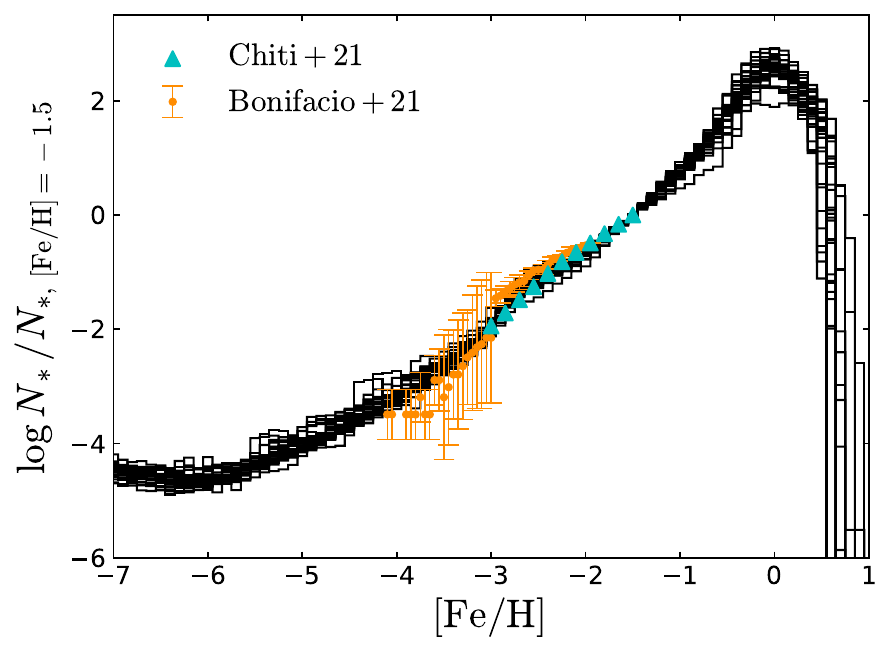}
\caption{Top: Differential CEMP fractions of the 31 MW analogues (coloured lines). Datapoints show the observed fractions of \citet{Placco2014} and \citet{Yong13} for Galactic halo stars. Both are corrected for surface-carbon depletion along the red-giant branch, the latter adopting the luminosity criterion of \cite{Aoki2007}. 
Shaded areas show the CEMP fraction of SAGA stars in LTE and non-LTE (see text). The width of the shaded areas reflects uncertainties in barium measurements: the lower boundary includes only stars with [Ba/Fe]$<0$, while the upper one also includes stars with upper limits or lacking Ba measurements. In both cases, evolutionary corrections for carbon have been applied following \citet{Placco2014}. 
Bottom: MDFs of the 31 Caterpillar MW-analogues as predicted by {\sc NEFERTITI} (black lines), compared to the observed MDFs of thick-disc and inner-halo MP stars from \cite{Bonifacio2021} and \cite{chiti21}, shown with colored symbols; normalized at ${\rm [Fe/H]}=-1.5$. In both panels, the curves correspond to averages over five realizations for each merger tree.}
\label{fig: MDF-CEMP}
\end{center}
\end{figure}



Fig.~\ref{fig: MDF-CEMP} shows $F_{\rm CEMP}$ as a function of metallicity (top) and the MDFs (bottom panel) predicted for the 31 Caterpillar merger trees. These results refer to the MW-analogues as a whole; since the Caterpillar simulations provide information only on DM halos and not on individual DM particles, we cannot use particle tagging to distinguish between different Galactic components, as done in our previous works \citep{Koutsouridou2023, Koutsouridou2024}. 

In the top panel of Fig.~\ref{fig: MDF-CEMP}, we compare to observational estimates of the halo CEMP fraction, considering only CEMP-no stars (no enhancement in s-process elements such as Sr and Ba), whose abundances are commonly interpreted as representative of their natal gas. CEMP-s stars, which are thought to acquire their carbon and s-process enhancement through binary mass transfer from an AGB companion, are lacking from our model, which does not account for binaries. 

The {\sc NEFERTITI} model, with our assumed IMF and EDF, reproduces the observed trend of $F_{\rm CEMP}$ with [Fe/H] from \cite{Yong13}, as well as the CEMP fractions of MW stars in the SAGA\footnote{\url{http://sagadatabase.jp/}} \citep{Suda2008, Suda2011, Yamada2013, Suda2017} database, after correcting their C and Fe abundances for non-Local Thermodynamic Equilibrium (non-LTE) effects (see \citealt{Koutsouridou2025}). However, as shown in Fig.~\ref{fig: MDF-CEMP}, different observational estimates of these trends are not always consistent. Our results are also in agreement with the observed metal-poor MDFs (Fig.~\ref{fig: MDF-CEMP}, bottom panel). Interestingly, we find no strong dependence of our predictions on the underlying merger history. For example, the peak of the normalized MDF lies within $-0.2\leq {\rm [Fe/H]} \leq 0.2$ across all merger trees, with a total variation of at most 0.8 dex. Consequently, given the current observational uncertainties, these diagnostics alone are unlikely to uniquely constrain the assembly history of our Galaxy.

\section{From Pop~III to second-generation stars}

Having validated {\sc NEFERTITI} against present-day MW observables (Sec.~\ref{Calibration}), we now use it to predict the properties of ancient stars in the MW and their host systems. For each Caterpillar merger tree we run five realizations to account for the stochasticity in the sampling of the masses, SN explosion energies and mixing efficiencies of Pop~III stars, as well as Pop~II/I stellar masses and rotation velocities (Secs.~\ref{SF-SSPs} and \ref{sec: metals}). Unless stated otherwise, in what follows we present results averaged over these realizations.

\subsection{Pop~III star formation across cosmic times}
\label{pop3_across_times}

\begin{figure}
\begin{center}
\includegraphics[width=0.99\hsize]{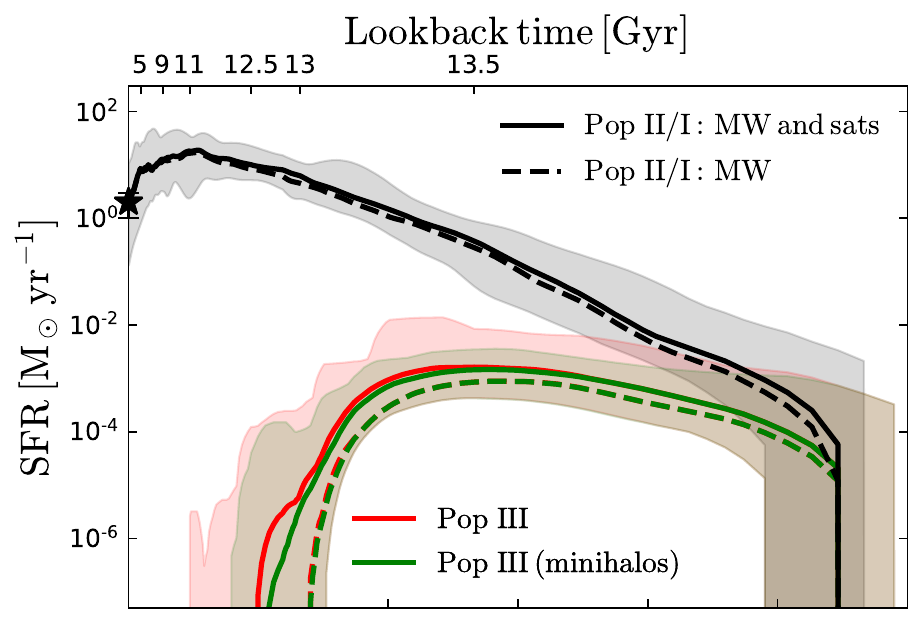}\\
\includegraphics[width=0.99\hsize]{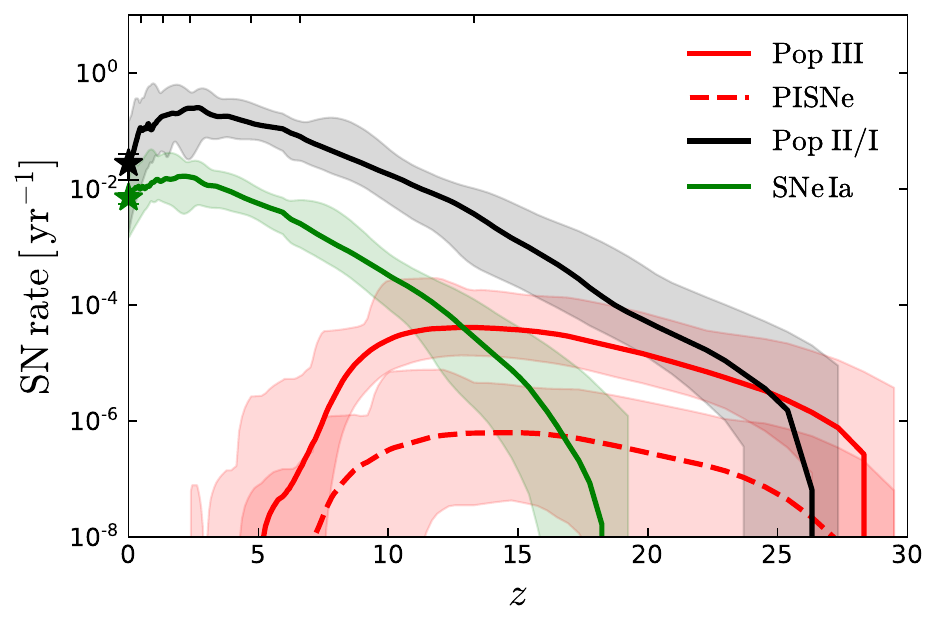}
\caption{Top: Median star formation rate across the 31 Caterpillar boxes, as a function of redshift, for Pop~II/I stars (black), all Pop~III stars (red) and Pop~III stars formed in minihalos ($T_{\rm vir}<2 \times 10^4\:$K; green). Solid lines show the SFR integrated over all halos in each volume, while dashed lines show only the SFR in the progenitors of the main halo. Bottom: Median SN rate across the 31 Caterpillar boxes, as a function of redshift, for Pop~II/I stars (black), SNe~Ia (green), all Pop~III stars (solid red) and Pop~III PISNe (dashed red line).
Shaded areas in both panels indicate the spread between the minimum and maximum values at each redshift among the simulations, considering all halos in each box. Star symbols at $z=0$ mark the observed present-day values for our Galaxy (Table~\ref{table: global_properties}).}
\label{fig: SFR}
\end{center}
\end{figure}

The top panel of Fig.~\ref{fig: SFR} displays the median SFR across the 31 Caterpillar boxes as a function of redshift. We find that Pop~III SF begins shortly after the start of the simulations, typically around $z=27$. Pop~II SF follows promptly and rises steadily down to $z=2.5$, reaching a peak of $17\: {\rm M_\odot\, yr^{-1}}$. Below this point, it gradually declines to the present-day value of $2\: {\rm M_\odot\, yr^{-1}}$. 

The median Pop~III SFR shows a broad peak around $z=10-15$ 
and then declines, reaching zero by $z \simeq 5$. At early times, Pop~III SF takes place exclusively in molecular cooling minihalos, and later extends to atomic-cooling halos in ionized regions. The latter are unaffected by the accretion suppression induced by IGM heating, and can continue forming Pop~III stars after reionization is complete (see Fig.~\ref{fig: filling}). We further find that Pop~III SF persists to lower redshifts in halos that survive as MW satellites (solid lines) than in those that eventually merge into the main halo (dashed lines in Fig.~\ref{fig: SFR}), which tend to reside in denser regions and are therefore more likely to be externally enriched. Typically, below $z\sim7.5$ (lookback times of $\lesssim13\:$Gyr) only less than $\sim10\%$ of Pop~III SF takes place in the MW progenitors, and this ceases entirely at $z<7$. However, there exist merger histories where Pop~III SF continues to later times: down to $z\sim4.8$ in MW progenitors and $z \sim 2.4$ in satellite progenitors (shaded areas) that remain in pristine environments.

The corresponding SN rates for Pop~III and Pop~II/I stars, the latter separated into ccSNe and Type~Ia SNe, are shown in the bottom panel of Fig.~\ref{fig: SFR}. We find that the first SNe~Ia typically begin exploding at $z\approx18$. Yet, their contribution to the MW MDF becomes significant only at $\rm[Fe/H]>-0.7$ (see Appendix~\ref{sec: SNIa}).
Given our adopted Pop~III IMF, the PISN rate accounts for only 2-3$\%$ of the total Pop~III SN rate at all redshifts, reaching a maximum around $z\approx 15$. Still, as we will see in the following Sections, their large metal yields and high explosion energies make them the dominant source of external enrichment at early times.

\begin{figure*}
\begin{center}
\includegraphics[width=0.495\textwidth]{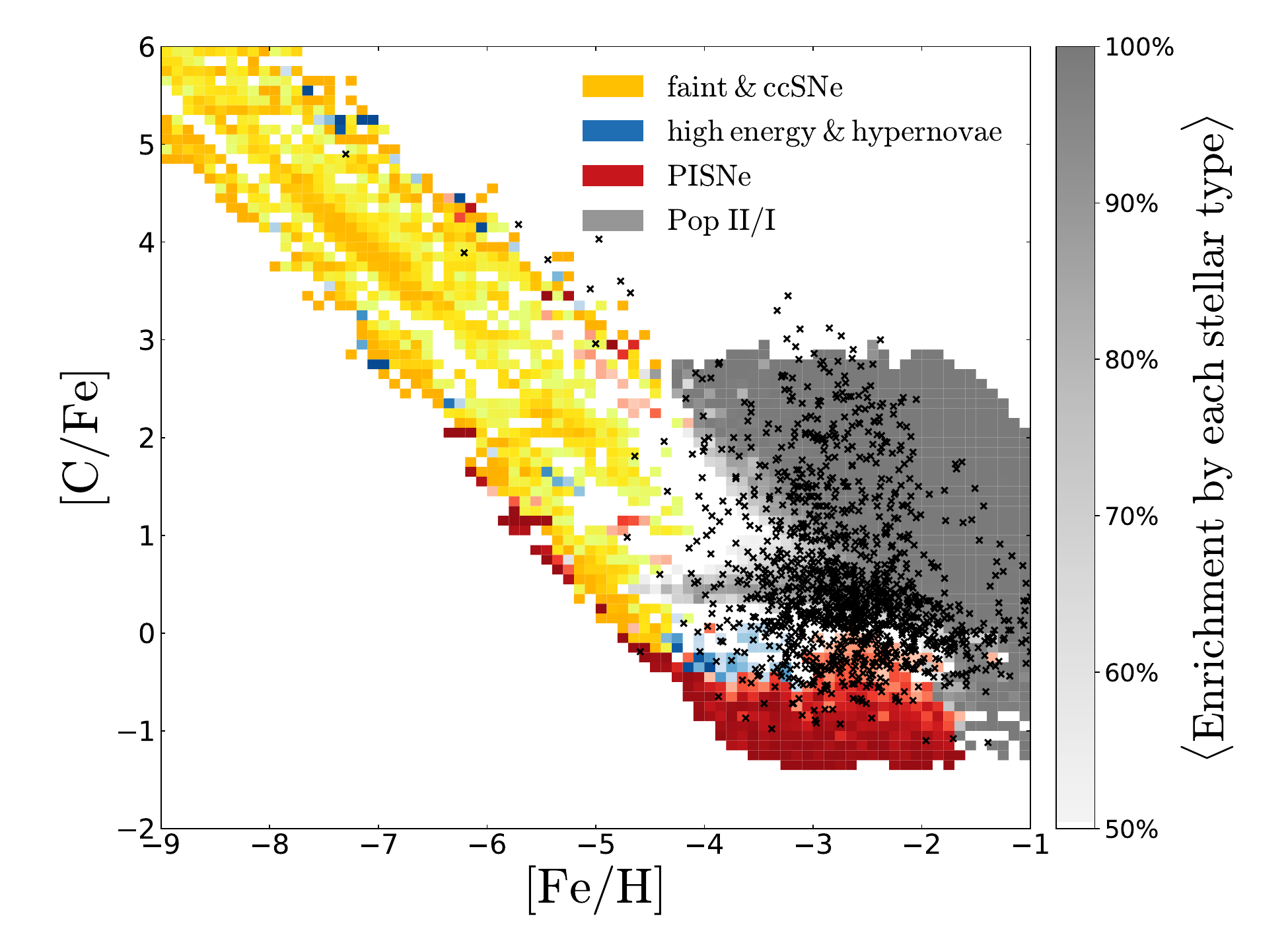}
\hfill
\includegraphics[width=0.495\textwidth]{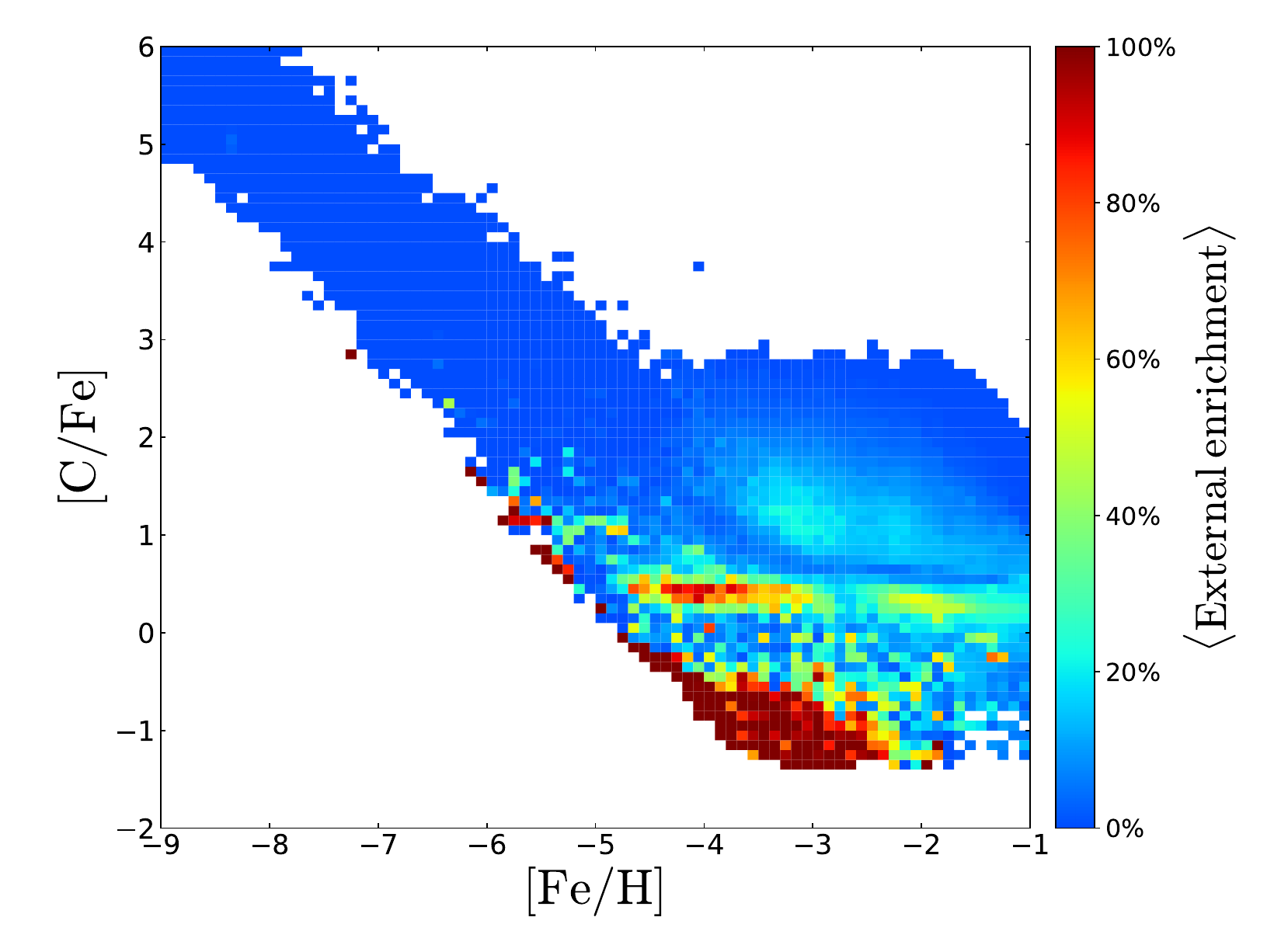} 
\\
\includegraphics[width=0.495\textwidth]{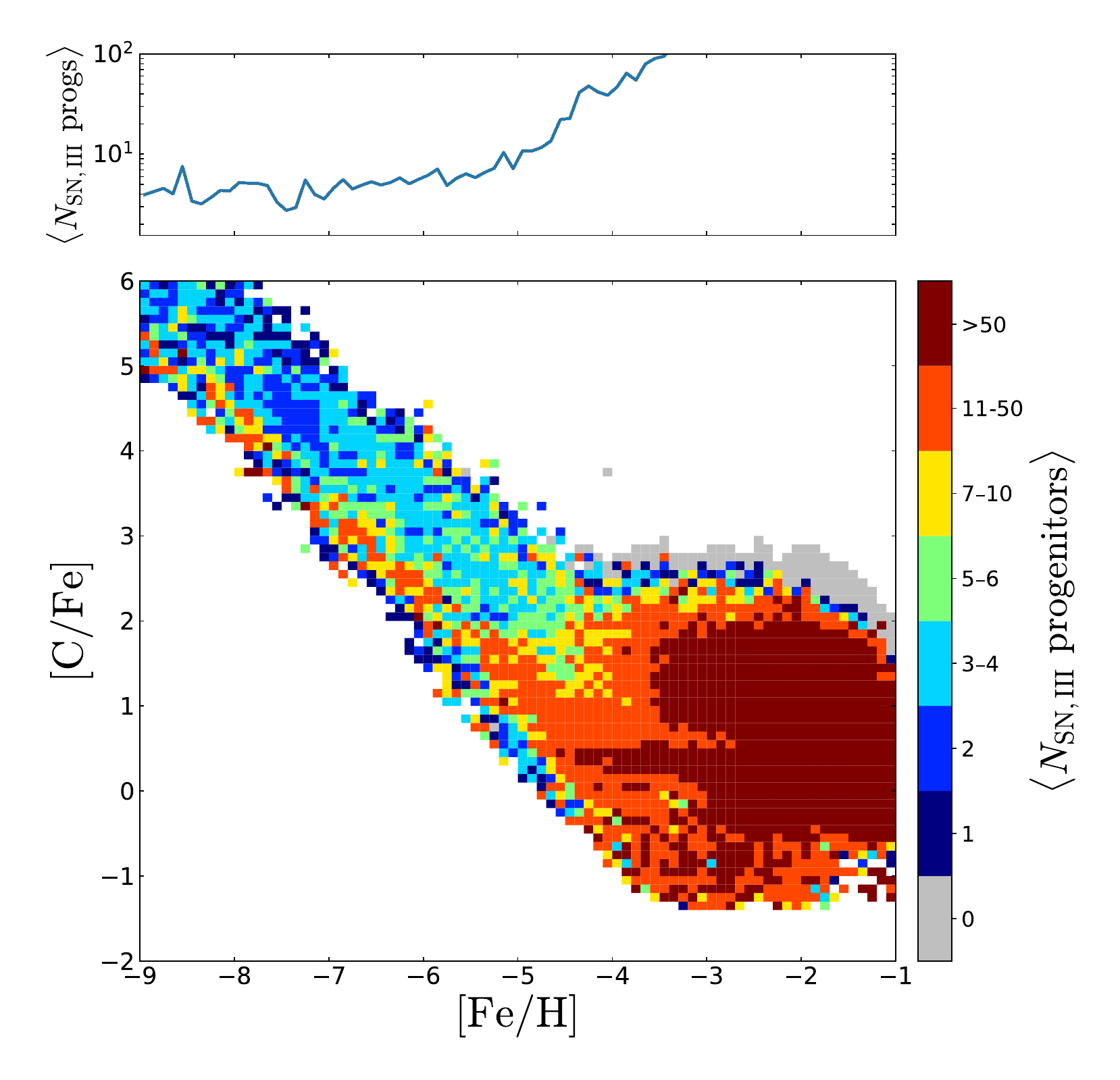}
\hfill
\includegraphics[width=0.495\hsize]
{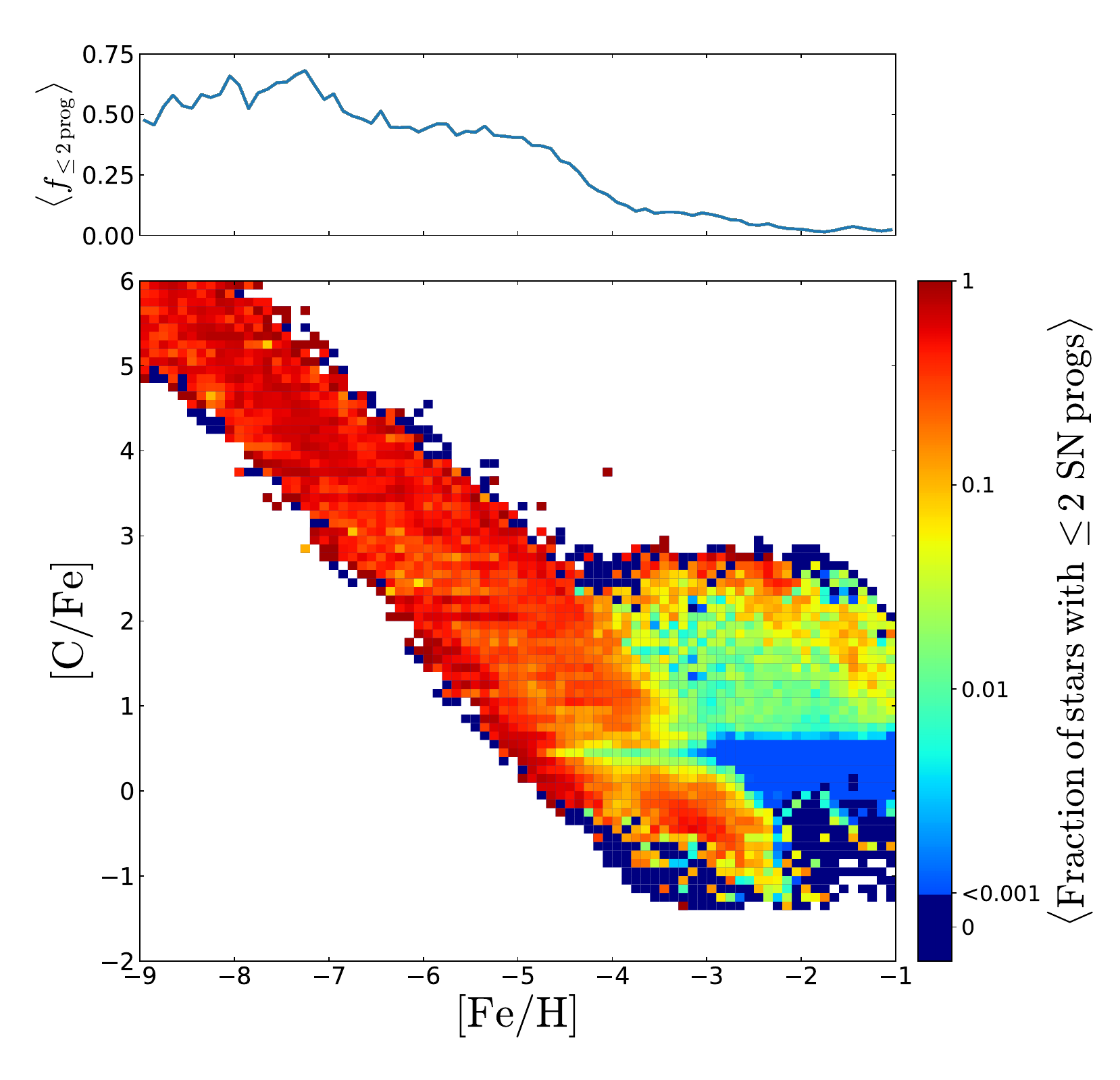}

\caption{Distribution of all metal-poor stars at $z=0$ in our MW analogues on the [C/Fe]--[Fe/H] diagram. Top left: bins where more than 50$\%$ of metals, on average, have been inherited by either Pop~III faint SNe and ccSNe (yellow), high energy SNe and hypernovae (blue), PISNe (red) or Pop~II/I SNe progenitors (including both ccSNe and Type Ia SNe; gray). Top right: fraction of metals contributed from external enrichment by neighbouring halos. Bottom left: average number of Pop~III SNe progenitors in each bin. Bottom right: fraction of stars in each bin that have only 2 or less SNe progenitors. All distributions represent the number-weighted average of the 31 Caterpillar merger trees (5 realizations for each tree). The datapoints in the top left panel show all SAGA MW data in LTE. }
\label{fig: CFe2}
\end{center}
\end{figure*} 

\subsection{Pop~III descendants in the Milky Way}

Fig.~\ref{fig: CFe2} presents the [C/Fe]--[Fe/H] distribution of all stars surviving to $z=0$ in our MW analogues. The top-left panel highlights regions where stars are primarily enriched by different types of Pop~III/II/I SNe. Note that here we show the number-weighted average enrichment; so each bin marks the most likely descendants in that part of the diagram. However, overlap exists, and stars enriched by other SN types can also be found in regions where a different source dominates. For example, there exist stars with $>50\%$ Pop~III enrichment in our realizations that reach ${\rm [Fe/H]}=-1.2$. For the full range of [C/Fe]--[Fe/H] spanned by the different descendants, see \citet{Vanni23}.

We find that the bulk of the observed C-normal population, as well as CEMP stars with $\rm[Fe/H]>-3$, lie in regions mainly enriched by PopII/I stars. In contrast, C-poor stars ($\rm[C/Fe]\lesssim0$) occupy regions associated with Pop~III high energy SNe, hypernovae and PISNe at ${\rm [Fe/H]\leq-2}$, and Pop~II Type Ia SNe at higher metallicities. This reflects the fact that theoretical yields for metal-poor Pop~II ccSNe rarely produce sub-solar [C/Fe] abundances (see discussion in \citealt{Koutsouridou2023}). However, when considering their full abundance patterns, these C-poor stars are not necessarily consistent with enrichment by those sources, indicating a tension between theoretical yields and observed abundances. For example, no PISN descendant has yet been unambiguously identified in the MW \citep{Koutsouridou2024,Skuladottir2024,Thibodeaux2024}.

The bottom-left panel of Fig.~\ref{fig: CFe2} shows the mean number of Pop~III SN progenitors in each bin. This has been computed including both the contribution of metals that have been produced internally and retained within the halo, as well as metals accreted from the halo's own bubble and/or neighbouring ones. 
Most observational studies infer the progenitors of metal-poor stars by matching their abundance patterns to theoretical SN yields, assuming enrichment by only one or two progenitors.
In our model, however, ultra metal-poor stars with [Fe/H]$\lesssim-4.5$---typically enriched by low-energy Pop~III SNe---generally have 3-20 SN progenitors. Toward higher metallicities, where PISNe and Pop~II descendants are found (Fig.~\ref{fig: CFe2}, top left), the number  of progenitors increases and exceeds 100 at $\rm[Fe/H]>-3.5$. This is because PISNe preferentially form in massive SF bursts that sample the Pop~III IMF more completely and therefore produce more SNe (see Fig.~\ref{fig: pop3_halos}), whereas Pop~II descendants trace later generations enriched by multiple preceding SN events. 

However, this does not mean that mono- or oligo-enriched (i.e., enriched just be a few progenitors) stars are absent from these regions. Indeed, the bottom-right panel of Fig.~\ref{fig: CFe2} shows that the probability of finding stars with only one or two SN progenitors (Pop~III and/or Pop~II) is in the range 30-70$\%$ at $\rm[Fe/H]<-4.5$ and then decreases to $<5\%$ at $\rm [Fe/H]>-2.5$. Moreover, even multi-enriched stars may have inherited most of their metals from only one or two main progenitors, in which case they would preserve their chemical imprint. This is often the case for PISNe descendants; despite typically having $>10$ contributing progenitors, they can show near-total PISN enrichment (up to $\sim$100\%) as even one PISN releases a huge amount of metals ($y_{\rm Z, PISNe} \approx 63-127\:{\rm M_\odot}$), far exceeding normal ccSN yields ($y_{\rm Z, Pop~III}\approx 1^{-10}-42  \:{\rm M_\odot}$ and $y_{\rm Z, Pop~II/I}\approx 0.63-8.5  \:{\rm M_\odot}$).

\begin{figure}
\begin{center}
\includegraphics[width=0.99\hsize]{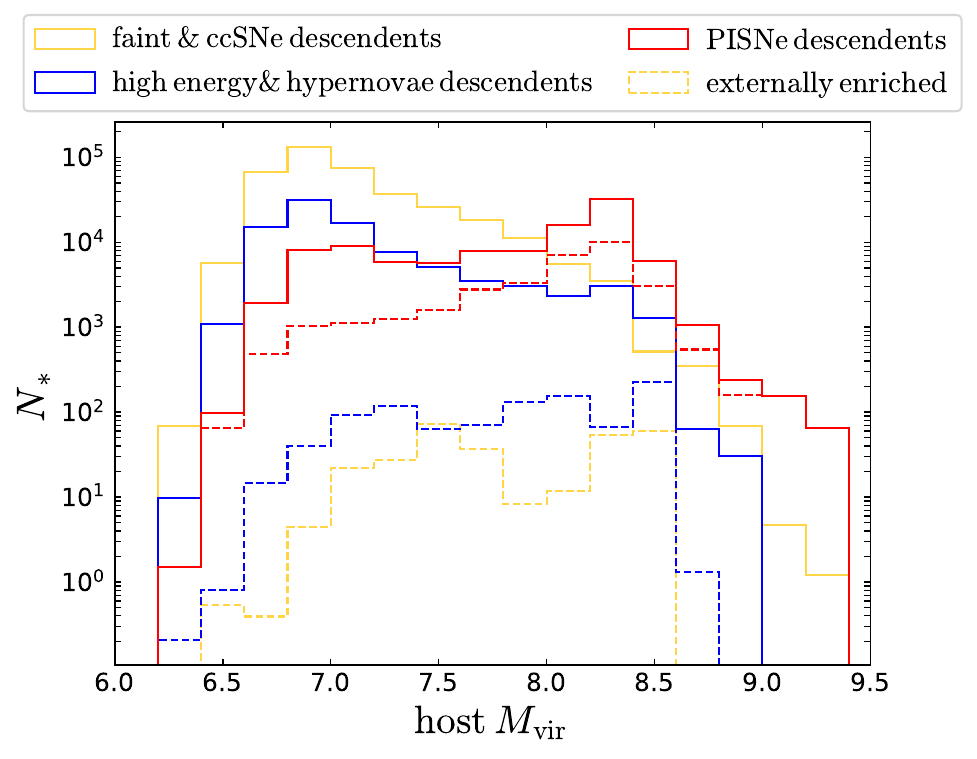}
\caption{Average number of \textit{pure} Pop~III descendants, i.e., stars inheriting $>90\%$ of their metals from Pop III progenitors, as a function of the mass of the host halo in which they formed. Different colors denote enrichment by faint and core-collapse SNe (yellow), high-energy SNe and hypernovae (blue), and PISNe (red). Dashed lines show the number of pure descendants for which  $>50\%$ of their metals came from external enrichment from neighboring halos.}
\label{fig: descendents}
\end{center}
\end{figure}

As shown in the top-right panel of Fig.~\ref{fig: CFe2}, regions dominated by PISN enrichment also tend to coincide with regions where external enrichment---i.e., metals originating from neighbouring halos---dominates, particularly at $\rm[Fe/H]\lesssim-2.5$.
Quantitatively, we find that $\sim30\%$ of \textit{pure} PISN descendants (defined here as stars inheriting $>90\%$ of their metals from PISNe) form in externally enriched halos with $M_{\rm vir}=10^{6.5-9.5}\:{\rm M_\odot}$ (Fig.~\ref{fig: descendents}).

Hydrodynamical simulations by \cite{Smith2015} and \cite{Hicks2021}, similarly find that external enrichment by metal-free stars is the dominant enrichment channel for halos with $M_{\rm vir}<10^6\:{\rm M_\odot}$, but even when the enriching events are ccSNe with $E_{\rm SN} \sim 10^{51}\:$erg. Since we do not form Pop~III stars in halos below $10^6\,{\rm M_\odot}$ (see Sec.~\ref{LW-Ionization} and Fig.~\ref{fig: pop3_halos}), our comparison is restricted to more massive systems; in this regime we do not find a strong external-enrichment signature for descendants of less energetic Pop~III SNe. In particular, the externally enriched fractions are only 0.08\%  for faint and ccSNe, and 1\% for high-energy SNe/hypernovae (Fig.~\ref{fig: descendents}). \textit{Pure} descendants of all Pop~III SN types form across virtually the full host-halo mass range. However PISNe descendants are the most numerous at the highest masses, $M_{\rm vir}\gtrsim 10^8\:{\rm M_\odot}$, while low-energy faint and ccSNe descendants dominate at lower halo masses.

\begin{figure}
\begin{center}
\includegraphics[width=0.99\hsize]{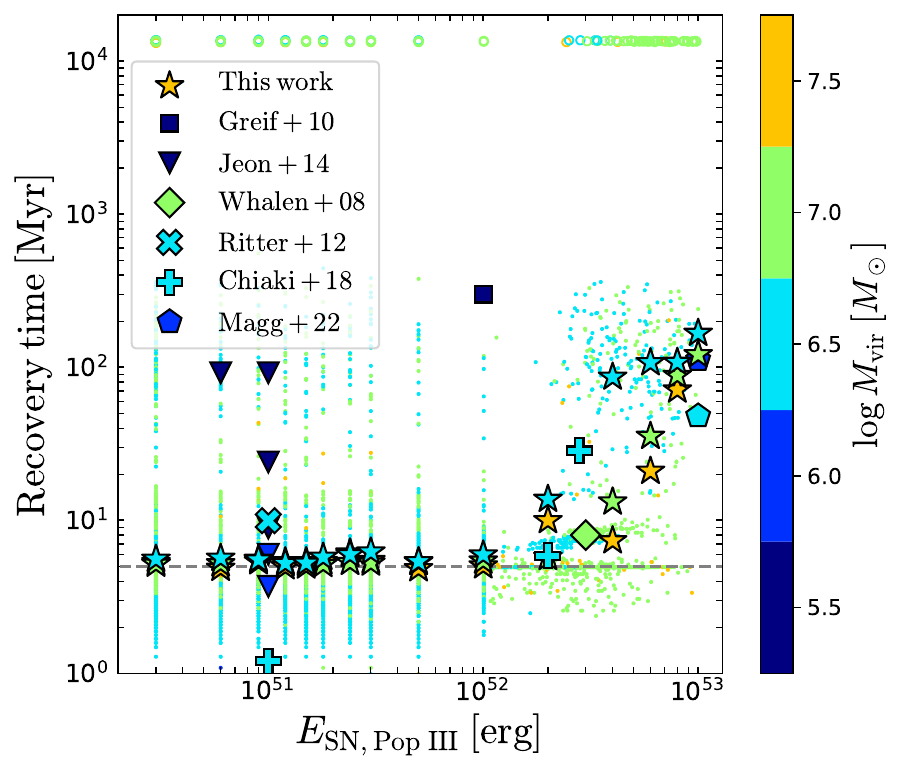}
\caption{Recovery time, i.e., time delay between the explosion of a Pop~III star and the formation of second generation stars in the same halo, as a function of the Pop~III explosion energy (colored dots), with color indicating the virial mass of the host halo. Mean values for each host halo mass and energy are shown with star symbols. Empty points at $>10^4\:$Myr indicate halos that never recovered, i.e., never formed second generation stars. The gray dashed line marks our adopted SF timestep of 5 Myr (see text). Squares, trianges, diamonds, X-marks, pluses and pentagons show the results of \cite{Greif2010, Jeon2014, Whalen2008, Ritter2012, Chiaki2018} and \cite{Magg2022}, respectively.}
\label{fig: Delay_time}
\end{center}
\end{figure}

The link between PISN descendants and external enrichment can be understood from the strong feedback of PISNe. A single PISN explosion releases such high energy ($10^{52-53}\:{\rm erg}$) that in many cases all gas is expelled from the host halo. At the same time, ionizing radiation from the massive progenitor(s) quenches star formation and supresses subsequent gas accretion. The neighbouring halos may also be affected by the same radiation (ionized bubbles expand faster than metal-enriched SN bubbles). However, as recombination takes over and the ionized bubble shrinks, neighboring halos may fall out of its radius, receive the enriched material and resume star formation earlier than the original host halo can recover.


This picture is consistent with the recovery time between a Pop~III SN explosion and the onset of second-generation star formation in the same halo. In Fig.~\ref{fig: Delay_time}, we show recovery times for halos where only one Pop~III SN occured before the onset of Pop~II SF, as a function of explosion energy and host virial mass. We see that across all SNe types, recovery times range from $\sim1\:$ to $>100\:$Myr (small datapoints). Yet for Pop~III stars with $m_\star \leq 100\: {\rm M_\odot}$ (corresponding to $E_{\rm SN}\leq 10^{52}\:$erg) they are on average $\sim$5 Myr for all  halo masses, equal to our SF timestep\footnote{Shorter recovery times arise because the SAM sub-timesteps must align with the (variable) timestep of the DM simulation; when the DM timestep is not an integer multiple of 5 Myr, the final SAM sub-step is shorter to match the remaining interval.} (Sec.~\ref{SF-SSPs}), meaning that in most cases the ISM is not evacuated and Pop~II SF proceeds immediately. Note, however, that these recovery times may be lower limits, since in our model SN feedback does not heat the retained gas.

Contrarily, when PISNe are involved, the mean recovery times increase to $\sim 70-200\:$Myr for the most energetic explosions, with the longest delays occuring in lower mass halos with shallower potential wells. Notice that for all types of Pop~III SNe, there exist cases where their host halos never recover (shown as empty points at recovery times $>10^4\:$Myr). These correspond to systems that lose all of their gas in the explosion and never accrete again, because, e.g., they remain within an ionized region of a neighboring halo or become satellites and stop growing (Sec.~\ref{Gas accretion}).


\section{Assembly of the metal-poor Milky Way}

\subsection{The Age-Metallicity Relation}

\begin{figure}
\begin{center}
\includegraphics[width=0.99\hsize]{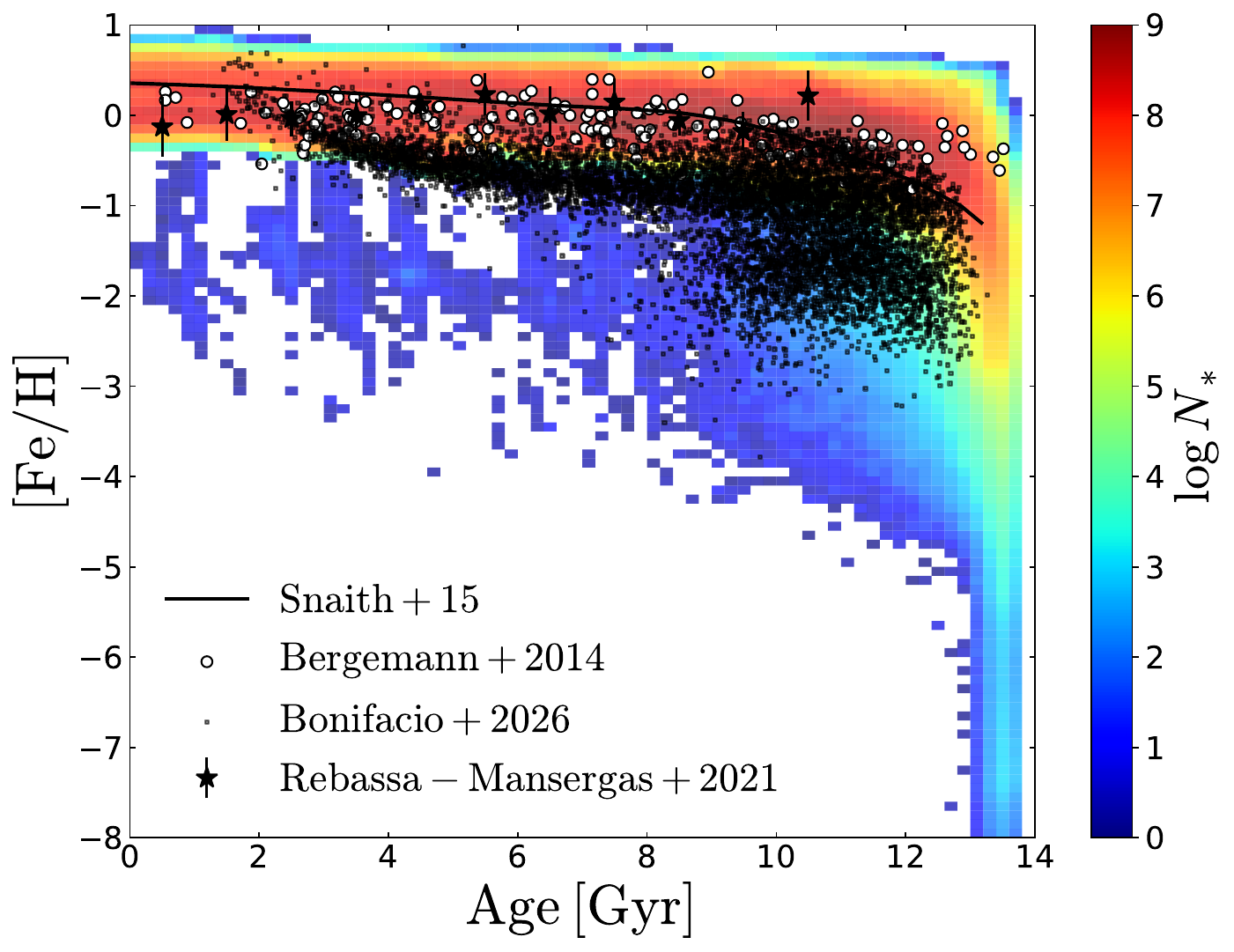}
\caption{Age-metallicity relation of all MW stars at $z=0$, averaged over the 31 Caterpillar merger trees, color-coded by the number of stars in each bin. Included are observational data from \citet[][white circles]{Bergemann2014} for Galactic disc stars (galactocentric distances 6-9 kpc and vertical distances from plane <1.5 kpc), \citet[][black stars]{Rebassa-Mansergas2021} for stars in the solar neighborhood, Bonifacio et al. (submitted, black dots) for disc and halo stars, 
and the AMR from \citet[][black line]{Snaith2015}, which provides the best fit to the thin and thick disc data of \citet{Haywood2013}.}
\label{fig: AGES}
\end{center}
\end{figure}

Fig.~\ref{fig: AGES} displays the age-metallicity relation (AMR) of all MW stars at $z=0$, averaged across the 31 Caterpillar merger trees. We find that the majority of stars, which formed along the MW's main progenitor branch (dark red area), exhibit a nearly flat AMR at lookback times $\lesssim 11-11.5 \:$Gyr ($z\lesssim2.5-3$),  spanning [Fe/H] abundances of approximately $\pm 0.7\:$dex around solar at all ages. This is consistent with the observed AMR for Galactic disc stars \citep{Bergemann2014, Snaith2015, Rebassa-Mansergas2021}.

Coeval stars with lower metallicities are formed in smaller halos, $\log (M_{\rm vir}/M_{\odot})\sim 8-11$, that are later accreted onto the main branch (see also Sec.~\ref{sec: origin metal-poor}).  These coincide with the low-metallicity AMR (at $\rm[Fe/H]\lesssim-1$) from Bonifacio et al. (submitted), which comprises stars dynamically classified as belonging to the thick disc and Galactic halo. Note that the Bonifacio et al. (submitted) sample is highly biased towards metal-poor stars, therefore their high AMR, consistent with thin disc kinematics, lies toward the lower edge of our predicted metallicity range for stars formed in the main progenitor.

\begin{figure}
\begin{center}
\includegraphics[width=0.99\hsize]{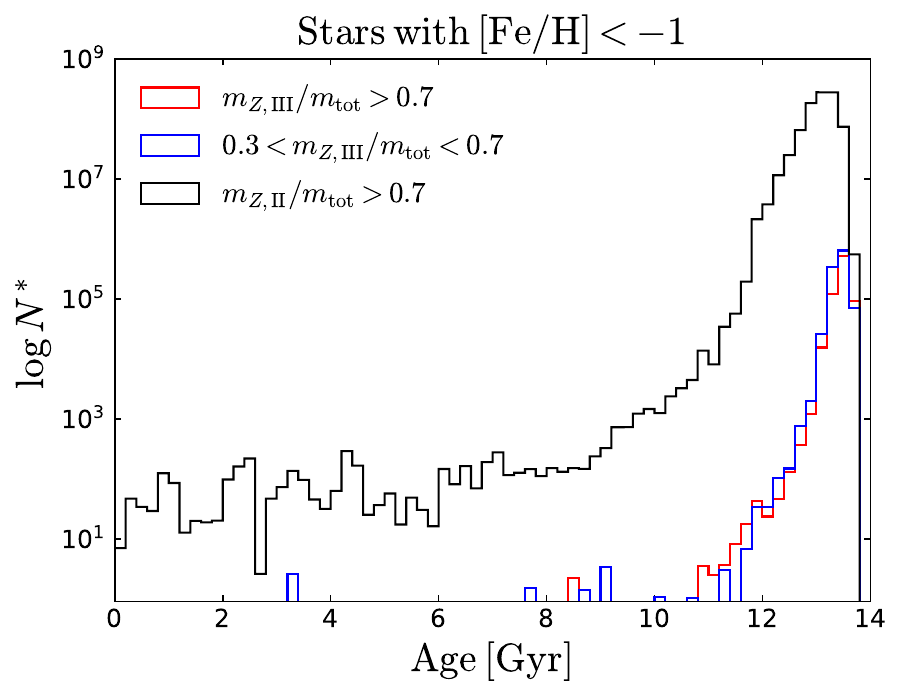}
\caption{Average age distribution of metal-poor stars (${\rm [Fe/H]<-1}$) with different fractions of Pop~III enrichment.}
\label{fig: hist_ages}
\end{center}
\end{figure}

Bonifacio et al. (submitted) report that the metallicity dispersion decreases drastically at ages younger than 8 Gyr; for younger ages their sample seems to be entirely composed of disc stars with [Fe/H]$\gtrsim-1$. 
In our model, a small number of stars with $-4 \lesssim {\rm [Fe/H]} \leq -1$ continue to form at lookback times $<8\:$Gyr. However, their contribution is negligible, corresponding to only $\sim10^{-6}$ per cent of the total metal-poor MW population, or $\sim10^{-7}$ per cent of the population with ages $<8 \:$Gyr.
These stars form in halos predominantly enriched by Pop~II stars (Fig.~\ref{fig: hist_ages}). Contrarily, metal-poor Pop~III descendants have typically ages $>11\:$Gyr, which is also shown in Fig.~\ref{fig: hist_ages}.

The main formation channel we identify for the young ($<8 \:$Gyr), Pop~II enriched metal-poor stars is the following. 
A minihalo that has retained some metal-enriched gas stops growing, and therefore its virial temperature decreases\footnote{Owing to the redshift dependence of the $M_{\rm vir}$--$T_{\rm vir}$ relation.}, strongly suppressing star formation ($\propto T_{\rm vir}^{3}$; Sec.~\ref{SF-SSPs}).
This halo then merges, at lower redshift, with a more massive but gas-devoid companion, whose gas reservoir has been removed either by photoevaporation in an ionized region or by SN feedback. Following the merger, the combined system reaches a higher virial temperature, allowing the metal-poor gas brought in by the first halo to resume forming stars.\\

\begin{figure}[ht]
    \centering
    \begin{minipage}{0.48\textwidth}
        \centering
        \includegraphics[width=\linewidth]{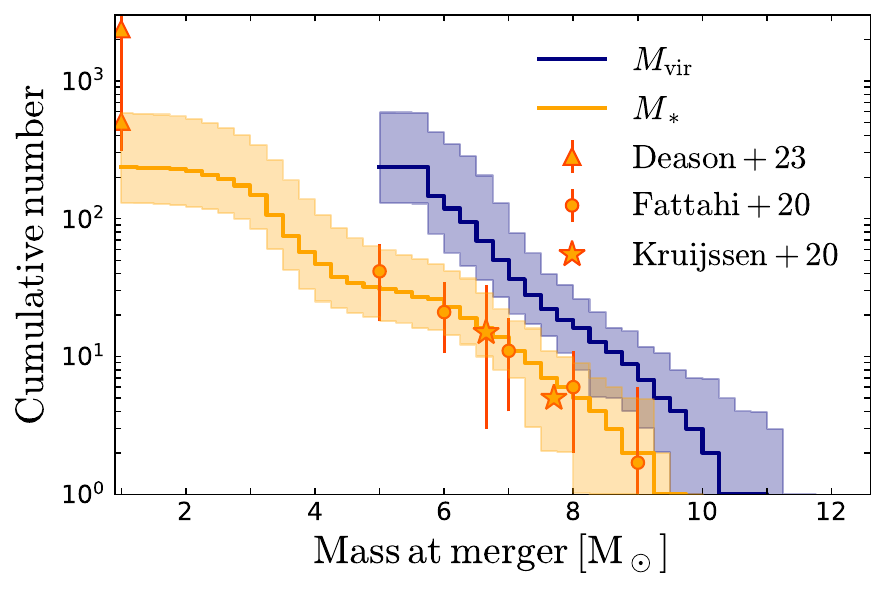}
        \caption{Median cumulative number of luminous ($M_*>0$) galaxies accreted onto the MW as a function of their stellar mass (orange) and virial mass (blue) at the time of merger. The shaded area indicates the full range spanned by the 31 merger trees. Included are the estimates of \cite{Kruijssen2019, Kruijssen2020}, \cite{Fattahi2020} and \cite{Deason2023}.}
        \label{fig: accreted}
    \end{minipage}
    \hfill
    \begin{minipage}{0.48\textwidth}
        \centering
        \includegraphics[width=\linewidth]{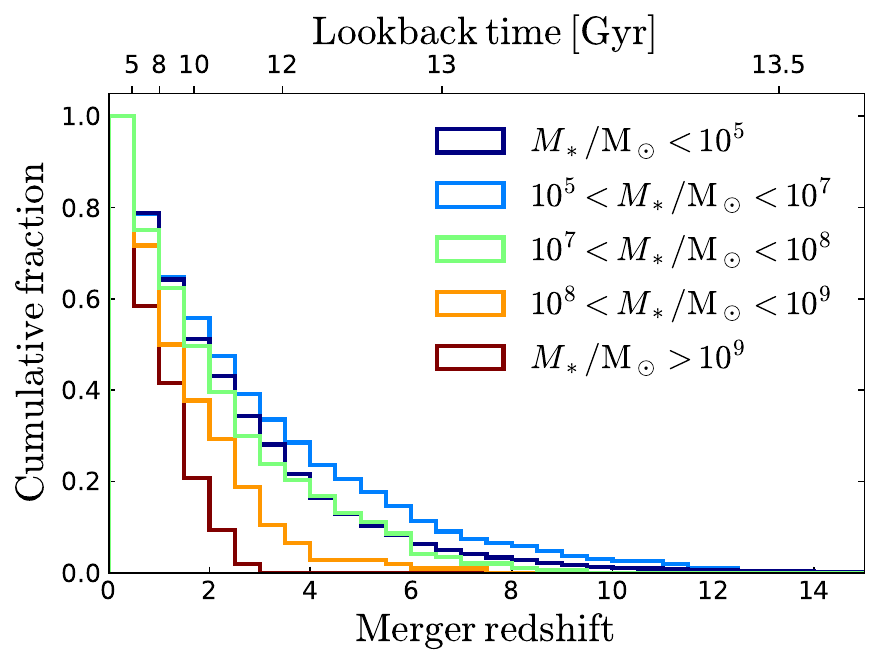}
        \caption{Cumulative fraction of accreted galaxies as a function of merger redshift, split by their stellar mass as indicated by the colour.}
        \label{fig: accreted_z}
    \end{minipage}
\end{figure}

\newpage

\subsection{The Milky Way Assembly History}
\label{sec: MW assembly}

In Fig.~\ref{fig: accreted}, we show the number of galaxies accreted onto the MW as a function of their virial (blue) and stellar mass (orange). Our stellar mass function is in good agreement with the results of \cite{Fattahi2020}, who used the Auriga simulations and with the estimates of \cite{Kruijssen2019,Kruijssen2020}, who reconstructed the MW merger history from its globular cluster population.
Overall, our MW analogues merge with 130-590 (median 237) luminous satellites (with stellar masses down to $\sim10\:{\rm M_\odot}$) during their formation history. \cite{Deason2023} who inferred the progenitor mass spectrum by modelling the Galactic halo MDF as a superposition of disrupted dwarf galaxies, found totals higher by a factor of $2-10$, depending on the assumed mass-metallicity relation for destroyed satellites. They cautioned, however, that their estimates at the low-mass end are likely biased high, and may decrease with larger samples and/or better  accounting of the metal-poor tails of massive progenitors.


In Fig.~\ref{fig: accreted_z} we show that the merger redshifts of the accreted galaxies in our simulations span a wide range, with lower-mass satellites typically being accreted earlier: 50$\%$ of the smallest ultra faint dwarfs (UFDs: $M_* <10^5\:{\rm M_\odot}$) and  dwarfs with $10^5<M_*/{\rm M_\odot} <10^8$, have already been accreted by $z=1.6$ and $z=1.7$, respectively, compared to $z\approx1$ for galaxies with $10^8<M_*/{\rm M_\odot} <10^9$ and $z\approx0.6$ for those with $M_* >10^9\:{\rm M_\odot}$. 

In 25 out of 31 merger trees, at least one merger with a galaxy of $M_*>5\times10^8,{\rm M_\odot}$ occurs at $z<1$ (the most massive accreted halo reaches $M_{\rm vir}=7.3\times10^{11},{\rm M_\odot}$, comparable to the MW virial mass). If the last massive merger of the MW was with GSE ($M_*\sim10^{8-9},{\rm M_\odot}$) at $z\sim1-2$ \citep[e.g.,][]{Helmi2018, Belokurov2018, Naidu2021}, then the recently quiet assembly history of the MW appears relatively rare among galaxies of similar mass ($\approx 20\%$). Considering additional evidence, such as the high stellar eccentricities in the Galactic halo \citep{Mackereth2019}, the ongoing infall of the Large Magellanic Cloud \citep{Evans2020, Buch2024}, and the possibility of earlier massive mergers (\citealt{Kruijssen2020}, the \textit{Kraken}; \citealt{Horta2021}, \textit{Heracles}) make the MW’s overall accretion history even more unusual.

\begin{figure}
\begin{center}
\includegraphics[width=0.99\hsize]
{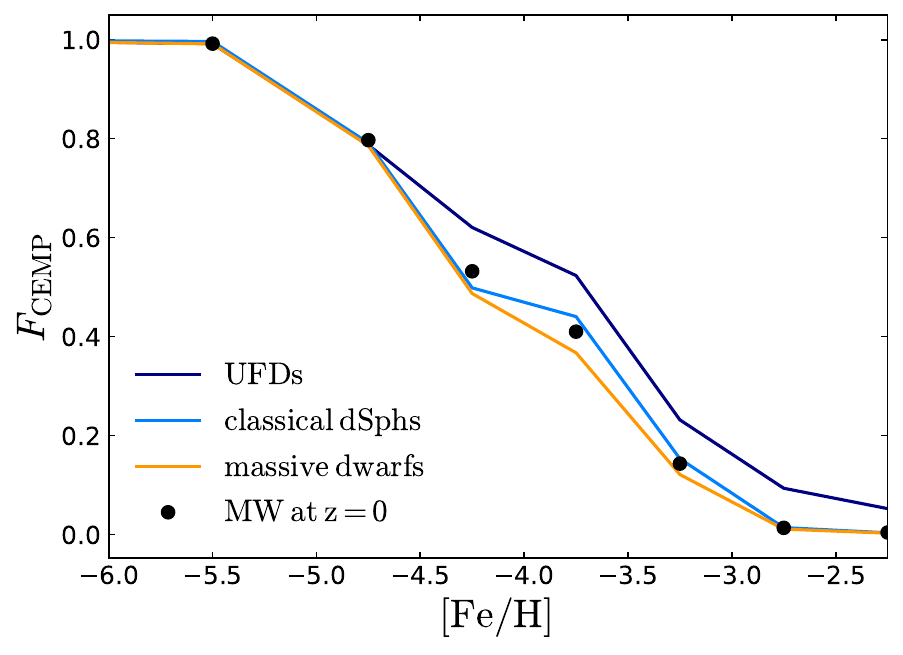}
\caption{Median CEMP fractions of galaxies that merged with the MW major branch. Colours indicate the stellar mass of the accreted galaxies at the time of merger: $M_*<10^5\:{\rm M_\odot}$ for UFDs, $10^5<M_*/{\rm M_\odot}<10^7$ for classical dSphs and $M_*>10^7\:{\rm M_\odot}$ for massive dwarfs. Black points show the median CEMP fraction of the 31 MW-analogues at $z=0$. Less massive accreted systems show higher CEMP fractions, similar to what is observed in surviving satellites of the MW (see text).}
\label{fig: accreted_Fcemp}
\end{center}
\end{figure}

\newpage

\subsection{Origin of the Metal-Poor Stars}
\label{sec: origin metal-poor}

\begin{figure}
\begin{center}
\includegraphics[width=0.99\hsize]{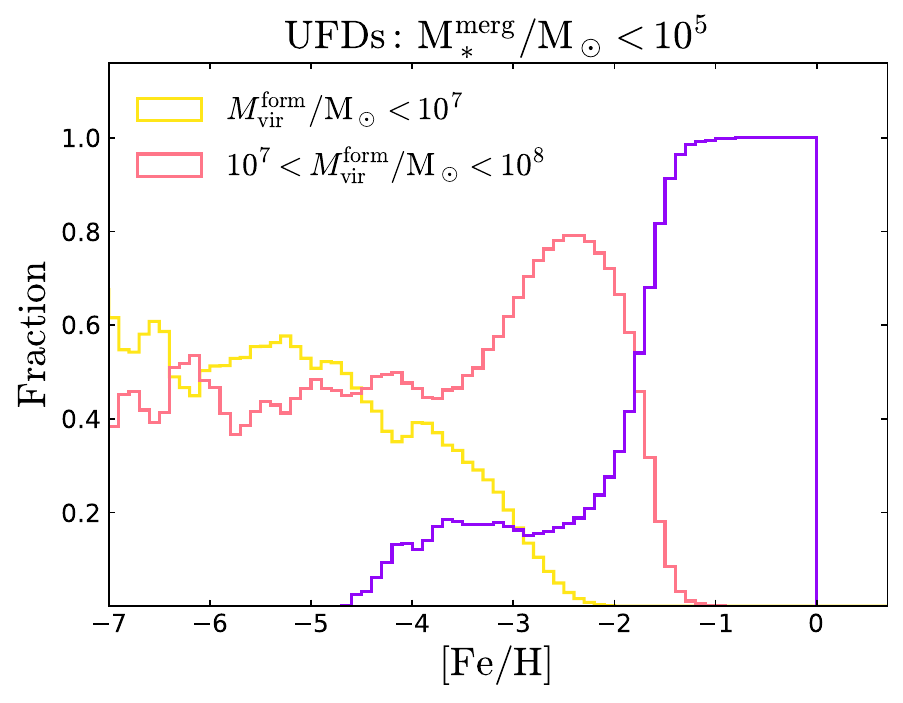}
\\
\includegraphics[width=0.99\hsize]{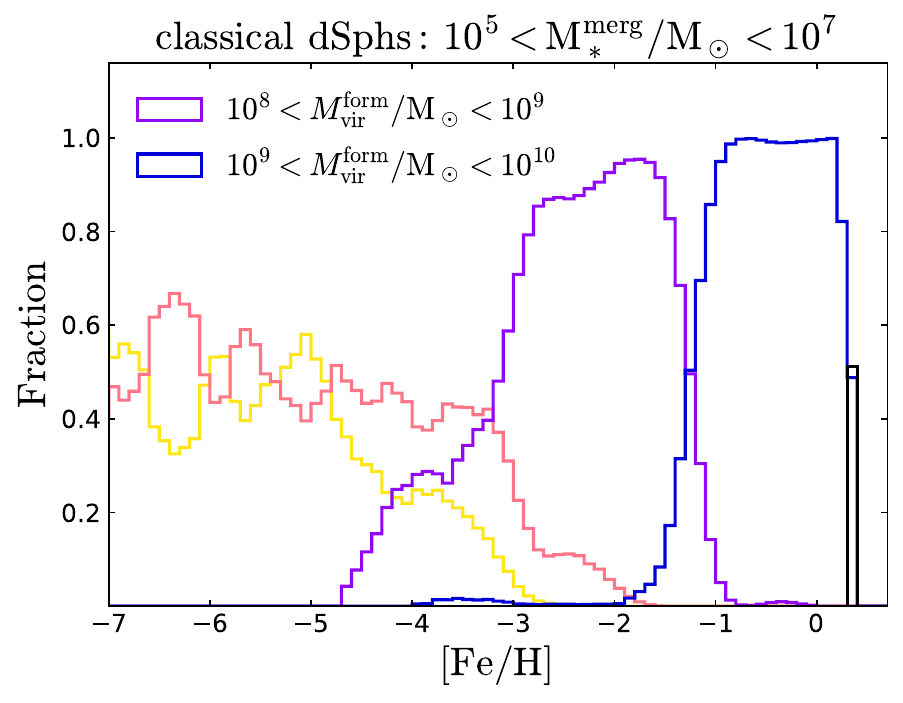} 
\\
\includegraphics[width=0.48\textwidth]{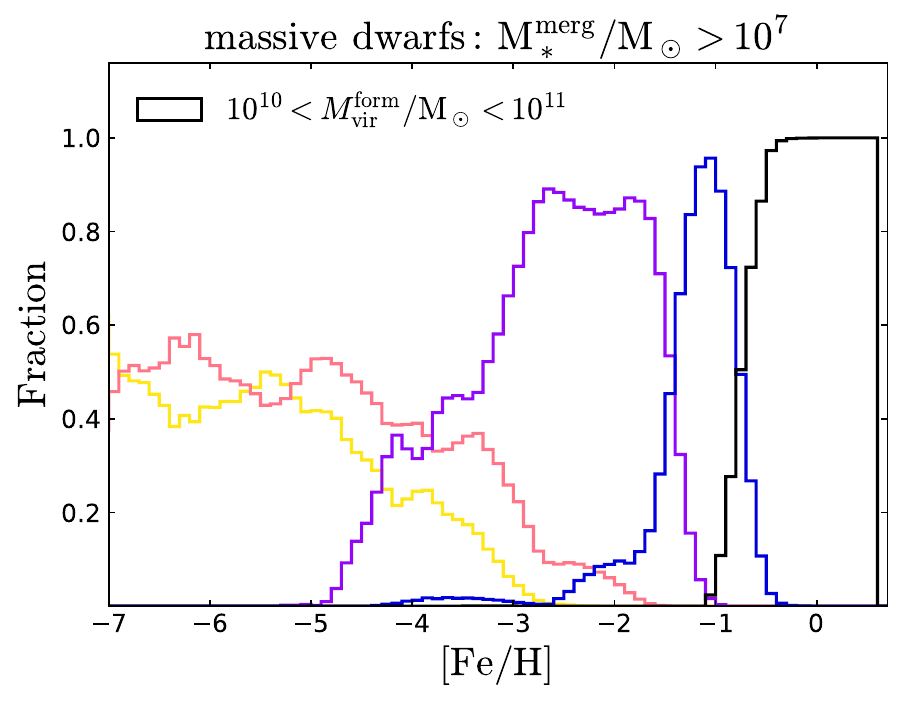}
\caption{Formation of stars in the accreted galaxies: analogues of UFDs (top), dSphs (middle), and massive dwarfs (bottom). Colored lines show the fraction of stars that formed in halos of different virial masses, $M_{\rm vir}^{\rm form}$, as a function of [Fe/H]. 
}
\label{fig: Mvir_form}
\end{center}
\end{figure}

Let us now focus on the chemical imprint of these accreted systems to the MW population.

Fig.~\ref{fig: accreted_Fcemp} shows the CEMP fractions, $F_{\rm CEMP}$, of accreted systems at the time of merger with the MW major branch. Although the trends converge at ${\rm[Fe/H]} \leq -4.75$, a dependence on galaxy mass appears at higher metallicities, with smaller systems exhibiting higher $F_{\rm CEMP}$. The difference is modest: for example at ${\rm [Fe/H]}=-3.5$ UFDs have CEMP fractions $\sim50 \%$ higher than those of massive, $M_*>10^7\:{\rm M_\odot}$, accreted dwarfs. The MW VMP population at $z=0$, exhibits an intermediate CEMP fraction, close to that of the most massive accreted systems. This trend, albeit more pronounced, has been observed among surviving UFDs, dSphs and the MW halo (see, e.g., Fig.~14 in \citealt{Lucchesi2024}).

To better understand the behavior in Fig.~\ref{fig: accreted_Fcemp}, it is necessary to consider that the accreted galaxies also have an assembly history of their own.
Fig.~\ref{fig: Mvir_form} shows where the stars in the accreted galaxies were formed, as a function of metallicity. Regardless of the stellar mass at the time of merger, all stars with $\rm[Fe/H]<-5$ are typically formed in similar environments, namely the lowest-mass DM halos: $\sim 50\%$ in halos with 
$M_{\rm vir}^{\rm form}<10^7\:{\rm M_\odot}$ and $\sim 50\%$ in halos with $10^7<M_{\rm vir}^{\rm form}/{\rm M_\odot}<10^8$. This explains why the CEMP fractions in Fig.~\ref{fig: accreted_Fcemp} converge at low $\rm[Fe/H]\leq-4.75$. At higher metallicities, however, the formation histories of the accreted systems diverge more clearly. For example, at $\rm[Fe/H]=-3.5$, the fraction of stars formed in halos with $M_{\rm vir}>10^8\:{\rm M_\odot}$, increases from 18$\%$ for UFD analogues, to 37$\%$ for classical dSphs, to 46$\%$ for the most massive accreted systems. Because more massive DM halos have deeper gravitational potential wells, they are better able to retain the heavy-element-rich ejecta of energetic Pop~III stars (hypernovae and PISNe; see Figs.~\ref{fig: descendents} and ~\ref{fig: Delay_time}, and also \citealt{Rossi2025}), which have low C yields and thereby exhibit lower CEMP fractions. As a consequence, the CEMP fraction in accreted galaxies of different masses diverges at higher metallicities, as is shown in Fig.~\ref{fig: accreted_Fcemp}.

\begin{figure}
\begin{center}
\includegraphics[width=0.99\hsize]
{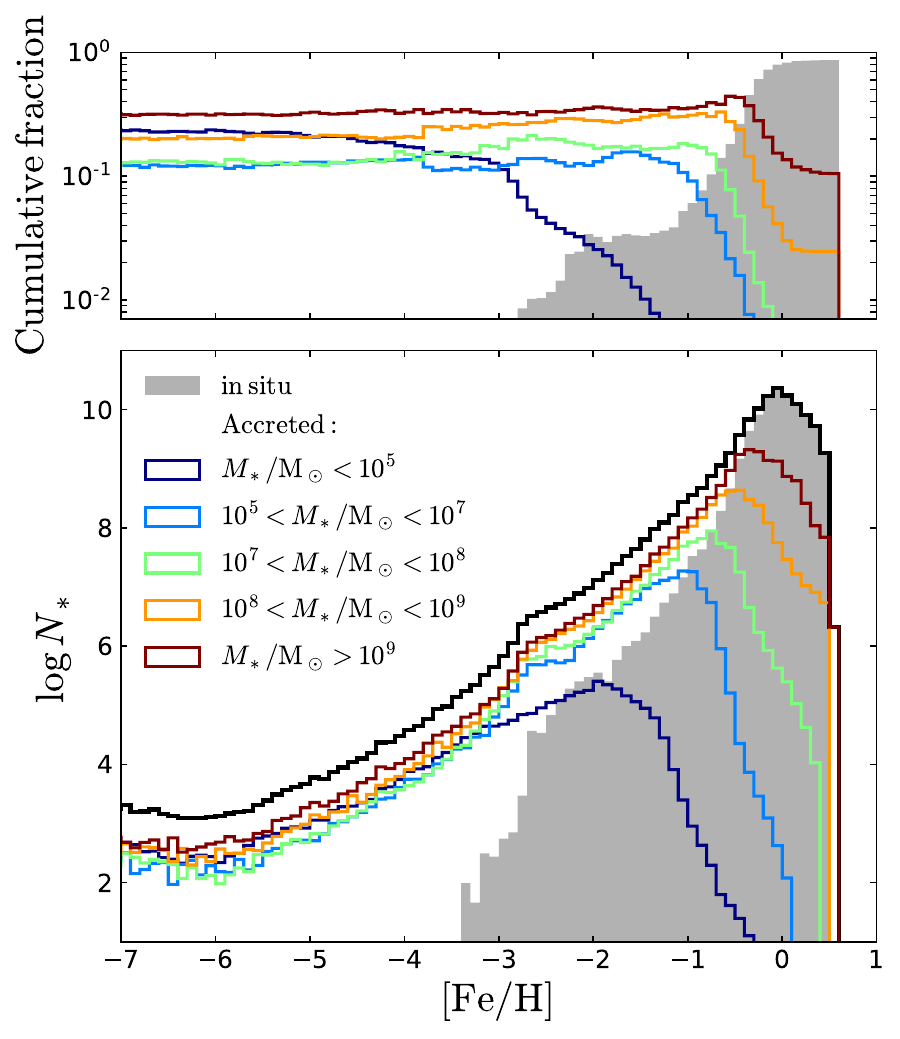}
\caption{Bottom: Median metallicity distribution function of the 31 MW analogues at $z=0$ (black line), along with the median contributions from stars formed in-situ (gray area) and accreted from mergers with galaxies of different stellar masses (coloured lines). Top: Cumulative fractional contribution of the in-situ stars and accreted components to the total MDF.}
\label{fig: accreted_MDFs}
\end{center}
\end{figure}

Finally, Fig.~\ref{fig: accreted_MDFs} illustrates the contribution of the merged galaxies to the total MDF of the MW.
A key result is that stars with $\rm [Fe/H]\lesssim -3.5$ form almost entirely in accreted galaxies, with the in situ contribution having a median of only $\sim 1\%$ across our MW analogues. Overall, stars formed in situ comprise 68--96$\%$ of all stars, with a median of 87$\%$, but account for only  $5^{+13}_{-4}\%$ of the metal-poor population. 

When the accreted component is split into different mass bins, a clear trend emerges: the lower the mass of the accreted galaxy, the lower the metallicity at which its MDF peaks (see also \citealt{Salvadori2015}). Nevertheless, the contribution of the most massive systems ($M_*>10^9\:{\rm M_\odot}$) dominates at all metallicities. 
Lower-mass dwarfs with $10^5<M_*/{\rm M_\odot}<10^7$ and $10^7<M_*/{\rm M_\odot}<10^8$ still make a substantial contribution to the metal-poor tail, providing 10$\%$ and 18$\%$ of the total, respectively at $\rm [Fe/H]<-1$ (top panel of Fig.~\ref{fig: accreted_MDFs}). UFDs ($M_*<10^5\:{\rm M_\odot}$), by contrast, become significant only below $\rm[Fe/H]=-3$, where their contribution rises above that of brighter dwarf galaxies with $10^5<M_*/{\rm M_\odot}<10^9$, from $\sim13\%$ to $20\%$ at $\rm[Fe/H]<-4.5$.

Our findings are compared with the previous works of \cite{Deason2016} and \cite{Fattahi2020} in Table~\ref{table: Contribution_mergers}.

\begin{table}[t]
\centering
\footnotesize
\caption{Fraction of MW stars with $\rm[Fe/H]\leq-2$ and $\rm[Fe/H]\leq-3$ that were accreted from mergers with galaxies of different stellar mass. Our predictions (median, maximum and minimum values) are compared with those of \cite{Deason2016} and \cite{Fattahi2020}; note that the former did not extend to stars with $\rm[Fe/H]<-3$, while the latter did not resolve UFDs ($M_*/{\rm M_\odot}<10^5)$.}
\tabcolsep=0.25cm
\renewcommand{\arraystretch}{1.3}
\begin{tabular}{l c c c}
\hline
\hline
\multicolumn{4}{c}{$\rm[Fe/H]\leq-2$}\\
\hline
Mass at merger & This work & Deason+16 & Fattahi+20 \\
\hline
$M_*/{\rm M_\odot}<10^5$ & $3^{+2}_{-2}\, \%$  & 2-5$\%$ & - \\
$10^5<M_*/{\rm M_\odot}<10^7$ & $12^{+18}_{-8} \,\%$ & \multirow{2}{*}{40-80$\%$} & $\leq9\%$\\
$10^7<M_*/{\rm M_\odot}<10^8$ & $18^{+25}_{-8}\,\%$ & & $\leq17\%$\\[3pt]
\hline
\hline
\multicolumn{4}{c}{$\rm[Fe/H]\leq-3$}\\
\hline
Mass at merger & This work & Deason+16 & Fattahi+20 \\
\hline
$M_*/{\rm M_\odot}<10^5$ &  $13^{+7}_{-6}\, \%$  & - & - \\
$10^5<M_*/{\rm M_\odot}<10^7$ & $11^{+43}_{-5} \,\%$ & - & $\leq8\%$\\
$10^7<M_*/{\rm M_\odot}<10^8$ & $17^{+26}_{-14} \,\%$ &  - & $\leq15\%$\\[3pt]
\hline
\end{tabular}
\label{table: Contribution_mergers}
\end{table}



\section{Implications for JWST findings}

\begin{figure}
\begin{center}
\includegraphics[width=0.99\hsize]{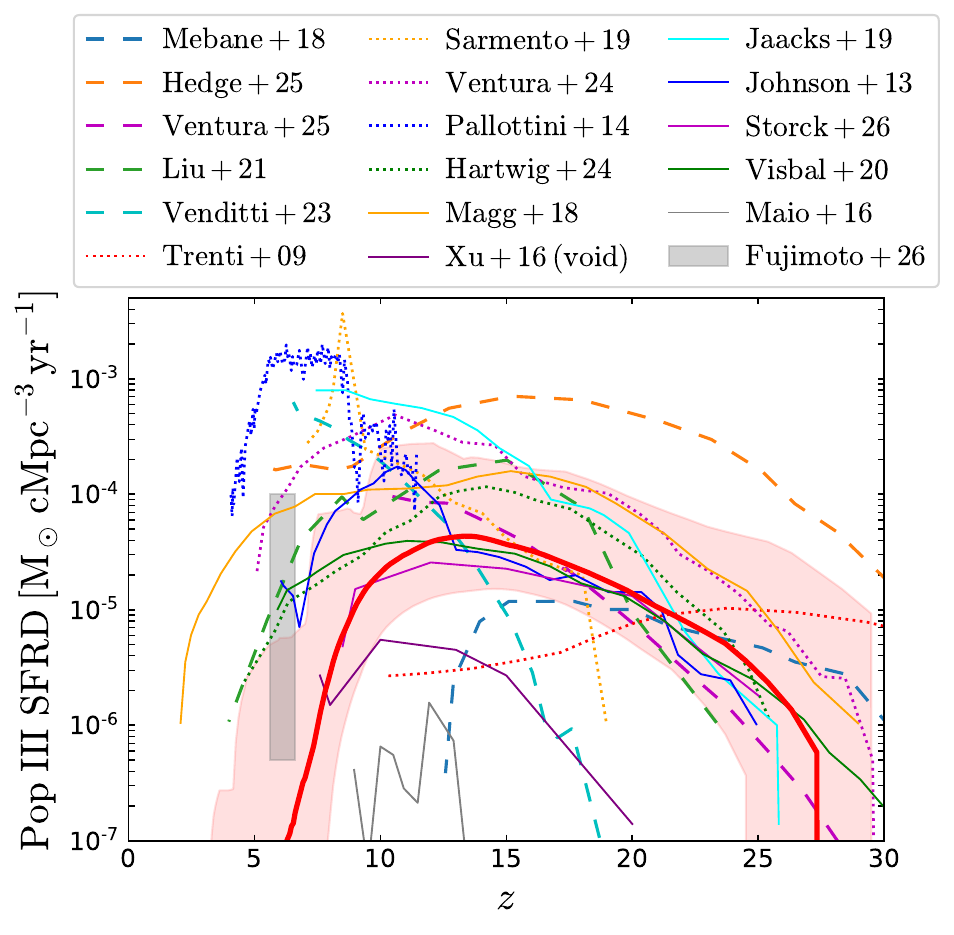}
\caption{Median comoving Pop~III SFR density in the 31 Caterpillar boxes (thick red line), compared with predictions from \cite{Mebane2018, Hegde2025, Ventura2025, Liu2021, Venditti2023, Trenti2009, Sarmento2019, Ventura2024, Pallottini2014, Hartwig2024, Magg2018, Xu2016, Jaacks2019,Johnson2013, Storck2026,Visbal2020} and \cite{Maio2016}, shown as thin lines and ordered by decreasing simulated volume; dashed lines denote cosmologically representative boxes ($L>50\:{\rm Mpc}/h$), dotted lines indicate intermediate volumes ($L=8-12\:{\rm Mpc}/h$) and solid lines small volumes ($L<6\:{\rm Mpc}/h$), the smallest being the one of \citep[$0.5\:{\rm Mpc}/h$]{Maio2016}. The observational constraint from  \cite{Fujimoto2025} is shown as a gray square.
Shaded areas indicate the full range spanned by our simulations at each redshift. All models, except ours and those of \cite{Liu2021} and \cite{Magg2018} were not run down to $z=0$.} 
\label{fig:SFRD}
\end{center}
\end{figure} 

Our locally calibrated {\sc NEFERTITI} model represents a powerful tool to study early galaxy formation processes, from the first stars to the present day, while accounting for the most relevant feedback effects shaping galaxy evolution. Once coupled with the 31 Caterpillar simulations of a MW-like analogue, we calibrated {\sc NEFERTITI} to reproduce a number of observed properties of our Galaxy, including those of ancient and metal-poor stars, such as the MDF and the fraction of CEMP-no stars (Fig.~\ref{fig: MDF-CEMP}). These observables are directly linked to the early stages of MW formation and are therefore highly sensitive to the uncertain rate of Pop~III star formation, their IMF, and the EDF of Pop~III SNe.

Armed with {\sc NEFERTITI}, which also naturally reproduces the observed age–metallicity relation, we can now compare our predictions with recent JWST observations that are providing an unprecedented glimpse into the early Universe.

\subsection{Pop III star formation rate density}

In Fig.~\ref{fig:SFRD}, we compare the evolution of the Pop III star formation rate density (SFRD) predicted by {\sc NEFERTITI} with existing theoretical models and a tentative observational constraint from JWST at $z\approx 6$ \citep{Fujimoto2025}. Due to the lack of direct observational constraints at high-z, the Pop III SFRD remains highly uncertain. Predictions from different studies, which probe a range of cosmic volumes and adopt different prescriptions for Pop III star formation, span a wide range of shapes and normalizations. Peak redshifts vary from $z=24$ to $z=8.5$ while peak SFRD values span more than three orders of magnitude, from $1.5 \times 10^{-6}\: {\rm M_\odot \, cMpc^{-3} \, yr^{-1}}$ to $\sim4\times 10^{-3}\: {\rm M_\odot \, cMpc^{-3} \, yr^{-1}}$.

The median SFRD derived with {\sc NEFERTITI} lies on the lower side of this range, reaching a peak value of $4.4\times 10^{-5}\: {\rm M_\odot \, cMpc^{-3} \, yr^{-1}}$ at $z=13.5$. As shown in Fig.~\ref{fig:SFRD}, our prediction is in closest agreement with the models of \cite{Johnson2013} and \cite{Visbal2020}
 down to $z \sim 12$ and then declines similarly to the simulations of \cite{Storck2026}, which also focus on a MW-like environment (see also \citealt{Katz2026}). Integrating over time, {\sc NEFERTITI} yields a median cumulative Pop~III stellar mass density of $1.2^{+0.5}_{-0.4} \times 10^4 \, {\rm M_\odot \,Mpc^{-3}}$  (plus/minus 5th-95th percentile).

Our simulated volumes are not representative of an average cosmic region, but rather of an overdense environment from which the MW may have formed. Although the \textit{total} SFRD is expected to be enhanced in such regions, the Pop~III SFRD need not be. In fact, it may be comparable to—or even lower than—the cosmic average, since dense environments are enriched earlier and therefore retain fewer pristine pockets at low redshift. More broadly, however, the comparison across models  in Fig.~\ref{fig:SFRD} reveals no clear correlation between Pop~III SFRD and the simulated volume, suggesting that the differences are mainly driven by  model assumptions rather than box size alone (see also discussion in \citealt{Venditti2023}).

In addition, our results exhibit significant scatter across different realizations, with the upper end reaching values consistent with the estimate at $z=6$ from \cite{Fujimoto2025}. The spread reflects both the diversity of MW assembly histories (31 merger trees) and the stochastic sampling of the Pop~III stellar properties. This highlights that the Pop~III SFRD is extremely sensitive not only to the simulated environment, but also to the Pop~III IMF and the EDF of Pop~III SNe, which regulate feedback processes—particularly metal enrichment—ultimately triggering the transition to Pop~II star formation and suppressing further Pop~III activity.

\subsection{Pop III galaxy candidates}
\begin{figure}
\begin{center}
\includegraphics[width=0.99\hsize]{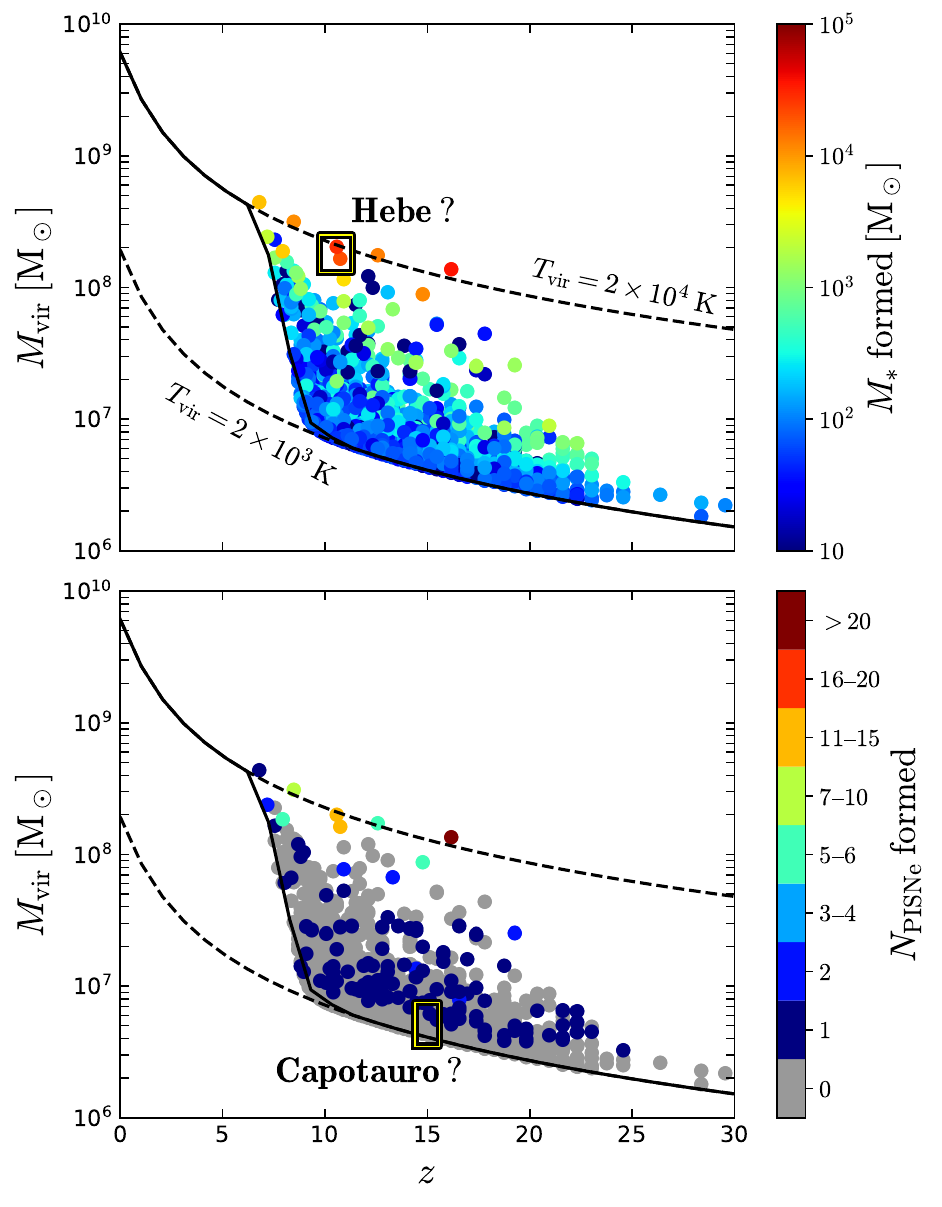}
\caption{Pop~III star-forming halos in one realization of a Caterpillar merger tree, as a function of virial mass and redshift.  The color denotes the total Pop~III stellar mass (top) and the number of Pop~III stars formed in the PISNe mass range (bottom). Note that the same halo can experience multiple Pop~III SF episodes. In this example, approximately $2\%$ of Pop~III galaxies across the full redshift range reach stellar masses $> 10^4\:{\rm M_\odot}$. The square in the top panel highlights two of those at $z = 10.7$ and $z = 10.6$, with total Pop~III stellar masses of $\approx 2 \times 10^4\,{\rm M_\odot}$ and $\approx 4 \times 10^4\,{\rm M_\odot}$, respectively, consistent with the redshift and inferred stellar mass of the Hebe galaxy \citep{Maiolino2026, Rusta2026}. The square in the bottom panel marks the range of virial mass and redshift inferred by \cite{Ferrara2026}, interpreting Capotauro as a PISN explosion.}
\label{fig: pop3_halos}
\end{center}
\end{figure} 

While the cosmic evolution of the Pop III SFRD remains highly uncertain, JWST is beginning to provide complementary constraints on primordial star formation from individual systems. The search for Pop III galaxies has so far yielded a few promising, though still unconfirmed Pop~III candidates at $z\sim5-7$ with inferred stellar masses ranging from $< 10^3$ to a few $10^{6}\:{\rm M_\odot}$ \citep{Vanzella2023, Morishita2025, Nakajima2025, Vanzella2026}. 
These systems, however, lack the strong He\,{\sc ii} emission lines expected for Pop~III stars \citep[e.g.,][]{Nakajima2022}, and most exhibit non-negligible metallicities. They may therefore represent extremely metal-poor Pop~II galaxies, hybrid Pop~III systems, or self-polluted Pop~III galaxies \citep[see definitions in][]{Rusta2025}. In this context, the recently reported source ``Hebe” located near GN-z11 \citep{Bunker2023} at $z = 10.6$, is of particular interest. It shows no detectable metal lines and a strong He\,II emission with $\rm EW(HeII\lambda1640) > 20$\,\AA\ \citep{Maiolino2026}, consistent with a pure Pop III system with a total stellar mass $M_* \approx (2-60)\times 10^4\:{\rm M_\odot}$ \citep{Rusta2026}.

In our model, accounting for inhomogeneous ionization enables Pop III star formation across a wide range of host halo masses already at $z\gtrsim 15$ (Fig.~\ref{fig: pop3_halos}). In ionized regions, Pop~III formation shifts to atomic-cooling halos ($T_{\rm vir}>2\times 10^4$~K), where star formation bursts are systematically stronger than in H$_2$-cooling minihalos, which typically form $< 10^4\:{\rm M_\odot}$ in Pop~III stars (Fig.~\ref{fig: pop3_halos}, top). However, massive Pop~III bursts can occasionally occur in halos below the Ly-$\alpha$ cooling threshold, as illustrated in the examples marked by the square in the top panel of Fig.~\ref{fig: pop3_halos}.
This can happen when previously sterile minihalos undergo episodes of rapid gas accretion, e.g.,~following mergers, or after exiting an ionized region or the virial radius of a larger host (Sec.~\ref{Gas accretion}). A similar outcome may also arise in the presence of enhanced Lyman–Werner radiation near a more massive galaxy (such as in the observed case of Hebe), where Pop III systems are predicted to reach stellar masses up to $M_* \leq 10^6 \:{\rm M_{\odot}}$ \citep{Jeong2026}. Although our model treats LW radiation as a uniform background, i.e., by imposing a minimum halo mass for Pop III star formation (solid black line in Fig.~\ref{fig: pop3_halos}), the resulting Pop III stellar masses are consistent with those found in simulations that include spatially inhomogeneous LW fields \citep[e.g.,][]{Storck2026, Jeon2026_hebe}. 

Overall, we find that 89$\%$ of our realizations\footnote{We remind the reader that for each merger tree we ran 5 realizations to account for the random sampling of stellar properties.} contain galaxies with total Pop~III stellar masses in the range $\approx (1-50)\times 10^{4}\:{\rm M_\odot}$, forming at $z\sim 18-7.5$. Pop~III systems with $M_* > 10^5\:{\rm M_\odot}$ are exclusively associated to massive (atomic-cooling) halos with $M_{\rm vir}\gtrsim 1.5 \times 10^8\:{\rm M_\odot}$ and are rarer, occurring in 10 out of 31 MW merger trees, and even in those cases, only in about one third of the corresponding realizations.
Ultimately, the Pop~III galaxy population in our Local Group progenitors spans a broad mass range $(10-5\times 10^5)\:{\rm M_\odot}$.
These results are fully consistent with the range of stellar masses inferred from current JWST observations, including both typical systems $M_* < 10^4\:{\rm M_\odot}$ and the rarer $M_* \approx (1-50)\times  10^4\:{\rm M_\odot}$ objects comparable to the "Hebe" galaxy at $z\approx 10.6$ \citep{Rusta2026}, which resides in the peculiar environment near GN-z11, one of the brightest high-z galaxies.


In the bottom panel of Fig.~\ref{fig: pop3_halos}, we show, for each Pop~III–hosting halo, the number of PISNe formed.
Minihalos typically host zero to two PISN progenitors (gray and blue points), while the most massive among them can host up to $\sim 15$ (green and orange points). In ionized regions, where Pop~III star formation shifts to atomic cooling halos, we find cases in which $> 20$ ($>100$) PISN progenitors form within SF bursts of $>10^4\:{\rm M_\odot}$ ($>10^5 \:{\rm M_\odot}$).
Naturally, these results depend sensitively on the adopted Pop III IMF (Eq.~\ref{eq:imf}).

Recently, \cite{Ferrara2026} proposed that the high-z source "Capotauro" could be a metal-free PISN explosion at $z\simeq 15$, rather than a luminous galaxy at $z\sim30$ \citep{Gandolfi2026}. In Fig.~\ref{fig: pop3_halos} (bottom panel) we highlight the corresponding redshift and virial-mass range for systems hosting such an event, as inferred by \cite{Ferrara2026}. This region is consistent with our predictions for the least massive minihalos hosting PISN progenitors, while in our model such progenitors can arise over a broad range of virial masses. Notably, $z \approx 15$ coincides with the epoch at which we predict the PISN rate to peak (see Sec.~\ref{pop3_across_times}).
Although theoretical models predict that such explosions are plausible \citep{Klessen23}, the distinctive chemical signatures of PISNe \citep{SS19} have not yet been unambiguously identified in present-day stellar relics \citep{Skuladottir2024,Thibodeaux2024}.
A confirmed detection of a PISN event—either direct, as a highly energetic explosion at high-$z$ \citep{Jeon2026}, or indirect, through an unambiguous chemical signature in present-day stellar relics \citep{Aguado2023} or more distant gas clouds \citep{Vanni2024}—would provide valuable insight into the nature of the first Pop~III stars \citep[e.g.,][]{Koutsouridou2024}.


\section{Conclusions}
We have developed the {\sc NEFERTITI} model of galaxy formation \citep{Koutsouridou2023, Koutsouridou2025} to self-consistently trace the Galaxy's assembly across cosmic times, from the formation of the first Pop~III stars to the present day. The model follows all individual Pop~III and metal-poor stars and now incorporates a treatment of inhomogeneous ionization and metal enrichment (Secs.~\ref{Ionization} and \ref{Inhomogeneous_mixing}), enabling a physically motivated description of early feedback and chemical enrichment. By coupling {\sc NEFERTITI} to the Caterpillar suite of 31 high-resolution simulations of Milky Way analogues \citep{Griffen2016, Griffen2018}, we resolve the minihalos that host the first stars and explore a wide range of plausible Milky Way assembly histories. Calibrated to reproduce both global Galactic properties (Table~\ref{table: global_properties}, Fig.~\ref{fig: AGES}) and key observables of ancient metal-poor stars, such as the MDF and CEMP fraction (Fig.~\ref{fig: MDF-CEMP}), {\sc NEFERTITI} provides a direct link between the earliest star-forming systems and the Milky Way stellar fossil record at $z=0$. Our main findings are as follows.\\


\noindent For Pop~III stars and their descendants:
\vspace{0.5em}
\begin{enumerate}
\item{Pop~III star-formation in our Milky Way volumes typically starts at $z=27$, reaches a broad maximum at $z=10-15$ with rates of about $\sim5 \times 10^3 \:{\rm M_\odot \:yr^{-1}}$, and  drops to zero by $z\sim5$ (Fig.~\ref{fig: SFR}).}

\item{Pop~III descendants span a wide metallicity range, from ${\rm [Fe/H] < -9}$ up to ${\rm [Fe/H] \approx -1}$.  The lowest metallicity regime (${\rm [Fe/H]}< -4$) is characterized by ${\rm [C/Fe]} > 0$, driven by the low-energy faint and ccSNe, whereas descendants of hypernovae and PISNe dominate at higher metallicities and show sub-solar carbonicities  (Fig.~\ref{fig: CFe2}).}

\item{While the average number of Pop~III progenitors for metal-poor stars increases from $\sim 4$ at ${\rm [Fe/H] < -6}$ to over $100$ at ${\rm [Fe/H] > -3.5}$, a significant fraction ($30$–$70\%$) of stars at ${\rm [Fe/H] < -4.5}$ are enriched by only one or two Pop~III SNe  (Fig.~\ref{fig: CFe2}).}

\item{Pure descendants of all Pop~III SN types form across a broad halo mass range, from our minimum star-forming halo mass of $M_{\rm vir}=10^6\,{\rm M_\odot}$ up to $\sim10^{9}\,{\rm M_\odot}$. Yet, PISNe descendants are the most numerous at the highest masses ($>10^8\:{\rm M_\odot}$), while faint and ccSNe descendants dominate in
lower-mass systems   (Fig.~\ref{fig: descendents}).}

\item{Energetic Pop~III explosions are associated with longer recovery times and more frequent external enrichment of neighboring halos. Roughly $30\%$ of pure PISN descendants form in externally enriched halos, whereas this occurs for only  1$\%$ of high-energy SNe/hypernovae descendants and a mere $0.1\%$ for those of faint and ccSNe (Fig.~\ref{fig: Delay_time}).}\\

\end{enumerate}

\noindent For the Milky Way assembly:
\vspace{0.5em}
\begin{enumerate}
\item{The metal-poor population is remarkably old; $99.9\%$ of ${\rm [Fe/H]<-1}$ stars at $z=0$ formed before $z=3$, making them older than $11.5$ Gyr (Fig.~\ref{fig: hist_ages}).}
\item{Our Milky Way analogues merge with 130-590 (median 237) luminous satellites during their lifetimes, spanning stellar masses down to $\sim10\:{\rm M_\odot}$. Lower-mass satellites are typically accreted earlier (Figs.~\ref{fig: accreted} and \ref{fig: accreted_z}).}
\item{While $\sim 90\%$ of the Milky Way's total stellar mass formed
\textit{in situ}, these stars represent only $5\%$ of the metal-poor population. Extremely metal-poor stars ($\rm [Fe/H] \lesssim -3.5$) formed almost exclusively ($99\%$) in accreted galaxies (Fig.~\ref{fig: accreted_MDFs}).}
\item{The contribution of the most massive ($M_*/{\rm M_\odot}>10^9$) accreted dwarfs dominates the Milky Way metal-poor tail at all metallicities. Analogues of classical dSphs ($10^5< M_*/{\rm M_\odot}<10^7\:$) contribute approximately 10$\%$ of stars at $\rm [Fe/H]=-1$, while the contribution from UFDs becomes increasingly important (13-20$\%$) at the extremely metal-poor end (Fig.~\ref{fig: accreted_MDFs}).}
\item{The chemical imprint of the accreted galaxies depends on their mass. UFDs form their very metal-poor stars in smaller dark matter halos than dSphs and more massive dwarfs (Fig.~\ref{fig: Mvir_form}). Because these shallower potential wells are less able to retain the ejecta of energetic Pop~III explosions, UFDs are preferentially enriched by low-energy SNe and therefore generally exhibit higher CEMP fractions. The Milky Way at $z=0$ shows an intermediate CEMP fraction, but closer to that of the most massive accreted systems, which contribute the majority of its metal-poor stars (Fig.~\ref{fig: accreted_Fcemp}).}\\

\end{enumerate}

\noindent For the implications for JWST and early galaxy formation:
\vspace{0.5em}
\begin{enumerate}
\item{Our predicted Pop~III star formation rate density peaks at $z=13.5$ with a median value of $4.4 \times 10^{-5}\: {\rm M_\odot \, cMpc^{-3} \, yr^{-1}}$, on the lower side of current theoretical predictions (Fig.~\ref{fig:SFRD}). Despite this, the upper tail of our model reaches values compatible with the tentative JWST observational constraint at $z=6$ reported by \cite{Fujimoto2025}.} 

\item{The large scatter in the Pop~III star formation rate density indicates that this quantity is highly sensitive to both the assembly history and the stochastic sampling of the Pop~III IMF, which sets the radiative, mechanical and chemical feedback of Pop III stars and regulates the transition to Pop~II star formation (Fig.~\ref{fig:SFRD}).}

\item{Pop~III stars form preferentially in halos with masses of $\sim 10^6-5 \times 10^8 \:{\rm M_\odot}$, initially in minihalos and later, typically at $z<14-17$, also in atomic-cooling halos within ionized regions. The resulting Pop~III stellar masses span a wide range ($10 - 5 \times 10^5 \: {\rm M_\odot}$), with typical values from a few tens to a few hundreds solar masses (Fig.~\ref{fig: pop3_halos}). Pop~III galaxies with $M_* \geq 10^4\:{\rm M_\odot}$ are nearly universal across our realizations (occurring at least once in $90\%$ of them), while the most massive ones ($M_*>10^5 \: M_\odot$) are rare ($\sim 10\%$ of realizations) and confined to atomic-cooling halos.} 

\item{The redshift and inferred stellar mass \citep{Rusta2026} of the Pop~III candidate "Hebe" observed by JWST \citep{Ubler2026, Maiolino2026}, align with those of pure Pop~III galaxies in our model, supporting the interpretation of "Hebe" as a pure Pop~III system (Fig.~\ref{fig: pop3_halos}, top).}

\item{The abundance of Pair-Instability SNe scales with host halo mass, typically ranging from zero to two in minihalos to over 20 in massive atomic-cooling halos (Fig.~\ref{fig: pop3_halos}, bottom). Our model predicts a peak in the PISN rate at $z \approx 15$ (Fig.~\ref{fig: SFR}), consistent with both the environment and redshift of a recently proposed PISN candidate \citep{Ferrara2026}.}\\
\end{enumerate}

In the upcoming years, the rapidly growing samples of metal-poor stars in the Milky Way and its satellite galaxies, enabled by the new generation of wide-field, spectroscopic surveys such as WEAVE \citep{Jin2024} and 4MOST \citep{Christlieb2019, Skuladottir2023}, will open a new window onto the initial stages of chemical enrichment and the build-up of galaxies like our own. These data, together with increasingly stringent constraints from JWST on the earliest galaxies and stellar populations, will allow us to probe the properties of Pop~III stars, the onset of galaxy formation, and the assembly of the Milky Way down to scales that remain inaccessible to current survey samples. 
As we move into this data-rich era, cosmological models like {\sc NEFERTITI} will be key to interpret these observations, providing the bridge between the unknown nature of the first stars and early galaxy formation to the stellar fossil record we now observe at $z=0$.

\section{Acknowledgments}.
\begin{acknowledgments}
This project has received funding from the European Research Council (ERC) under the European Union’s Horizon Europe research and innovation programme (grant agreements No. 101117455 [TREASURES, PI: Skúladóttir] and No. 101221278 [OUTLIERS, PI: Agudao]) and the Horizon 2020 research and innovation programme (grant agreement No. 804240 [NEFERTITI, PI: Salvadori]). D.A. also acknowledges financial support from the Spanish Ministry of Science and Innovation (MICINN) under the 2021 Ramón y Cajal program MICINN RYC2021-032609. We thank Alexander P. Ji for providing the Caterpillar simulation data and for helpful discussions.
\end{acknowledgments}

%




\bibliography{mybib}{}

@ARTICLE{Vanni2024,
       author = {{Vanni}, Irene and {Salvadori}, Stefania and {D'Odorico}, Valentina and {Becker}, George D. and {Cupani}, Guido},
        title = "{Chemical Diagnostics to Unveil Environments Enriched by First Stars}",
      journal = {\apjl},
         year = 2024,
        month = jun,
       volume = {967},
       number = {2},
          eid = {L22},
        pages = {L22},
          doi = {10.3847/2041-8213/ad46fa},
archivePrefix = {arXiv},
       eprint = {2402.18640},
 primaryClass = {astro-ph.GA},
       adsurl = {https://ui.adsabs.harvard.edu/abs/2024ApJ...967L..22V}
}

@ARTICLE{Jeong2026,
       author = {{Jeong}, Tae Bong and {Venditti}, Alessandra and {Bromm}, Volker and {Jeon}, Myoungwon and {Hsiao}, Tiger Yu-Yang and {Finkelstein}, Steven L. and {Chisholm}, John},
        title = "{How Massive Can a Population III Starburst Be? Simulating the First Galaxies with High Lyman-Werner Background}",
      journal = {arXiv e-prints},
         year = 2026,
        month = mar,
          eid = {arXiv:2603.23209},
        pages = {arXiv:2603.23209},
          doi = {10.48550/arXiv.2603.23209},
archivePrefix = {arXiv},
       eprint = {2603.23209},
 primaryClass = {astro-ph.GA},
       adsurl = {https://ui.adsabs.harvard.edu/abs/2026arXiv260323209J}
}

@ARTICLE{Bromm2013,
       author = {{Bromm}, Volker},
        title = "{Formation of the first stars}",
      journal = {Reports on Progress in Physics},
         year = 2013,
        month = nov,
       volume = {76},
       number = {11},
          eid = {112901},
        pages = {112901},
          doi = {10.1088/0034-4885/76/11/112901},
archivePrefix = {arXiv},
       eprint = {1305.5178},
 primaryClass = {astro-ph.CO},
       adsurl = {https://ui.adsabs.harvard.edu/abs/2013RPPh...76k2901B}
}

@ARTICLE{Jean-Baptiste2017,
       author = {{Jean-Baptiste}, I. and {Di Matteo}, P. and {Haywood}, M. and {G{\'o}mez}, A. and {Montuori}, M. and {Combes}, F. and {Semelin}, B.},
        title = "{On the kinematic detection of accreted streams in the Gaia era: a cautionary tale}",
      journal = {\aap},
         year = 2017,
        month = aug,
       volume = {604},
          eid = {A106},
        pages = {A106},
          doi = {10.1051/0004-6361/201629691},
archivePrefix = {arXiv},
       eprint = {1611.07193},
 primaryClass = {astro-ph.GA},
       adsurl = {https://ui.adsabs.harvard.edu/abs/2017A&A...604A.106J}
}

@ARTICLE{Khoperskov2023,
       author = {{Khoperskov}, Sergey and {Minchev}, Ivan and {Libeskind}, Noam and {Haywood}, Misha and {Di Matteo}, Paola and {Belokurov}, Vasily and {Steinmetz}, Matthias and {Gomez}, Facundo A. and {Grand}, Robert J.~J. and {Hoffman}, Yehuda and {Knebe}, Alexander and {Sorce}, Jenny G. and {Spaare}, Martin and {Tempel}, Elmo and {Vogelsberger}, Mark},
        title = "{The stellar halo in Local Group Hestia simulations. II. The accreted component}",
      journal = {\aap},
         year = 2023,
        month = sep,
       volume = {677},
          eid = {A90},
        pages = {A90},
          doi = {10.1051/0004-6361/202244233},
archivePrefix = {arXiv},
       eprint = {2206.04522},
 primaryClass = {astro-ph.GA},
       adsurl = {https://ui.adsabs.harvard.edu/abs/2023A&A...677A..90K}
}

@ARTICLE{Cunningham2022,
       author = {{Cunningham}, Emily C. and {Sanderson}, Robyn E. and {Johnston}, Kathryn V. and {Panithanpaisal}, Nondh and {Ness}, Melissa K. and {Wetzel}, Andrew and {Loebman}, Sarah R. and {Escala}, Ivanna and {Horta}, Danny and {Faucher-Gigu{\`e}re}, Claude-Andr{\'e}},
        title = "{Reading the CARDs: The Imprint of Accretion History in the Chemical Abundances of the Milky Way's Stellar Halo}",
      journal = {\apj},
         year = 2022,
        month = aug,
       volume = {934},
       number = {2},
          eid = {172},
        pages = {172},
          doi = {10.3847/1538-4357/ac78ea},
archivePrefix = {arXiv},
       eprint = {2110.02957},
 primaryClass = {astro-ph.GA},
       adsurl = {https://ui.adsabs.harvard.edu/abs/2022ApJ...934..172C}
}

@ARTICLE{Katz2026,
       author = {{Katz}, Harley and {Rey}, Martin P. and {Cadiou}, Corentin and {Kimm}, Taysun and {Agertz}, Oscar},
        title = "{The Impact of Star Formation and Feedback Recipes on the Stellar Mass and Interstellar Medium of High-Redshift Galaxies}",
      journal = {The Open Journal of Astrophysics},
         year = 2026,
        month = feb,
       volume = {9},
        pages = {56097},
          doi = {10.33232/001c.156097},
archivePrefix = {arXiv},
       eprint = {2411.07282},
 primaryClass = {astro-ph.GA},
       adsurl = {https://ui.adsabs.harvard.edu/abs/2026OJAp....956097K}
}

@ARTICLE{Monachesi2019,
       author = {{Monachesi}, Antonela and {G{\'o}mez}, Facundo A. and {Grand}, Robert J.~J. and {Simpson}, Christine M. and {Kauffmann}, Guinevere and {Bustamante}, Sebasti{\'a}n and {Marinacci}, Federico and {Pakmor}, R{\"u}diger and {Springel}, Volker and {Frenk}, Carlos S. and {White}, Simon D.~M. and {Tissera}, Patricia B.},
        title = "{The Auriga stellar haloes: connecting stellar population properties with accretion and merging history}",
      journal = {\mnras},
         year = 2019,
        month = may,
       volume = {485},
       number = {2},
        pages = {2589-2616},
          doi = {10.1093/mnras/stz538},
archivePrefix = {arXiv},
       eprint = {1804.07798},
 primaryClass = {astro-ph.GA},
       adsurl = {https://ui.adsabs.harvard.edu/abs/2019MNRAS.485.2589M}
}

@ARTICLE{DSouza2018,
       author = {{D'Souza}, Richard and {Bell}, Eric F.},
        title = "{The masses and metallicities of stellar haloes reflect galactic merger histories}",
      journal = {\mnras},
         year = 2018,
        month = mar,
       volume = {474},
       number = {4},
        pages = {5300-5318},
          doi = {10.1093/mnras/stx3081},
archivePrefix = {arXiv},
       eprint = {1705.08442},
 primaryClass = {astro-ph.GA},
       adsurl = {https://ui.adsabs.harvard.edu/abs/2018MNRAS.474.5300D}
}

@ARTICLE{Griffen2018,
       author = {{Griffen}, Brendan F. and {Dooley}, Gregory A. and {Ji}, Alexander P. and {O'Shea}, Brian W. and {G{\'o}mez}, Facundo A. and {Frebel}, Anna},
        title = "{Tracing the first stars and galaxies of the Milky Way}",
      journal = {\mnras},
         year = 2018,
        month = feb,
       volume = {474},
       number = {1},
        pages = {443-459},
          doi = {10.1093/mnras/stx2749},
archivePrefix = {arXiv},
       eprint = {1611.00759},
 primaryClass = {astro-ph.GA},
       adsurl = {https://ui.adsabs.harvard.edu/abs/2018MNRAS.474..443G}
}

@ARTICLE{Fujimoto2025,
       author = {{Fujimoto}, Seiji and {Asada}, Yoshihisa and {Naidu}, Rohan P. and {Chisholm}, John and {Atek}, Hakim and {Brammer}, Gabriel and {Berg}, Danielle A. and {Schaerer}, Daniel and {Kokorev}, Vasily and {Furtak}, Lukas J. and {Richard}, Johan and {Venditti}, Alessandra and {Bromm}, Volker and {Adamo}, Angela and {Claeyssens}, Adelaide and {Dessauges-Zavadsky}, Miroslava and {Fei}, Qinyue and {Hsiao}, Tiger Yu-Yang and {Korber}, Damien and {Munoz}, Julian B. and {Pan}, Richard and {Saldana-Lopez}, Alberto},
        title = "{GLIMPSE-D: An Exotic Balmer-Jump Object at z=6.20? Revisiting Photometric Selection and the Cosmic Abundance of Pop III Galaxies}",
      journal = {arXiv e-prints},
         year = 2025,
        month = dec,
          eid = {arXiv:2512.11790},
        pages = {arXiv:2512.11790},
          doi = {10.48550/arXiv.2512.11790},
archivePrefix = {arXiv},
       eprint = {2512.11790},
 primaryClass = {astro-ph.GA},
       adsurl = {https://ui.adsabs.harvard.edu/abs/2025arXiv251211790F}
}

@ARTICLE{Pallottini2014,
       author = {{Pallottini}, A. and {Ferrara}, A. and {Gallerani}, S. and {Salvadori}, S. and {D'Odorico}, V.},
        title = "{Simulating cosmic metal enrichment by the first galaxies}",
      journal = {\mnras},
         year = 2014,
        month = may,
       volume = {440},
       number = {3},
        pages = {2498-2518},
          doi = {10.1093/mnras/stu451},
archivePrefix = {arXiv},
       eprint = {1403.1261},
 primaryClass = {astro-ph.CO},
       adsurl = {https://ui.adsabs.harvard.edu/abs/2014MNRAS.440.2498P}
}

@ARTICLE{Kruijssen2019,
       author = {{Kruijssen}, J.~M. Diederik and {Pfeffer}, Joel L. and {Reina-Campos}, Marta and {Crain}, Robert A. and {Bastian}, Nate},
        title = "{The formation and assembly history of the Milky Way revealed by its globular cluster population}",
      journal = {\mnras},
         year = 2019,
        month = jul,
       volume = {486},
       number = {3},
        pages = {3180-3202},
          doi = {10.1093/mnras/sty1609},
archivePrefix = {arXiv},
       eprint = {1806.05680},
 primaryClass = {astro-ph.GA},
       adsurl = {https://ui.adsabs.harvard.edu/abs/2019MNRAS.486.3180K}
}

@ARTICLE{Kruijssen2020,
       author = {{Kruijssen}, J.~M. Diederik and {Pfeffer}, Joel L. and {Chevance}, M{\'e}lanie and {Bonaca}, Ana and {Trujillo-Gomez}, Sebastian and {Bastian}, Nate and {Reina-Campos}, Marta and {Crain}, Robert A. and {Hughes}, Meghan E.},
        title = "{Kraken reveals itself - the merger history of the Milky Way reconstructed with the E-MOSAICS simulations}",
      journal = {\mnras},
         year = 2020,
        month = oct,
       volume = {498},
       number = {2},
        pages = {2472-2491},
          doi = {10.1093/mnras/staa2452},
archivePrefix = {arXiv},
       eprint = {2003.01119},
 primaryClass = {astro-ph.GA},
       adsurl = {https://ui.adsabs.harvard.edu/abs/2020MNRAS.498.2472K}
}

@ARTICLE{Davison2020,
       author = {{Davison}, Thomas A. and {Norris}, Mark A. and {Pfeffer}, Joel L. and {Davies}, Jonathan J. and {Crain}, Robert A.},
        title = "{An EAGLE's view of ex situ galaxy growth}",
      journal = {\mnras},
         year = 2020,
        month = sep,
       volume = {497},
       number = {1},
        pages = {81-93},
          doi = {10.1093/mnras/staa1816},
archivePrefix = {arXiv},
       eprint = {2006.08590},
 primaryClass = {astro-ph.GA},
       adsurl = {https://ui.adsabs.harvard.edu/abs/2020MNRAS.497...81D}
}

@ARTICLE{Rodriguez-Gomez2016,
       author = {{Rodriguez-Gomez}, Vicente and {Pillepich}, Annalisa and {Sales}, Laura V. and {Genel}, Shy and {Vogelsberger}, Mark and {Zhu}, Qirong and {Wellons}, Sarah and {Nelson}, Dylan and {Torrey}, Paul and {Springel}, Volker and {Ma}, Chung-Pei and {Hernquist}, Lars},
        title = "{The stellar mass assembly of galaxies in the Illustris simulation: growth by mergers and the spatial distribution of accreted stars}",
      journal = {\mnras},
         year = 2016,
        month = may,
       volume = {458},
       number = {3},
        pages = {2371-2390},
          doi = {10.1093/mnras/stw456},
archivePrefix = {arXiv},
       eprint = {1511.08804},
 primaryClass = {astro-ph.GA},
       adsurl = {https://ui.adsabs.harvard.edu/abs/2016MNRAS.458.2371R}
}

@ARTICLE{Horta2023,
       author = {{Horta}, Danny and {Cunningham}, Emily C. and {Sanderson}, Robyn E. and {Johnston}, Kathryn V. and {Panithanpaisal}, Nondh and {Arora}, Arpit and {Necib}, Lina and {Wetzel}, Andrew and {Bailin}, Jeremy and {Faucher-Gigu{\`e}re}, Claude-Andr{\'e}},
        title = "{The Observable Properties of Galaxy Accretion Events in Milky Way-like Galaxies in the FIRE-2 Cosmological Simulations}",
      journal = {\apj},
         year = 2023,
        month = feb,
       volume = {943},
       number = {2},
          eid = {158},
        pages = {158},
          doi = {10.3847/1538-4357/acae87},
archivePrefix = {arXiv},
       eprint = {2211.05799},
 primaryClass = {astro-ph.GA},
       adsurl = {https://ui.adsabs.harvard.edu/abs/2023ApJ...943..158H}
}

@ARTICLE{Koppelman2020,
       author = {{Koppelman}, Helmer H. and {Bos}, Roy O.~Y. and {Helmi}, Amina},
        title = "{The messy merger of a large satellite and a Milky Way-like galaxy}",
      journal = {\aap},
         year = 2020,
        month = oct,
       volume = {642},
          eid = {L18},
        pages = {L18},
          doi = {10.1051/0004-6361/202038652},
archivePrefix = {arXiv},
       eprint = {2006.07620},
 primaryClass = {astro-ph.GA},
       adsurl = {https://ui.adsabs.harvard.edu/abs/2020A&A...642L..18K}
}

@ARTICLE{Fu2024,
       author = {{Fu}, Hao and {Shankar}, Francesco and {Ayromlou}, Mohammadreza and {Koutsouridou}, Ioanna and {Cattaneo}, Andrea and {Bertemes}, Caroline and {Bellstedt}, Sabine and {Mart{\'\i}n-Navarro}, Ignacio and {Leja}, Joel and {Allevato}, Viola and {Bernardi}, Mariangela and {Boco}, Lumen and {Dimauro}, Paola and {Gruppioni}, Carlotta and {Lapi}, Andrea and {Menci}, Nicola and {Rodr{\'\i}guez}, Iv{\'a}n Mu{\~n}oz and {Puglisi}, Annagrazia and {Alonso-Tetilla}, Alba V.},
        title = "{Unveiling the (in)consistencies among the galaxy stellar mass function, star formation histories, satellite abundances, and intracluster light from a semi-empirical perspective}",
      journal = {\mnras},
         year = 2024,
        month = jul,
       volume = {532},
       number = {1},
        pages = {177-197},
          doi = {10.1093/mnras/stae1492},
archivePrefix = {arXiv},
       eprint = {2406.07605},
 primaryClass = {astro-ph.GA},
       adsurl = {https://ui.adsabs.harvard.edu/abs/2024MNRAS.532..177F}
}

@ARTICLE{rusta2024,
       author = {{Rusta}, Elka and {Salvadori}, Stefania and {Gelli}, Viola and {Koutsouridou}, Ioanna and {Marconi}, Alessandro},
        title = "{Linking High-z and Low-z: Are We Observing the Progenitors of the Milky Way with JWST?}",
      journal = {\apjl},
         year = 2024,
        month = oct,
       volume = {974},
       number = {2},
          eid = {L35},
        pages = {L35},
          doi = {10.3847/2041-8213/ad833d},
archivePrefix = {arXiv},
       eprint = {2407.06255},
 primaryClass = {astro-ph.GA},
       adsurl = {https://ui.adsabs.harvard.edu/abs/2024ApJ...974L..35R}
}

@ARTICLE{Salvadori2015,
       author = {{Salvadori}, Stefania and {Sk{\'u}lad{\'o}ttir}, {\'A}sa and {Tolstoy}, Eline},
        title = "{Carbon-enhanced metal-poor stars in dwarf galaxies}",
      journal = {\mnras},
         year = 2015,
        month = dec,
       volume = {454},
       number = {2},
        pages = {1320-1331},
          doi = {10.1093/mnras/stv1969},
archivePrefix = {arXiv},
       eprint = {1506.03451},
 primaryClass = {astro-ph.GA},
       adsurl = {https://ui.adsabs.harvard.edu/abs/2015MNRAS.454.1320S}
}

@ARTICLE{Salvadori2010,
       author = {{Salvadori}, S. and {Ferrara}, A. and {Schneider}, R. and {Scannapieco}, E. and {Kawata}, D.},
        title = "{Mining the Galactic halo for very metal-poor stars}",
      journal = {\mnras},
         year = 2010,
        month = jan,
       volume = {401},
       number = {1},
        pages = {L5-L9},
          doi = {10.1111/j.1745-3933.2009.00772.x},
archivePrefix = {arXiv},
       eprint = {0908.4279},
 primaryClass = {astro-ph.CO},
       adsurl = {https://ui.adsabs.harvard.edu/abs/2010MNRAS.401L...5S}
}

@ARTICLE{Morishita2025,
       author = {{Morishita}, Takahiro and {Liu}, Zhaoran and {Stiavelli}, Massimo and {Treu}, Tommaso and {Bergamini}, Pietro and {Zhang}, Yechi},
        title = "{Pristine Massive Star Formation Caught at the Break of Cosmic Dawn}",
      journal = {arXiv e-prints},
         year = 2025,
        month = jul,
          eid = {arXiv:2507.10521},
        pages = {arXiv:2507.10521},
          doi = {10.48550/arXiv.2507.10521},
archivePrefix = {arXiv},
       eprint = {2507.10521},
 primaryClass = {astro-ph.CO},
       adsurl = {https://ui.adsabs.harvard.edu/abs/2025arXiv250710521M}
}

@ARTICLE{Mackereth2019,
       author = {{Mackereth}, J. Ted and {Schiavon}, Ricardo P. and {Pfeffer}, Joel and {Hayes}, Christian R. and {Bovy}, Jo and {Anguiano}, Borja and {Allende Prieto}, Carlos and {Hasselquist}, Sten and {Holtzman}, Jon and {Johnson}, Jennifer A. and {Majewski}, Steven R. and {O'Connell}, Robert and {Shetrone}, Matthew and {Tissera}, Patricia B. and {Fern{\'a}ndez-Trincado}, J.~G.},
        title = "{The origin of accreted stellar halo populations in the Milky Way using APOGEE, Gaia, and the EAGLE simulations}",
      journal = {\mnras},
         year = 2019,
        month = jan,
       volume = {482},
       number = {3},
        pages = {3426-3442},
          doi = {10.1093/mnras/sty2955},
archivePrefix = {arXiv},
       eprint = {1808.00968},
 primaryClass = {astro-ph.GA},
       adsurl = {https://ui.adsabs.harvard.edu/abs/2019MNRAS.482.3426M}
}

@ARTICLE{Jeon2026_hebe,
       author = {{Jeon}, Junehyoung and {Jeong}, Tae Bong and {Zhang}, Saiyang and {Bromm}, Volker},
        title = "{What is Powering the Enigmatic He II Emitter Hebe: The First Stars or Black Holes?}",
      journal = {arXiv e-prints},
         year = 2026,
        month = apr,
          eid = {arXiv:2604.19075},
        pages = {arXiv:2604.19075},
          doi = {10.48550/arXiv.2604.19075},
archivePrefix = {arXiv},
       eprint = {2604.19075},
 primaryClass = {astro-ph.GA},
       adsurl = {https://ui.adsabs.harvard.edu/abs/2026arXiv260419075J}
}

@ARTICLE{Haywood2018,
       author = {{Haywood}, M. and {Di Matteo}, P. and {Lehnert}, M.~D. and {Snaith}, O. and {Khoperskov}, S. and {G{\'o}mez}, A.},
        title = "{In Disguise or Out of Reach: First Clues about In Situ and Accreted Stars in the Stellar Halo of the Milky Way from Gaia DR2}",
      journal = {\apj},
         year = 2018,
        month = aug,
       volume = {863},
       number = {2},
          eid = {113},
        pages = {113},
          doi = {10.3847/1538-4357/aad235},
archivePrefix = {arXiv},
       eprint = {1805.02617},
 primaryClass = {astro-ph.GA},
       adsurl = {https://ui.adsabs.harvard.edu/abs/2018ApJ...863..113H}
}

@ARTICLE{Jeon2026,
       author = {{Jeon}, Junehyoung and {Bromm}, Volker and {Venditti}, Alessandra and {Finkelstein}, Steven L. and {Hsiao}, Tiger Yu-Yang},
        title = "{Hunting for the First Explosions at the High-Redshift Frontier}",
      journal = {arXiv e-prints},
         year = 2026,
        month = jan,
          eid = {arXiv:2601.02469},
        pages = {arXiv:2601.02469},
          doi = {10.48550/arXiv.2601.02469},
archivePrefix = {arXiv},
       eprint = {2601.02469},
 primaryClass = {astro-ph.GA},
       adsurl = {https://ui.adsabs.harvard.edu/abs/2026arXiv260102469J}
}

@ARTICLE{Aguado2023,
       author = {{Aguado}, D.~S. and {Salvadori}, S. and {Sk{\'u}lad{\'o}ttir}, {\'A}. and {Caffau}, E. and {Bonifacio}, P. and {Vanni}, I. and {Gelli}, V. and {Koutsouridou}, I. and {Amarsi}, A.~M.},
        title = "{PISN-explorer: hunting the descendants of very massive first stars}",
      journal = {\mnras},
         year = 2023,
        month = mar,
       volume = {520},
       number = {1},
        pages = {866-878},
          doi = {10.1093/mnras/stad164},
archivePrefix = {arXiv},
       eprint = {2301.03604},
 primaryClass = {astro-ph.GA},
       adsurl = {https://ui.adsabs.harvard.edu/abs/2023MNRAS.520..866A}
}

@ARTICLE{Rusta2025,
       author = {{Rusta}, Elka and {Salvadori}, Stefania and {Gelli}, Viola and {Schaerer}, Daniel and {Marconi}, Alessandro and {Koutsouridou}, Ioanna and {Carniani}, Stefano},
        title = "{Metal-polluted Population III Galaxies and How to Find Them}",
      journal = {\apjl},
         year = 2025,
        month = aug,
       volume = {989},
       number = {2},
          eid = {L32},
        pages = {L32},
          doi = {10.3847/2041-8213/adf4e3},
archivePrefix = {arXiv},
       eprint = {2506.17400},
 primaryClass = {astro-ph.GA},
       adsurl = {https://ui.adsabs.harvard.edu/abs/2025ApJ...989L..32R}
}

@ARTICLE{Vanzella2023,
       author = {{Vanzella}, E. and {Loiacono}, F. and {Bergamini}, P. and {Me{\v{s}}tri{\'c}}, U. and {Castellano}, M. and {Rosati}, P. and {Meneghetti}, M. and {Grillo}, C. and {Calura}, F. and {Mignoli}, M. and {Brada{\v{c}}}, M. and {Adamo}, A. and {Rihtar{\v{s}}i{\v{c}}}, G. and {Dickinson}, M. and {Gronke}, M. and {Zanella}, A. and {Annibali}, F. and {Willott}, C. and {Messa}, M. and {Sani}, E. and {Acebron}, A. and {Bolamperti}, A. and {Comastri}, A. and {Gilli}, R. and {Caputi}, K.~I. and {Ricotti}, M. and {Gruppioni}, C. and {Ravindranath}, S. and {Mercurio}, A. and {Strait}, V. and {Martis}, N. and {Pascale}, R. and {Caminha}, G.~B. and {Annunziatella}, M. and {Nonino}, M.},
        title = "{An extremely metal-poor star complex in the reionization era: Approaching Population III stars with JWST}",
      journal = {\aap},
         year = 2023,
        month = oct,
       volume = {678},
          eid = {A173},
        pages = {A173},
          doi = {10.1051/0004-6361/202346981},
archivePrefix = {arXiv},
       eprint = {2305.14413},
 primaryClass = {astro-ph.GA},
       adsurl = {https://ui.adsabs.harvard.edu/abs/2023A&A...678A.173V}
}

@ARTICLE{Vanzella2026,
       author = {{Vanzella}, E. and {Messa}, M. and {Zanella}, A. and {Bolamperti}, A. and {Castellano}, M. and {Loiacono}, F. and {Bergamini}, P. and {Roberts Borsani}, G. and {Adamo}, A. and {Fontana}, A. and {Treu}, T. and {Calura}, F. and {Grillo}, C. and {Lombardi}, M. and {Rosati}, P. and {Gilli}, R. and {Meneghetti}, M.},
        title = "{A pristine, star-forming complex at z = 4.19}",
      journal = {\aap},
         year = 2026,
        month = jan,
       volume = {705},
          eid = {L12},
        pages = {L12},
          doi = {10.1051/0004-6361/202557153},
archivePrefix = {arXiv},
       eprint = {2509.07073},
 primaryClass = {astro-ph.GA},
       adsurl = {https://ui.adsabs.harvard.edu/abs/2026A&A...705L..12V}
}

@ARTICLE{Nakajima2025,
       author = {{Nakajima}, Kimihiko and {Ouchi}, Masami and {Harikane}, Yuichi and {Vanzella}, Eros and {Ono}, Yoshiaki and {Isobe}, Yuki and {Nishigaki}, Moka and {Tsujimoto}, Takuji and {Nakamura}, Fumitaka and {Xu}, Yi and {Umeda}, Hiroya and {Zhang}, Yechi},
        title = "{An Ultra-Faint, Chemically Primitive Galaxy Forming in the Reionization Era}",
      journal = {arXiv e-prints},
         year = 2025,
        month = jun,
          eid = {arXiv:2506.11846},
        pages = {arXiv:2506.11846},
          doi = {10.48550/arXiv.2506.11846},
archivePrefix = {arXiv},
       eprint = {2506.11846},
 primaryClass = {astro-ph.GA},
       adsurl = {https://ui.adsabs.harvard.edu/abs/2025arXiv250611846N}
}

@ARTICLE{Nakajima2022,
       author = {{Nakajima}, K. and {Maiolino}, R.},
        title = "{Diagnostics for PopIII galaxies and direct collapse black holes in the early universe}",
      journal = {\mnras},
         year = 2022,
        month = jul,
       volume = {513},
       number = {4},
        pages = {5134-5147},
          doi = {10.1093/mnras/stac1242},
archivePrefix = {arXiv},
       eprint = {2204.11870},
 primaryClass = {astro-ph.GA},
       adsurl = {https://ui.adsabs.harvard.edu/abs/2022MNRAS.513.5134N}
}

@ARTICLE{Chiaki2018,
       author = {{Chiaki}, Gen and {Susa}, Hajime and {Hirano}, Shingo},
        title = "{Metal-poor star formation triggered by the feedback effects from Pop III stars}",
      journal = {\mnras},
         year = 2018,
        month = apr,
       volume = {475},
       number = {4},
        pages = {4378-4395},
          doi = {10.1093/mnras/sty040},
archivePrefix = {arXiv},
       eprint = {1801.01583},
 primaryClass = {astro-ph.GA},
       adsurl = {https://ui.adsabs.harvard.edu/abs/2018MNRAS.475.4378C}
}

@ARTICLE{Magg2022,
       author = {{Magg}, Mattis and {Schauer}, Anna T.~P. and {Klessen}, Ralf S. and {Glover}, Simon C.~O. and {Tress}, Robin G. and {Jaura}, Ondrej},
        title = "{Metal Mixing in Minihalos: The Descendants of Pair-instability Supernovae}",
      journal = {\apj},
         year = 2022,
        month = apr,
       volume = {929},
       number = {2},
          eid = {119},
        pages = {119},
          doi = {10.3847/1538-4357/ac5aac},
archivePrefix = {arXiv},
       eprint = {2110.15372},
 primaryClass = {astro-ph.GA},
       adsurl = {https://ui.adsabs.harvard.edu/abs/2022ApJ...929..119M}
}

@ARTICLE{Skuladottir2024,
       author = {{Sk{\'u}lad{\'o}ttir}, {\'A}sa and {Koutsouridou}, Ioanna and {Vanni}, Irene and {Amarsi}, Anish M. and {Lucchesi}, Romain and {Salvadori}, Stefania and {Aguado}, David S.},
        title = "{On the Pair-instability Supernova Origin of J1010+2358}",
      journal = {\apjl},
         year = 2024,
        month = jun,
       volume = {968},
       number = {2},
          eid = {L23},
        pages = {L23},
          doi = {10.3847/2041-8213/ad4b1a},
archivePrefix = {arXiv},
       eprint = {2404.19086},
 primaryClass = {astro-ph.SR},
       adsurl = {https://ui.adsabs.harvard.edu/abs/2024ApJ...968L..23S}
}

@ARTICLE{Thibodeaux2024,
       author = {{Thibodeaux}, Pierre and {Ji}, Alexander P. and {Cerny}, William and {Kirby}, Evan N. and {Simon}, Joshua D.},
        title = "{LAMOST J1010+2358 is not a Pair-Instability Supernova Relic}",
      journal = {The Open Journal of Astrophysics},
         year = 2024,
        month = aug,
       volume = {7},
          eid = {66},
        pages = {66},
          doi = {10.33232/001c.122335},
archivePrefix = {arXiv},
       eprint = {2404.17078},
 primaryClass = {astro-ph.SR},
       adsurl = {https://ui.adsabs.harvard.edu/abs/2024OJAp....7E..66T}
}

@ARTICLE{Gandolfi2026,
       author = {{Gandolfi}, G. and {Rodighiero}, G. and {Castellano}, M. and {Fontana}, A. and {Santini}, P. and {Dickinson}, M. and {Finkelstein}, S. and {Catone}, M. and {Calabr{\`o}}, A. and {Merlin}, E. and {Pentericci}, L. and {Bisigello}, L. and {Grazian}, A. and {Napolitano}, L. and {Vulcani}, B. and {Taylor}, A.~J. and {Arrabal Haro}, P. and {Kirkpatrick}, A. and {Backhaus}, B.~E. and {Holwerda}, B.~W. and {Giulietti}, M. and {Bianchetti}, A. and {Cassata}, P. and {Cleri}, N.~J. and {Daddi}, E. and {Ferguson}, H.~C. and {Girardi}, G. and {Hirschmann}, M. and {Koekemoer}, A.~M. and {Lapi}, A. and {Pacucci}, F. and {P{\'e}rez-Gonz{\'a}lez}, P.~G. and {de la Vega}, A. and {Vietri}, A. and {Wilkins}, S. and {Yung}, L.~Y.~A. and {Bagley}, M. and {Bhatawdekar}, R. and {Kartaltepe}, J. and {Papovich}, C. and {Pirzkal}, N.},
        title = "{Mysteries of Capotauro: Investigating the puzzling nature of an extreme F356W-dropout}",
      journal = {\aap},
         year = 2026,
        month = feb,
       volume = {706},
          eid = {A364},
        pages = {A364},
          doi = {10.1051/0004-6361/202557061},
archivePrefix = {arXiv},
       eprint = {2509.01664},
 primaryClass = {astro-ph.GA},
       adsurl = {https://ui.adsabs.harvard.edu/abs/2026A&A...706A.364G}
}

@ARTICLE{Ferrara2026,
       author = {{Ferrara}, Andrea and {Carniani}, Stefano and {Morishita}, Takahiro and {Stiavelli}, Massimo},
        title = "{Possible evidence for a pair-instability supernova nature of ultra-early JWST sources}",
      journal = {arXiv e-prints},
         year = 2026,
        month = jan,
          eid = {arXiv:2601.07374},
        pages = {arXiv:2601.07374},
          doi = {10.48550/arXiv.2601.07374},
archivePrefix = {arXiv},
       eprint = {2601.07374},
 primaryClass = {astro-ph.GA},
       adsurl = {https://ui.adsabs.harvard.edu/abs/2026arXiv260107374F}
}

@ARTICLE{Salvadori2007,
       author = {{Salvadori}, Stefania and {Schneider}, Raffaella and {Ferrara}, Andrea},
        title = "{Cosmic stellar relics in the Galactic halo}",
      journal = {\mnras},
         year = 2007,
        month = oct,
       volume = {381},
       number = {2},
        pages = {647-662},
          doi = {10.1111/j.1365-2966.2007.12133.x},
archivePrefix = {arXiv},
       eprint = {astro-ph/0611130},
 primaryClass = {astro-ph},
       adsurl = {https://ui.adsabs.harvard.edu/abs/2007MNRAS.381..647S}
}

@ARTICLE{Weaver1977,
       author = {{Weaver}, R. and {McCray}, R. and {Castor}, J. and {Shapiro}, P. and {Moore}, R.},
        title = "{Interstellar bubbles. II. Structure and evolution.}",
      journal = {\apj},
         year = 1977,
        month = dec,
       volume = {218},
        pages = {377-395},
          doi = {10.1086/155692},
       adsurl = {https://ui.adsabs.harvard.edu/abs/1977ApJ...218..377W}
}

@ARTICLE{Bland-Hawthorn2016,
       author = {{Bland-Hawthorn}, Joss and {Gerhard}, Ortwin},
        title = "{The Galaxy in Context: Structural, Kinematic, and Integrated Properties}",
      journal = {\araa},
         year = 2016,
        month = sep,
       volume = {54},
        pages = {529-596},
          doi = {10.1146/annurev-astro-081915-023441},
archivePrefix = {arXiv},
       eprint = {1602.07702},
 primaryClass = {astro-ph.GA},
       adsurl = {https://ui.adsabs.harvard.edu/abs/2016ARA&A..54..529B}
}

@ARTICLE{Maio2016,
       author = {{Maio}, Umberto and {Petkova}, Margarita and {De Lucia}, Gabriella and {Borgani}, Stefano},
        title = "{Radiative feedback and cosmic molecular gas: the role of different radiative sources}",
      journal = {\mnras},
         year = 2016,
        month = aug,
       volume = {460},
       number = {4},
        pages = {3733-3752},
          doi = {10.1093/mnras/stw1196},
archivePrefix = {arXiv},
       eprint = {1606.00006},
 primaryClass = {astro-ph.GA},
       adsurl = {https://ui.adsabs.harvard.edu/abs/2016MNRAS.460.3733M}
}

@ARTICLE{Adams2013,
       author = {{Adams}, Scott M. and {Kochanek}, C.~S. and {Beacom}, John F. and {Vagins}, Mark R. and {Stanek}, K.~Z.},
        title = "{Observing the Next Galactic Supernova}",
      journal = {\apj},
         year = 2013,
        month = dec,
       volume = {778},
       number = {2},
          eid = {164},
        pages = {164},
          doi = {10.1088/0004-637X/778/2/164},
archivePrefix = {arXiv},
       eprint = {1306.0559},
 primaryClass = {astro-ph.HE},
       adsurl = {https://ui.adsabs.harvard.edu/abs/2013ApJ...778..164A}
}

@ARTICLE{Rozwadowska2021,
       author = {{Rozwadowska}, Karolina and {Vissani}, Francesco and {Cappellaro}, Enrico},
        title = "{On the rate of core collapse supernovae in the milky way}",
      journal = {\na},
         year = 2021,
        month = feb,
       volume = {83},
          eid = {101498},
        pages = {101498},
          doi = {10.1016/j.newast.2020.101498},
archivePrefix = {arXiv},
       eprint = {2009.03438},
 primaryClass = {astro-ph.HE},
       adsurl = {https://ui.adsabs.harvard.edu/abs/2021NewA...8301498R}
}

@ARTICLE{Grauz2013,
       author = {{Graur}, Or and {Maoz}, Dan},
        title = "{Discovery of 90 Type Ia supernovae among 700 000 Sloan spectra: the Type Ia supernova rate versus galaxy mass and star formation rate at redshift {\ensuremath{\sim}}0.1}",
      journal = {\mnras},
         year = 2013,
        month = apr,
       volume = {430},
       number = {3},
        pages = {1746-1763},
          doi = {10.1093/mnras/sts718},
archivePrefix = {arXiv},
       eprint = {1209.0008},
 primaryClass = {astro-ph.CO},
       adsurl = {https://ui.adsabs.harvard.edu/abs/2013MNRAS.430.1746G}
}

@ARTICLE{Andrews2013,
       author = {{Andrews}, Brett H. and {Martini}, Paul},
        title = "{The Mass-Metallicity Relation with the Direct Method on Stacked Spectra of SDSS Galaxies}",
      journal = {\apj},
         year = 2013,
        month = mar,
       volume = {765},
       number = {2},
          eid = {140},
        pages = {140},
          doi = {10.1088/0004-637X/765/2/140},
archivePrefix = {arXiv},
       eprint = {1211.3418},
 primaryClass = {astro-ph.CO},
       adsurl = {https://ui.adsabs.harvard.edu/abs/2013ApJ...765..140A}
}

@ARTICLE{Pino2019,
       author = {{Rodr{\'\i}guez del Pino}, Bruno and {Arribas}, Santiago and {Piqueras L{\'o}pez}, Javier and {Villar-Mart{\'\i}n}, Montserrat and {Colina}, Luis},
        title = "{Properties of ionized outflows in MaNGA DR2 galaxies}",
      journal = {\mnras},
         year = 2019,
        month = jun,
       volume = {486},
       number = {1},
        pages = {344-359},
          doi = {10.1093/mnras/stz816},
archivePrefix = {arXiv},
       eprint = {1903.07432},
 primaryClass = {astro-ph.GA},
       adsurl = {https://ui.adsabs.harvard.edu/abs/2019MNRAS.486..344R}
}

@ARTICLE{Schreiber2019,
       author = {{F{\"o}rster Schreiber}, N.~M. and {{\"U}bler}, H. and {Davies}, R.~L. and {Genzel}, R. and {Wisnioski}, E. and {Belli}, S. and {Shimizu}, T. and {Lutz}, D. and {Fossati}, M. and {Herrera-Camus}, R. and {Mendel}, J.~T. and {Tacconi}, L.~J. and {Wilman}, D. and {Beifiori}, A. and {Brammer}, G.~B. and {Burkert}, A. and {Carollo}, C.~M. and {Davies}, R.~I. and {Eisenhauer}, F. and {Fabricius}, M. and {Lilly}, S.~J. and {Momcheva}, I. and {Naab}, T. and {Nelson}, E.~J. and {Price}, S.~H. and {Renzini}, A. and {Saglia}, R. and {Sternberg}, A. and {van Dokkum}, P. and {Wuyts}, S.},
        title = "{The KMOS$^{3D}$ Survey: Demographics and Properties of Galactic Outflows at z = 0.6-2.7}",
      journal = {\apj},
         year = 2019,
        month = apr,
       volume = {875},
       number = {1},
          eid = {21},
        pages = {21},
          doi = {10.3847/1538-4357/ab0ca2},
archivePrefix = {arXiv},
       eprint = {1807.04738},
 primaryClass = {astro-ph.GA},
       adsurl = {https://ui.adsabs.harvard.edu/abs/2019ApJ...875...21F}
}

@ARTICLE{Chisholm2017,
       author = {{Chisholm}, John and {Tremonti}, Christy A. and {Leitherer}, Claus and {Chen}, Yanmei},
        title = "{The mass and momentum outflow rates of photoionized galactic outflows}",
      journal = {\mnras},
         year = 2017,
        month = aug,
       volume = {469},
       number = {4},
        pages = {4831-4849},
          doi = {10.1093/mnras/stx1164},
archivePrefix = {arXiv},
       eprint = {1702.07351},
 primaryClass = {astro-ph.GA},
       adsurl = {https://ui.adsabs.harvard.edu/abs/2017MNRAS.469.4831C}
}

@ARTICLE{Fox2019,
       author = {{Fox}, Andrew J. and {Richter}, Philipp and {Ashley}, Trisha and {Heckman}, Timothy M. and {Lehner}, Nicolas and {Werk}, Jessica K. and {Bordoloi}, Rongmon and {Peeples}, Molly S.},
        title = "{The Mass Inflow and Outflow Rates of the Milky Way}",
      journal = {\apj},
         year = 2019,
        month = oct,
       volume = {884},
       number = {1},
          eid = {53},
        pages = {53},
          doi = {10.3847/1538-4357/ab40ad},
archivePrefix = {arXiv},
       eprint = {1909.05561},
 primaryClass = {astro-ph.GA},
       adsurl = {https://ui.adsabs.harvard.edu/abs/2019ApJ...884...53F}
}

@ARTICLE{Watkins2019,
       author = {{Watkins}, Laura L. and {van der Marel}, Roeland P. and {Sohn}, Sangmo Tony and {Evans}, N. Wyn},
        title = "{Evidence for an Intermediate-mass Milky Way from Gaia DR2 Halo Globular Cluster Motions}",
      journal = {\apj},
         year = 2019,
        month = mar,
       volume = {873},
       number = {2},
          eid = {118},
        pages = {118},
          doi = {10.3847/1538-4357/ab089f},
archivePrefix = {arXiv},
       eprint = {1804.11348},
 primaryClass = {astro-ph.GA},
       adsurl = {https://ui.adsabs.harvard.edu/abs/2019ApJ...873..118W}
}

@ARTICLE{Eadie2019,
       author = {{Eadie}, Gwendolyn and {Juri{\'c}}, Mario},
        title = "{The Cumulative Mass Profile of the Milky Way as Determined by Globular Cluster Kinematics from Gaia DR2}",
      journal = {\apj},
         year = 2019,
        month = apr,
       volume = {875},
       number = {2},
          eid = {159},
        pages = {159},
          doi = {10.3847/1538-4357/ab0f97},
archivePrefix = {arXiv},
       eprint = {1810.10036},
 primaryClass = {astro-ph.GA},
       adsurl = {https://ui.adsabs.harvard.edu/abs/2019ApJ...875..159E}
}

@ARTICLE{Posti2019,
       author = {{Posti}, Lorenzo and {Helmi}, Amina},
        title = "{Mass and shape of the Milky Way's dark matter halo with globular clusters from Gaia and Hubble}",
      journal = {\aap},
         year = 2019,
        month = jan,
       volume = {621},
          eid = {A56},
        pages = {A56},
          doi = {10.1051/0004-6361/201833355},
archivePrefix = {arXiv},
       eprint = {1805.01408},
 primaryClass = {astro-ph.GA},
       adsurl = {https://ui.adsabs.harvard.edu/abs/2019A&A...621A..56P}
}

@ARTICLE{Ventura2025,
       author = {{Ventura}, Emanuele M. and {Qin}, Yuxiang and {Balu}, Sreedhar and {Wyithe}, J. Stuart B.},
        title = "{Semi-analytical modelling of Pop. III star formation and metallicity evolution {\textendash} II. Impact on 21 cm power spectrum}",
      journal = {\mnras},
         year = 2025,
        month = jun,
       volume = {540},
       number = {1},
        pages = {483-497},
          doi = {10.1093/mnras/staf699},
archivePrefix = {arXiv},
       eprint = {2502.08971},
 primaryClass = {astro-ph.CO},
       adsurl = {https://ui.adsabs.harvard.edu/abs/2025MNRAS.540..483V}
}

@ARTICLE{Hegde2025,
       author = {{Hegde}, Sahil and {Furlanetto}, Steven R.},
        title = "{Efficient semi-analytic modelling of Pop III star formation from Cosmic Dawn to Reionization}",
      journal = {The Open Journal of Astrophysics},
         year = 2025,
        month = oct,
       volume = {8},
          eid = {147},
        pages = {147},
          doi = {10.33232/001c.145070},
archivePrefix = {arXiv},
       eprint = {2507.19581},
 primaryClass = {astro-ph.GA},
       adsurl = {https://ui.adsabs.harvard.edu/abs/2025OJAp....8E.147H}
}

@ARTICLE{Trenti2009,
       author = {{Trenti}, Michele and {Stiavelli}, Massimo},
        title = "{Formation Rates of Population III Stars and Chemical Enrichment of Halos during the Reionization Era}",
      journal = {\apj},
         year = 2009,
        month = apr,
       volume = {694},
       number = {2},
        pages = {879-892},
          doi = {10.1088/0004-637X/694/2/879},
archivePrefix = {arXiv},
       eprint = {0901.0711},
 primaryClass = {astro-ph.CO},
       adsurl = {https://ui.adsabs.harvard.edu/abs/2009ApJ...694..879T}
}

@ARTICLE{Mebane2018,
       author = {{Mebane}, Richard H. and {Mirocha}, Jordan and {Furlanetto}, Steven R.},
        title = "{The Persistence of Population III Star Formation}",
      journal = {\mnras},
         year = 2018,
        month = oct,
       volume = {479},
       number = {4},
        pages = {4544-4559},
          doi = {10.1093/mnras/sty1833},
archivePrefix = {arXiv},
       eprint = {1710.02528},
 primaryClass = {astro-ph.GA},
       adsurl = {https://ui.adsabs.harvard.edu/abs/2018MNRAS.479.4544M}
}

@ARTICLE{Johnson2013,
       author = {{Johnson}, Jarrett L. and {Dalla Vecchia}, Claudio and {Khochfar}, Sadegh},
        title = "{The First Billion Years project: the impact of stellar radiation on the co-evolution of Populations II and III}",
      journal = {\mnras},
         year = 2013,
        month = jan,
       volume = {428},
       number = {3},
        pages = {1857-1872},
          doi = {10.1093/mnras/sts011},
archivePrefix = {arXiv},
       eprint = {1206.5824},
 primaryClass = {astro-ph.CO},
       adsurl = {https://ui.adsabs.harvard.edu/abs/2013MNRAS.428.1857J}
}

@ARTICLE{Jaacks2019,
       author = {{Jaacks}, Jason and {Finkelstein}, Steven L. and {Bromm}, Volker},
        title = "{Legacy of star formation in the pre-reionization universe}",
      journal = {\mnras},
         year = 2019,
        month = sep,
       volume = {488},
       number = {2},
        pages = {2202-2221},
          doi = {10.1093/mnras/stz1529},
archivePrefix = {arXiv},
       eprint = {1804.07372},
 primaryClass = {astro-ph.GA},
       adsurl = {https://ui.adsabs.harvard.edu/abs/2019MNRAS.488.2202J}
}

@ARTICLE{Xu2016,
       author = {{Xu}, Hao and {Norman}, Michael L. and {O'Shea}, Brian W. and {Wise}, John H.},
        title = "{Late Pop III Star Formation During the Epoch of Reionization: Results from the Renaissance Simulations}",
      journal = {\apj},
         year = 2016,
        month = jun,
       volume = {823},
       number = {2},
          eid = {140},
        pages = {140},
          doi = {10.3847/0004-637X/823/2/140},
archivePrefix = {arXiv},
       eprint = {1604.03586},
 primaryClass = {astro-ph.GA},
       adsurl = {https://ui.adsabs.harvard.edu/abs/2016ApJ...823..140X}
}

@ARTICLE{Sarmento2019,
       author = {{Sarmento}, Richard and {Scannapieco}, Evan and {C{\^o}t{\'e}}, Benoit},
        title = "{Following the Cosmic Evolution of Pristine Gas. III. The Observational Consequences of the Unknown Properties of Population III Stars}",
      journal = {\apj},
         year = 2019,
        month = feb,
       volume = {871},
       number = {2},
          eid = {206},
        pages = {206},
          doi = {10.3847/1538-4357/aafa1a},
archivePrefix = {arXiv},
       eprint = {1901.03727},
 primaryClass = {astro-ph.GA},
       adsurl = {https://ui.adsabs.harvard.edu/abs/2019ApJ...871..206S}
}

@ARTICLE{Greig2017,
       author = {{Greig}, Bradley and {Mesinger}, Andrei},
        title = "{The global history of reionization}",
      journal = {\mnras},
         year = 2017,
        month = mar,
       volume = {465},
       number = {4},
        pages = {4838-4852},
          doi = {10.1093/mnras/stw3026},
archivePrefix = {arXiv},
       eprint = {1605.05374},
 primaryClass = {astro-ph.CO},
       adsurl = {https://ui.adsabs.harvard.edu/abs/2017MNRAS.465.4838G}
}

@ARTICLE{Graziani2015,
       author = {{Graziani}, L. and {Salvadori}, S. and {Schneider}, R. and {Kawata}, D. and {de Bennassuti}, M. and {Maselli}, A.},
        title = "{Galaxy formation with radiative and chemical feedback}",
      journal = {\mnras},
         year = 2015,
        month = may,
       volume = {449},
       number = {3},
        pages = {3137-3148},
          doi = {10.1093/mnras/stv494},
archivePrefix = {arXiv},
       eprint = {1502.07344},
 primaryClass = {astro-ph.GA},
       adsurl = {https://ui.adsabs.harvard.edu/abs/2015MNRAS.449.3137G}
}

@ARTICLE{Robertson2015,
       author = {{Robertson}, Brant E. and {Ellis}, Richard S. and {Furlanetto}, Steven R. and {Dunlop}, James S.},
        title = "{Cosmic Reionization and Early Star-forming Galaxies: A Joint Analysis of New Constraints from Planck and the Hubble Space Telescope}",
      journal = {\apjl},
         year = 2015,
        month = apr,
       volume = {802},
       number = {2},
          eid = {L19},
        pages = {L19},
          doi = {10.1088/2041-8205/802/2/L19},
archivePrefix = {arXiv},
       eprint = {1502.02024},
 primaryClass = {astro-ph.CO},
       adsurl = {https://ui.adsabs.harvard.edu/abs/2015ApJ...802L..19R}
}

@ARTICLE{Hartwig2024,
       author = {{Hartwig}, Tilman and {Lipatova}, Veronika and {Glover}, Simon C.~O. and {Klessen}, Ralf S.},
        title = "{A-SLOTH reveals the nature of the first stars}",
      journal = {\mnras},
         year = 2024,
        month = nov,
       volume = {535},
       number = {1},
        pages = {516-530},
          doi = {10.1093/mnras/stae2318},
archivePrefix = {arXiv},
       eprint = {2410.05393},
 primaryClass = {astro-ph.GA},
       adsurl = {https://ui.adsabs.harvard.edu/abs/2024MNRAS.535..516H}
}

@ARTICLE{Salvadori2009,
       author = {{Salvadori}, Stefania and {Ferrara}, Andrea},
        title = "{Ultra faint dwarfs: probing early cosmic star formation}",
      journal = {\mnras},
         year = 2009,
        month = may,
       volume = {395},
       number = {1},
        pages = {L6-L10},
          doi = {10.1111/j.1745-3933.2009.00627.x},
archivePrefix = {arXiv},
       eprint = {0812.3151},
 primaryClass = {astro-ph},
       adsurl = {https://ui.adsabs.harvard.edu/abs/2009MNRAS.395L...6S}
}

@ARTICLE{Placco2014,
       author = {{Placco}, Vinicius M. and {Frebel}, Anna and {Beers}, Timothy C. and {Stancliffe}, Richard J.},
        title = "{Carbon-enhanced Metal-poor Star Frequencies in the Galaxy: Corrections for the Effect of Evolutionary Status on Carbon Abundances}",
      journal = {\apj},
         year = 2014,
        month = dec,
       volume = {797},
       number = {1},
          eid = {21},
        pages = {21},
          doi = {10.1088/0004-637X/797/1/21},
archivePrefix = {arXiv},
       eprint = {1410.2223},
 primaryClass = {astro-ph.SR},
       adsurl = {https://ui.adsabs.harvard.edu/abs/2014ApJ...797...21P}
}

@ARTICLE{Tremonti2004,
       author = {{Tremonti}, Christy A. and {Heckman}, Timothy M. and {Kauffmann}, Guinevere and {Brinchmann}, Jarle and {Charlot}, St{\'e}phane and {White}, Simon D.~M. and {Seibert}, Mark and {Peng}, Eric W. and {Schlegel}, David J. and {Uomoto}, Alan and {Fukugita}, Masataka and {Brinkmann}, Jon},
        title = "{The Origin of the Mass-Metallicity Relation: Insights from 53,000 Star-forming Galaxies in the Sloan Digital Sky Survey}",
      journal = {\apj},
         year = 2004,
        month = oct,
       volume = {613},
       number = {2},
        pages = {898-913},
          doi = {10.1086/423264},
archivePrefix = {arXiv},
       eprint = {astro-ph/0405537},
 primaryClass = {astro-ph},
       adsurl = {https://ui.adsabs.harvard.edu/abs/2004ApJ...613..898T}
}

@ARTICLE{Gallazzi2005,
       author = {{Gallazzi}, Anna and {Charlot}, St{\'e}phane and {Brinchmann}, Jarle and {White}, Simon D.~M. and {Tremonti}, Christy A.},
        title = "{The ages and metallicities of galaxies in the local universe}",
      journal = {\mnras},
         year = 2005,
        month = sep,
       volume = {362},
       number = {1},
        pages = {41-58},
          doi = {10.1111/j.1365-2966.2005.09321.x},
archivePrefix = {arXiv},
       eprint = {astro-ph/0506539},
 primaryClass = {astro-ph},
       adsurl = {https://ui.adsabs.harvard.edu/abs/2005MNRAS.362...41G}
}

@ARTICLE{Callingham2019,
       author = {{Callingham}, Thomas M. and {Cautun}, Marius and {Deason}, Alis J. and {Frenk}, Carlos S. and {Wang}, Wenting and {G{\'o}mez}, Facundo A. and {Grand}, Robert J.~J. and {Marinacci}, Federico and {Pakmor}, Ruediger},
        title = "{The mass of the Milky Way from satellite dynamics}",
      journal = {\mnras},
         year = 2019,
        month = apr,
       volume = {484},
       number = {4},
        pages = {5453-5467},
          doi = {10.1093/mnras/stz365},
archivePrefix = {arXiv},
       eprint = {1808.10456},
 primaryClass = {astro-ph.GA},
       adsurl = {https://ui.adsabs.harvard.edu/abs/2019MNRAS.484.5453C}
}

@ARTICLE{Labini2023,
       author = {{Sylos Labini}, Francesco and {Chrob{\'a}kov{\'a}}, {\v{Z}}ofia and {Capuzzo-Dolcetta}, Roberto and {L{\'o}pez-Corredoira}, Mart{\'\i}n},
        title = "{Mass Models of the Milky Way and Estimation of Its Mass from the Gaia DR3 Data Set}",
      journal = {\apj},
         year = 2023,
        month = mar,
       volume = {945},
       number = {1},
          eid = {3},
        pages = {3},
          doi = {10.3847/1538-4357/acb92c},
archivePrefix = {arXiv},
       eprint = {2302.01379},
 primaryClass = {astro-ph.GA},
       adsurl = {https://ui.adsabs.harvard.edu/abs/2023ApJ...945....3S}
}

@ARTICLE{Chomiuk2011,
       author = {{Chomiuk}, Laura and {Povich}, Matthew S.},
        title = "{Toward a Unification of Star Formation Rate Determinations in the Milky Way and Other Galaxies}",
      journal = {\aj},
         year = 2011,
        month = dec,
       volume = {142},
       number = {6},
          eid = {197},
        pages = {197},
          doi = {10.1088/0004-6256/142/6/197},
archivePrefix = {arXiv},
       eprint = {1110.4105},
 primaryClass = {astro-ph.GA},
       adsurl = {https://ui.adsabs.harvard.edu/abs/2011AJ....142..197C}
}

@ARTICLE{Cautun2020,
       author = {{Cautun}, Marius and {Ben{\'\i}tez-Llambay}, Alejandro and {Deason}, Alis J. and {Frenk}, Carlos S. and {Fattahi}, Azadeh and {G{\'o}mez}, Facundo A. and {Grand}, Robert J.~J. and {Oman}, Kyle A. and {Navarro}, Julio F. and {Simpson}, Christine M.},
        title = "{The milky way total mass profile as inferred from Gaia DR2}",
      journal = {\mnras},
         year = 2020,
        month = may,
       volume = {494},
       number = {3},
        pages = {4291-4313},
          doi = {10.1093/mnras/staa1017},
archivePrefix = {arXiv},
       eprint = {1911.04557},
 primaryClass = {astro-ph.GA},
       adsurl = {https://ui.adsabs.harvard.edu/abs/2020MNRAS.494.4291C}
}

@ARTICLE{Elia2022,
       author = {{Elia}, D. and {Molinari}, S. and {Schisano}, E. and {Soler}, J.~D. and {Merello}, M. and {Russeil}, D. and {Veneziani}, M. and {Zavagno}, A. and {Noriega-Crespo}, A. and {Olmi}, L. and {Benedettini}, M. and {Hennebelle}, P. and {Klessen}, R.~S. and {Leurini}, S. and {Paladini}, R. and {Pezzuto}, S. and {Traficante}, A. and {Eden}, D.~J. and {Martin}, P.~G. and {Sormani}, M. and {Coletta}, A. and {Colman}, T. and {Plume}, R. and {Maruccia}, Y. and {Mininni}, C. and {Liu}, S.~J.},
        title = "{The Star Formation Rate of the Milky Way as Seen by Herschel}",
      journal = {\apj},
         year = 2022,
        month = dec,
       volume = {941},
       number = {2},
          eid = {162},
        pages = {162},
          doi = {10.3847/1538-4357/aca27d},
archivePrefix = {arXiv},
       eprint = {2211.05573},
 primaryClass = {astro-ph.GA},
       adsurl = {https://ui.adsabs.harvard.edu/abs/2022ApJ...941..162E}
}

@ARTICLE{Licquia2015,
       author = {{Licquia}, Timothy C. and {Newman}, Jeffrey A.},
        title = "{Improved Estimates of the Milky Way's Stellar Mass and Star Formation Rate from Hierarchical Bayesian Meta-Analysis}",
      journal = {\apj},
         year = 2015,
        month = jun,
       volume = {806},
       number = {1},
          eid = {96},
        pages = {96},
          doi = {10.1088/0004-637X/806/1/96},
archivePrefix = {arXiv},
       eprint = {1407.1078},
 primaryClass = {astro-ph.GA},
       adsurl = {https://ui.adsabs.harvard.edu/abs/2015ApJ...806...96L}
}

@ARTICLE{McMillan2017,
       author = {{McMillan}, Paul J.},
        title = "{The mass distribution and gravitational potential of the Milky Way}",
      journal = {\mnras},
         year = 2017,
        month = feb,
       volume = {465},
       number = {1},
        pages = {76-94},
          doi = {10.1093/mnras/stw2759},
archivePrefix = {arXiv},
       eprint = {1608.00971},
 primaryClass = {astro-ph.GA},
       adsurl = {https://ui.adsabs.harvard.edu/abs/2017MNRAS.465...76M}
}

@ARTICLE{McMillan2011,
       author = {{McMillan}, Paul J.},
        title = "{Mass models of the Milky Way}",
      journal = {\mnras},
         year = 2011,
        month = jul,
       volume = {414},
       number = {3},
        pages = {2446-2457},
          doi = {10.1111/j.1365-2966.2011.18564.x},
archivePrefix = {arXiv},
       eprint = {1102.4340},
 primaryClass = {astro-ph.GA},
       adsurl = {https://ui.adsabs.harvard.edu/abs/2011MNRAS.414.2446M}
}

@ARTICLE{Misiriotis2006,
       author = {{Misiriotis}, A. and {Xilouris}, E.~M. and {Papamastorakis}, J. and {Boumis}, P. and {Goudis}, C.~D.},
        title = "{The distribution of the ISM in the Milky Way. A three-dimensional large-scale model}",
      journal = {\aap},
         year = 2006,
        month = nov,
       volume = {459},
       number = {1},
        pages = {113-123},
          doi = {10.1051/0004-6361:20054618},
archivePrefix = {arXiv},
       eprint = {astro-ph/0607638},
 primaryClass = {astro-ph},
       adsurl = {https://ui.adsabs.harvard.edu/abs/2006A&A...459..113M}
}

@ARTICLE{Nakanishi2016,
       author = {{Nakanishi}, Hiroyuki and {Sofue}, Yoshiaki},
        title = "{Three-dimensional distribution of the ISM in the Milky Way galaxy. III. The total neutral gas disk}",
      journal = {\pasj},
         year = 2016,
        month = feb,
       volume = {68},
       number = {1},
          eid = {5},
        pages = {5},
          doi = {10.1093/pasj/psv108},
archivePrefix = {arXiv},
       eprint = {1511.08877},
 primaryClass = {astro-ph.GA},
       adsurl = {https://ui.adsabs.harvard.edu/abs/2016PASJ...68....5N}
}

@ARTICLE{Bovy2013,
       author = {{Bovy}, Jo and {Rix}, Hans-Walter},
        title = "{A Direct Dynamical Measurement of the Milky Way's Disk Surface Density Profile, Disk Scale Length, and Dark Matter Profile at 4 kpc <\raisebox{-0.5ex}\textasciitilde R <\raisebox{-0.5ex}\textasciitilde 9 kpc}",
      journal = {\apj},
         year = 2013,
        month = dec,
       volume = {779},
       number = {2},
          eid = {115},
        pages = {115},
          doi = {10.1088/0004-637X/779/2/115},
archivePrefix = {arXiv},
       eprint = {1309.0809},
 primaryClass = {astro-ph.GA},
       adsurl = {https://ui.adsabs.harvard.edu/abs/2013ApJ...779..115B}
}

@ARTICLE{Shapiro2004,
       author = {{Shapiro}, Paul R. and {Iliev}, Ilian T. and {Raga}, Alejandro C.},
        title = "{Photoevaporation of cosmological minihaloes during reionization}",
      journal = {\mnras},
         year = 2004,
        month = mar,
       volume = {348},
       number = {3},
        pages = {753-782},
          doi = {10.1111/j.1365-2966.2004.07364.x},
archivePrefix = {arXiv},
       eprint = {astro-ph/0307266},
 primaryClass = {astro-ph},
       adsurl = {https://ui.adsabs.harvard.edu/abs/2004MNRAS.348..753S}
}

@ARTICLE{Aoki2007,
       author = {{Aoki}, Wako and {Beers}, Timothy C. and {Christlieb}, Norbert and {Norris}, John E. and {Ryan}, Sean G. and {Tsangarides}, Stelios},
        title = "{Carbon-enhanced Metal-poor Stars. I. Chemical Compositions of 26 Stars}",
      journal = {\apj},
         year = 2007,
        month = jan,
       volume = {655},
       number = {1},
        pages = {492-521},
          doi = {10.1086/509817},
archivePrefix = {arXiv},
       eprint = {astro-ph/0609702},
 primaryClass = {astro-ph},
       adsurl = {https://ui.adsabs.harvard.edu/abs/2007ApJ...655..492A}
}

@ARTICLE{Hartwig2022,
       author = {{Hartwig}, Tilman and {Magg}, Mattis and {Chen}, Li-Hsin and {Tarumi}, Yuta and {Bromm}, Volker and {Glover}, Simon C.~O. and {Ji}, Alexander P. and {Klessen}, Ralf S. and {Latif}, Muhammad A. and {Volonteri}, Marta and {Yoshida}, Naoki},
        title = "{Public Release of A-SLOTH: Ancient Stars and Local Observables by Tracing Halos}",
      journal = {\apj},
         year = 2022,
        month = sep,
       volume = {936},
       number = {1},
          eid = {45},
        pages = {45},
          doi = {10.3847/1538-4357/ac7150},
archivePrefix = {arXiv},
       eprint = {2206.00223},
 primaryClass = {astro-ph.GA},
       adsurl = {https://ui.adsabs.harvard.edu/abs/2022ApJ...936...45H}
}

@ARTICLE{Magg2018,
       author = {{Magg}, Mattis and {Hartwig}, Tilman and {Agarwal}, Bhaskar and {Frebel}, Anna and {Glover}, Simon C.~O. and {Griffen}, Brendan F. and {Klessen}, Ralf S.},
        title = "{Predicting the locations of possible long-lived low-mass first stars: importance of satellite dwarf galaxies}",
      journal = {\mnras},
         year = 2018,
        month = feb,
       volume = {473},
       number = {4},
        pages = {5308-5323},
          doi = {10.1093/mnras/stx2729},
archivePrefix = {arXiv},
       eprint = {1706.07054},
 primaryClass = {astro-ph.GA},
       adsurl = {https://ui.adsabs.harvard.edu/abs/2018MNRAS.473.5308M}
}

@ARTICLE{Salvadori2014,
       author = {{Salvadori}, S. and {Tolstoy}, E. and {Ferrara}, A. and {Zaroubi}, S.},
        title = "{Metals and ionizing photons from dwarf galaxies}",
      journal = {\mnras},
         year = 2014,
        month = jan,
       volume = {437},
       number = {1},
        pages = {L26-L30},
          doi = {10.1093/mnrasl/slt132},
archivePrefix = {arXiv},
       eprint = {1309.7058},
 primaryClass = {astro-ph.CO},
       adsurl = {https://ui.adsabs.harvard.edu/abs/2014MNRAS.437L..26S}
}

@ARTICLE{Mesinger2009,
       author = {{Mesinger}, Andrei and {Furlanetto}, Steven},
        title = "{The inhomogeneous ionizing background following reionization}",
      journal = {\mnras},
         year = 2009,
        month = dec,
       volume = {400},
       number = {3},
        pages = {1461-1471},
          doi = {10.1111/j.1365-2966.2009.15547.x},
archivePrefix = {arXiv},
       eprint = {0906.3020},
 primaryClass = {astro-ph.CO},
       adsurl = {https://ui.adsabs.harvard.edu/abs/2009MNRAS.400.1461M}
}

@ARTICLE{Rahmati2018,
       author = {{Rahmati}, Alireza and {Schaye}, Joop},
        title = "{The mean free path of hydrogen ionizing photons during the epoch of reionization}",
      journal = {\mnras},
         year = 2018,
        month = aug,
       volume = {478},
       number = {4},
        pages = {5123-5134},
          doi = {10.1093/mnras/sty1382},
archivePrefix = {arXiv},
       eprint = {1708.04238},
 primaryClass = {astro-ph.CO},
       adsurl = {https://ui.adsabs.harvard.edu/abs/2018MNRAS.478.5123R}
}

@BOOK{Draine2011,
       author = {{Draine}, Bruce T.},
        title = "{Physics of the Interstellar and Intergalactic Medium}",
         year = 2011,
       adsurl = {https://ui.adsabs.harvard.edu/abs/2011piim.book.....D}
}

@ARTICLE{Dijkstra2008,
       author = {{Dijkstra}, Mark and {Haiman}, Zolt{\'a}n and {Mesinger}, Andrei and {Wyithe}, J. Stuart B.},
        title = "{Fluctuations in the high-redshift Lyman-Werner background: close halo pairs as the origin of supermassive black holes}",
      journal = {\mnras},
         year = 2008,
        month = dec,
       volume = {391},
       number = {4},
        pages = {1961-1972},
          doi = {10.1111/j.1365-2966.2008.14031.x},
archivePrefix = {arXiv},
       eprint = {0810.0014},
 primaryClass = {astro-ph},
       adsurl = {https://ui.adsabs.harvard.edu/abs/2008MNRAS.391.1961D}
}

@ARTICLE{Agarwal2012,
       author = {{Agarwal}, Bhaskar and {Khochfar}, Sadegh and {Johnson}, Jarrett L. and {Neistein}, Eyal and {Dalla Vecchia}, Claudio and {Livio}, Mario},
        title = "{Ubiquitous seeding of supermassive black holes by direct collapse}",
      journal = {\mnras},
         year = 2012,
        month = oct,
       volume = {425},
       number = {4},
        pages = {2854-2871},
          doi = {10.1111/j.1365-2966.2012.21651.x},
archivePrefix = {arXiv},
       eprint = {1205.6464},
 primaryClass = {astro-ph.CO},
       adsurl = {https://ui.adsabs.harvard.edu/abs/2012MNRAS.425.2854A}
}

@ARTICLE{Bromm2004,
       author = {{Bromm}, Volker and {Larson}, Richard B.},
        title = "{The First Stars}",
      journal = {\araa},
         year = 2004,
        month = sep,
       volume = {42},
       number = {1},
        pages = {79-118},
          doi = {10.1146/annurev.astro.42.053102.134034},
archivePrefix = {arXiv},
       eprint = {astro-ph/0311019},
 primaryClass = {astro-ph},
       adsurl = {https://ui.adsabs.harvard.edu/abs/2004ARA&A..42...79B}
}

@ARTICLE{Smith2024,
       author = {{Smith}, Britton D. and {O'Shea}, Brian W. and {Khochfar}, Sadegh and {Turk}, Matthew J. and {Wise}, John H. and {Norman}, Michael L.},
        title = "{Why does the Milky Way have a metallicity floor?}",
      journal = {\mnras},
         year = 2024,
        month = aug,
       volume = {532},
       number = {4},
        pages = {3797-3807},
          doi = {10.1093/mnras/stae1725},
archivePrefix = {arXiv},
       eprint = {2406.08199},
 primaryClass = {astro-ph.GA},
       adsurl = {https://ui.adsabs.harvard.edu/abs/2024MNRAS.532.3797S}
}

@ARTICLE{Cooper2010,
       author = {{Cooper}, A.~P. and {Cole}, S. and {Frenk}, C.~S. and {White}, S.~D.~M. and {Helly}, J. and {Benson}, A.~J. and {De Lucia}, G. and {Helmi}, A. and {Jenkins}, A. and {Navarro}, J.~F. and {Springel}, V. and {Wang}, J.},
        title = "{Galactic stellar haloes in the CDM model}",
      journal = {\mnras},
         year = 2010,
        month = aug,
       volume = {406},
       number = {2},
        pages = {744-766},
          doi = {10.1111/j.1365-2966.2010.16740.x},
archivePrefix = {arXiv},
       eprint = {0910.3211},
 primaryClass = {astro-ph.GA},
       adsurl = {https://ui.adsabs.harvard.edu/abs/2010MNRAS.406..744C}
}

@ARTICLE{Dopcke2013,
       author = {{Dopcke}, Gustavo and {Glover}, Simon C.~O. and {Clark}, Paul C. and {Klessen}, Ralf S.},
        title = "{On the Initial Mass Function of Low-metallicity Stars: The Importance of Dust Cooling}",
      journal = {\apj},
         year = 2013,
        month = apr,
       volume = {766},
       number = {2},
          eid = {103},
        pages = {103},
          doi = {10.1088/0004-637X/766/2/103},
archivePrefix = {arXiv},
       eprint = {1203.6842},
 primaryClass = {astro-ph.SR},
       adsurl = {https://ui.adsabs.harvard.edu/abs/2013ApJ...766..103D}
}

@ARTICLE{Tegmark1997,
       author = {{Tegmark}, Max and {Silk}, Joseph and {Rees}, Martin J. and {Blanchard}, Alain and {Abel}, Tom and {Palla}, Francesco},
        title = "{How Small Were the First Cosmological Objects?}",
      journal = {\apj},
         year = 1997,
        month = jan,
       volume = {474},
        pages = {1},
          doi = {10.1086/303434},
archivePrefix = {arXiv},
       eprint = {astro-ph/9603007},
 primaryClass = {astro-ph},
       adsurl = {https://ui.adsabs.harvard.edu/abs/1997ApJ...474....1T}
}

@ARTICLE{Smith2015,
       author = {{Smith}, Britton D. and {Wise}, John H. and {O'Shea}, Brian W. and {Norman}, Michael L. and {Khochfar}, Sadegh},
        title = "{The first Population II stars formed in externally enriched mini-haloes}",
      journal = {\mnras},
         year = 2015,
        month = sep,
       volume = {452},
       number = {3},
        pages = {2822-2836},
          doi = {10.1093/mnras/stv1509},
archivePrefix = {arXiv},
       eprint = {1504.07639},
 primaryClass = {astro-ph.GA},
       adsurl = {https://ui.adsabs.harvard.edu/abs/2015MNRAS.452.2822S}
}

@ARTICLE{Omukai2008,
       author = {{Omukai}, K. and {Schneider}, R. and {Haiman}, Z.},
        title = "{Can Supermassive Black Holes Form in Metal-enriched High-Redshift Protogalaxies?}",
      journal = {\apj},
         year = 2008,
        month = oct,
       volume = {686},
       number = {2},
        pages = {801-814},
          doi = {10.1086/591636},
archivePrefix = {arXiv},
       eprint = {0804.3141},
 primaryClass = {astro-ph},
       adsurl = {https://ui.adsabs.harvard.edu/abs/2008ApJ...686..801O}
}

@ARTICLE{Safranek2010,
       author = {{Safranek-Shrader}, Chalence and {Bromm}, Volker and {Milosavljevi{\'c}}, Milo{\v{s}}},
        title = "{Fragmentation in the First Galaxies}",
      journal = {\apj},
         year = 2010,
        month = nov,
       volume = {723},
       number = {2},
        pages = {1568-1582},
          doi = {10.1088/0004-637X/723/2/1568},
archivePrefix = {arXiv},
       eprint = {1004.0267},
 primaryClass = {astro-ph.CO},
       adsurl = {https://ui.adsabs.harvard.edu/abs/2010ApJ...723.1568S}
}

@ARTICLE{Holzbauer2012,
       author = {{Holzbauer}, Lauren N. and {Furlanetto}, Steven R.},
        title = "{Fluctuations in the high-redshift Lyman-Werner and Ly{\ensuremath{\alpha}} radiation backgrounds}",
      journal = {\mnras},
         year = 2012,
        month = jan,
       volume = {419},
       number = {1},
        pages = {718-731},
          doi = {10.1111/j.1365-2966.2011.19752.x},
archivePrefix = {arXiv},
       eprint = {1105.5648},
 primaryClass = {astro-ph.CO},
       adsurl = {https://ui.adsabs.harvard.edu/abs/2012MNRAS.419..718H}
}

@ARTICLE{Ahn2009,
       author = {{Ahn}, Kyungjin and {Shapiro}, Paul R. and {Iliev}, Ilian T. and {Mellema}, Garrelt and {Pen}, Ue-Li},
        title = "{The Inhomogeneous Background Of H$_{2}$-Dissociating Radiation During Cosmic Reionization}",
      journal = {\apj},
         year = 2009,
        month = apr,
       volume = {695},
       number = {2},
        pages = {1430-1445},
          doi = {10.1088/0004-637X/695/2/1430},
archivePrefix = {arXiv},
       eprint = {0807.2254},
 primaryClass = {astro-ph},
       adsurl = {https://ui.adsabs.harvard.edu/abs/2009ApJ...695.1430A}
}

@ARTICLE{Vogelsberger2013,
       author = {{Vogelsberger}, Mark and {Genel}, Shy and {Sijacki}, Debora and {Torrey}, Paul and {Springel}, Volker and {Hernquist}, Lars},
        title = "{A model for cosmological simulations of galaxy formation physics}",
      journal = {\mnras},
         year = 2013,
        month = dec,
       volume = {436},
       number = {4},
        pages = {3031-3067},
          doi = {10.1093/mnras/stt1789},
archivePrefix = {arXiv},
       eprint = {1305.2913},
 primaryClass = {astro-ph.CO},
       adsurl = {https://ui.adsabs.harvard.edu/abs/2013MNRAS.436.3031V}
}

@ARTICLE{Maoz2012,
       author = {{Maoz}, Dan and {Mannucci}, Filippo and {Brandt}, Timothy D.},
        title = "{The delay-time distribution of Type Ia supernovae from Sloan II}",
      journal = {\mnras},
         year = 2012,
        month = nov,
       volume = {426},
       number = {4},
        pages = {3282-3294},
          doi = {10.1111/j.1365-2966.2012.21871.x},
archivePrefix = {arXiv},
       eprint = {1206.0465},
 primaryClass = {astro-ph.CO},
       adsurl = {https://ui.adsabs.harvard.edu/abs/2012MNRAS.426.3282M}
}

@ARTICLE{deBen2017,
       author = {{de Bennassuti}, M. and {Salvadori}, S. and {Schneider}, R. and {Valiante}, R. and {Omukai}, K.},
        title = "{Limits on Population III star formation with the most iron-poor stars}",
      journal = {\mnras},
         year = 2017,
        month = feb,
       volume = {465},
       number = {1},
        pages = {926-940},
          doi = {10.1093/mnras/stw2687},
archivePrefix = {arXiv},
       eprint = {1610.05777},
 primaryClass = {astro-ph.GA},
       adsurl = {https://ui.adsabs.harvard.edu/abs/2017MNRAS.465..926D}
}

@article{raiteri1996simulations,
  title={Simulations of Galactic chemical evolution. I. O and Fe abundances in a simple collapse model.},
  author={Raiteri, CM and Villata, MNJF and Navarro, JF},
  journal={Astronomy and Astrophysics},
  volume={315},
  pages={105--115},
  year={1996}
}

@ARTICLE{Limongi2018,
       author = {{Limongi}, Marco and {Chieffi}, Alessandro},
        title = "{Presupernova Evolution and Explosive Nucleosynthesis of Rotating Massive Stars in the Metallicity Range -3 {\ensuremath{\leq}} [Fe/H] {\ensuremath{\leq}} 0}",
      journal = {\apjs},
         year = 2018,
        month = jul,
       volume = {237},
       number = {1},
          eid = {13},
        pages = {13},
          doi = {10.3847/1538-4365/aacb24},
archivePrefix = {arXiv},
       eprint = {1805.09640},
 primaryClass = {astro-ph.SR},
       adsurl = {https://ui.adsabs.harvard.edu/abs/2018ApJS..237...13L}
}

@ARTICLE{Rizzuti2021,
       author = {{Rizzuti}, F. and {Cescutti}, G. and {Matteucci}, F. and {Chieffi}, A. and {Hirschi}, R. and {Limongi}, M. and {Saro}, A.},
        title = "{Constraints on stellar rotation from the evolution of Sr and Ba in the Galactic halo}",
      journal = {\mnras},
         year = 2021,
        month = apr,
       volume = {502},
       number = {2},
        pages = {2495-2507},
          doi = {10.1093/mnras/stab158},
archivePrefix = {arXiv},
       eprint = {2101.05345},
 primaryClass = {astro-ph.GA},
       adsurl = {https://ui.adsabs.harvard.edu/abs/2021MNRAS.502.2495R}
}

@ARTICLE{Prantzos2018,
       author = {{Prantzos}, N. and {Abia}, C. and {Limongi}, M. and {Chieffi}, A. and {Cristallo}, S.},
        title = "{Chemical evolution with rotating massive star yields - I. The solar neighbourhood and the s-process elements}",
      journal = {\mnras},
         year = 2018,
        month = may,
       volume = {476},
       number = {3},
        pages = {3432-3459},
          doi = {10.1093/mnras/sty316},
archivePrefix = {arXiv},
       eprint = {1802.02824},
 primaryClass = {astro-ph.GA},
       adsurl = {https://ui.adsabs.harvard.edu/abs/2018MNRAS.476.3432P}
}

@ARTICLE{Karakas2010,
       author = {{Karakas}, A.~I.},
        title = "{Updated stellar yields from asymptotic giant branch models}",
      journal = {\mnras},
         year = 2010,
        month = apr,
       volume = {403},
       number = {3},
        pages = {1413-1425},
          doi = {10.1111/j.1365-2966.2009.16198.x},
archivePrefix = {arXiv},
       eprint = {0912.2142},
 primaryClass = {astro-ph.SR},
       adsurl = {https://ui.adsabs.harvard.edu/abs/2010MNRAS.403.1413K}
}

@ARTICLE{Heger2002,
       author = {{Heger}, A. and {Woosley}, S.~E.},
        title = "{The Nucleosynthetic Signature of Population III}",
      journal = {\apj},
         year = 2002,
        month = mar,
       volume = {567},
       number = {1},
        pages = {532-543},
          doi = {10.1086/338487},
archivePrefix = {arXiv},
       eprint = {astro-ph/0107037},
 primaryClass = {astro-ph},
       adsurl = {https://ui.adsabs.harvard.edu/abs/2002ApJ...567..532H}
}

@ARTICLE{Meynet2002,
       author = {{Meynet}, G. and {Maeder}, A.},
        title = "{Stellar evolution with rotation. VIII. Models at Z = 10$^{-5}$ and CNO yields for early galactic evolution}",
      journal = {\aap},
         year = 2002,
        month = aug,
       volume = {390},
        pages = {561-583},
          doi = {10.1051/0004-6361:20020755},
archivePrefix = {arXiv},
       eprint = {astro-ph/0205370},
 primaryClass = {astro-ph},
       adsurl = {https://ui.adsabs.harvard.edu/abs/2002A&A...390..561M}
}

@ARTICLE{Schaerer2002,
       author = {{Schaerer}, D.},
        title = "{On the properties of massive Population III stars and metal-free stellar populations}",
      journal = {\aap},
         year = 2002,
        month = jan,
       volume = {382},
        pages = {28-42},
          doi = {10.1051/0004-6361:20011619},
archivePrefix = {arXiv},
       eprint = {astro-ph/0110697},
 primaryClass = {astro-ph},
       adsurl = {https://ui.adsabs.harvard.edu/abs/2002A&A...382...28S}
}

@ARTICLE{Li2011,
       author = {{Li}, Weidong and {Chornock}, Ryan and {Leaman}, Jesse and {Filippenko}, Alexei V. and {Poznanski}, Dovi and {Wang}, Xiaofeng and {Ganeshalingam}, Mohan and {Mannucci}, Filippo},
        title = "{Nearby supernova rates from the Lick Observatory Supernova Search - III. The rate-size relation, and the rates as a function of galaxy Hubble type and colour}",
      journal = {\mnras},
         year = 2011,
        month = apr,
       volume = {412},
       number = {3},
        pages = {1473-1507},
          doi = {10.1111/j.1365-2966.2011.18162.x},
archivePrefix = {arXiv},
       eprint = {1006.4613},
 primaryClass = {astro-ph.SR},
       adsurl = {https://ui.adsabs.harvard.edu/abs/2011MNRAS.412.1473L}
}

@ARTICLE{Koutsouridou2024,
       author = {{Koutsouridou}, Ioanna and {Salvadori}, Stefania and {Sk{\'u}lad{\'o}ttir}, {\'A}sa},
        title = "{True Pair-instability Supernova Descendant: Implications for the First Stars' Mass Distribution}",
      journal = {\apjl},
         year = 2024,
        month = feb,
       volume = {962},
       number = {2},
          eid = {L26},
        pages = {L26},
          doi = {10.3847/2041-8213/ad2466},
archivePrefix = {arXiv},
       eprint = {2312.05309},
 primaryClass = {astro-ph.GA},
       adsurl = {https://ui.adsabs.harvard.edu/abs/2024ApJ...962L..26K}
}

@ARTICLE{Koutsouridou2025,
       author = {{Koutsouridou}, I. and {Sk{\'u}lad{\'o}ttir}, {\'A}. and {Salvadori}, S.},
        title = "{Large databases of metal-poor stars corrected for three-dimensional and/or non-local thermodynamic equilibrium effects}",
      journal = {\aap},
         year = 2025,
        month = jul,
       volume = {699},
          eid = {A32},
        pages = {A32},
          doi = {10.1051/0004-6361/202554228},
archivePrefix = {arXiv},
       eprint = {2505.13607},
 primaryClass = {astro-ph.GA},
       adsurl = {https://ui.adsabs.harvard.edu/abs/2025A&A...699A..32K}
}

@ARTICLE{Asplund2009,
       author = {{Asplund}, Martin and {Grevesse}, Nicolas and {Sauval}, A. Jacques and {Scott}, Pat},
        title = "{The Chemical Composition of the Sun}",
      journal = {\araa},
         year = 2009,
        month = sep,
       volume = {47},
       number = {1},
        pages = {481-522},
          doi = {10.1146/annurev.astro.46.060407.145222},
archivePrefix = {arXiv},
       eprint = {0909.0948},
 primaryClass = {astro-ph.SR},
       adsurl = {https://ui.adsabs.harvard.edu/abs/2009ARA&A..47..481A}
}

@ARTICLE{Ubler2026,
       author = {{{\"U}bler}, Hannah and {Maiolino}, Roberto and {P{\'e}rez-Gonz{\'a}lez}, Pablo G. and {Isobe}, Yuki and {Jones}, Gareth C. and {Kumari}, Nimisha and {Charlot}, St{\'e}phane and {Rusta}, Elka and {Salvadori}, Stefania and {Nakajima}, Kimihiko and {Perna}, Michele and {Arribas}, Santiago and {Bunker}, Andrew J. and {Carniani}, Stefano and {D'Eugenio}, Francesco and {Rodr{\'\i}guez Del Pino}, Bruno and {Bertola}, Elena and {B{\"o}ker}, Torsten and {Chevallard}, Jacopo and {Circosta}, Chiara and {Cresci}, Giovanni and {Curti}, Mirko and {Curtis-Lake}, Emma and {Eisenstein}, Daniel J. and {Hainline}, Kevin and {Johnson}, Benjamin D. and {Parlanti}, Eleonora and {Rinaldi}, Pierluigi and {Robertson}, Brant and {Scholtz}, Jan and {Tacchella}, Sandro and {Venturi}, Giacomo and {Witstok}, Joris and {Zamora}, Sandra},
        title = "{GA-NIFS \& JADES: Confirmation of pristine gas near GN-z11}",
      journal = {arXiv e-prints},
         year = 2026,
        month = mar,
          eid = {arXiv:2603.20360},
        pages = {arXiv:2603.20360},
          doi = {10.48550/arXiv.2603.20360},
archivePrefix = {arXiv},
       eprint = {2603.20360},
 primaryClass = {astro-ph.GA},
       adsurl = {https://ui.adsabs.harvard.edu/abs/2026arXiv260320360U}
}

@ARTICLE{DiMatteo2019,
       author = {{Di Matteo}, P. and {Haywood}, M. and {Lehnert}, M.~D. and {Katz}, D. and {Khoperskov}, S. and {Snaith}, O.~N. and {G{\'o}mez}, A. and {Robichon}, N.},
        title = "{The Milky Way has no in-situ halo other than the heated thick disc. Composition of the stellar halo and age-dating the last significant merger with Gaia DR2 and APOGEE}",
      journal = {\aap},
         year = 2019,
        month = dec,
       volume = {632},
          eid = {A4},
        pages = {A4},
          doi = {10.1051/0004-6361/201834929},
archivePrefix = {arXiv},
       eprint = {1812.08232},
 primaryClass = {astro-ph.GA},
       adsurl = {https://ui.adsabs.harvard.edu/abs/2019A&A...632A...4D}
}

@ARTICLE{Sestito2021,
       author = {{Sestito}, Federico and {Buck}, Tobias and {Starkenburg}, Else and {Martin}, Nicolas F. and {Navarro}, Julio F. and {Venn}, Kim A. and {Obreja}, Aura and {Jablonka}, Pascale and {Macci{\`o}}, Andrea V.},
        title = "{Exploring the origin of low-metallicity stars in Milky-Way-like galaxies with the NIHAO-UHD simulations}",
      journal = {\mnras},
         year = 2021,
        month = jan,
       volume = {500},
       number = {3},
        pages = {3750-3762},
          doi = {10.1093/mnras/staa3479},
archivePrefix = {arXiv},
       eprint = {2009.14207},
 primaryClass = {astro-ph.GA},
       adsurl = {https://ui.adsabs.harvard.edu/abs/2021MNRAS.500.3750S}
}

@ARTICLE{Zackrisson2024,
       author = {{Zackrisson}, Erik and {Hultquist}, Adam and {Kordt}, Aron and {Diego}, Jose M. and {Nabizadeh}, Armin and {Vikaeus}, Anton and {Meena}, Ashish Kumar and {Zitrin}, Adi and {Volpato}, Guglielmo and {Lundqvist}, Emma and {Welch}, Brian and {Costa}, Guglielmo and {Windhorst}, Rogier A.},
        title = "{The detection and characterization of highly magnified stars with JWST: prospects of finding Population III}",
      journal = {\mnras},
         year = 2024,
        month = sep,
       volume = {533},
       number = {3},
        pages = {2727-2746},
          doi = {10.1093/mnras/stae1881},
archivePrefix = {arXiv},
       eprint = {2312.09289},
 primaryClass = {astro-ph.GA},
       adsurl = {https://ui.adsabs.harvard.edu/abs/2024MNRAS.533.2727Z}
}

@ARTICLE{Jin2024,
       author = {{Jin}, Shoko and {Trager}, Scott C. and {Dalton}, Gavin B. and {Aguerri}, J. Alfonso L. and {Drew}, J.~E. and {Falc{\'o}n-Barroso}, Jes{\'u}s and {G{\"a}nsicke}, Boris T. and {Hill}, Vanessa and {Iovino}, Angela and {Pieri}, Matthew M. and {Poggianti}, Bianca M. and {Smith}, D.~J.~B. and {Vallenari}, Antonella and {Abrams}, Don Carlos and {Aguado}, David S. and {Antoja}, Teresa and {Arag{\'o}n-Salamanca}, Alfonso and {Ascasibar}, Yago and {Babusiaux}, Carine and {Balcells}, Marc and {Barrena}, R. and {Battaglia}, Giuseppina and {Belokurov}, Vasily and {Bensby}, Thomas and {Bonifacio}, Piercarlo and {Bragaglia}, Angela and {Carrasco}, Esperanza and {Carrera}, Ricardo and {Cornwell}, Daniel J. and {Dom{\'\i}nguez-Palmero}, Lilian and {Duncan}, Kenneth J. and {Famaey}, Benoit and {Fari{\~n}a}, Cecilia and {Gonzalez}, Oscar A. and {Guest}, Steve and {Hatch}, Nina A. and {Hess}, Kelley M. and {Hoskin}, Matthew J. and {Irwin}, Mike and {Knapen}, Johan H. and {Koposov}, Sergey E. and {Kuchner}, Ulrike and {Laigle}, Clotilde and {Lewis}, Jim and {Longhetti}, Marcella and {Lucatello}, Sara and {M{\'e}ndez-Abreu}, Jairo and {Mercurio}, Amata and {Molaeinezhad}, Alireza and {Mongui{\'o}}, Maria and {Morrison}, Sean and {Murphy}, David N.~A. and {Peralta de Arriba}, Luis and {P{\'e}rez}, Isabel and {P{\'e}rez-R{\`a}fols}, Ignasi and {Pic{\'o}}, Sergio and {Raddi}, Roberto and {Romero-G{\'o}mez}, Merc{\`e} and {Royer}, Fr{\'e}d{\'e}ric and {Siebert}, Arnaud and {Seabroke}, George M. and {Som}, Debopam and {Terrett}, David and {Thomas}, Guillaume and {Wesson}, Roger and {Worley}, C. Clare and {Alfaro}, Emilio J. and {Allende Prieto}, Carlos and {Alonso-Santiago}, Javier and {Amos}, Nicholas J. and {Ashley}, Richard P. and {Balaguer-N{\'u}{\~n}ez}, Lola and {Balbinot}, Eduardo and {Bellazzini}, Michele and {Benn}, Chris R. and {Berlanas}, Sara R. and {Bernard}, Edouard J. and {Best}, Philip and {Bettoni}, Daniela and {Bianco}, Andrea and {Bishop}, Georgia and {Blomqvist}, Michael and {Boeche}, Corrado and {Bolzonella}, Micol and {Bonoli}, Silvia and {Bosma}, Albert and {Britavskiy}, Nikolay and {Busarello}, Gianni and {Caffau}, Elisabetta and {Cantat-Gaudin}, Tristan and {Castro-Ginard}, Alfred and {Couto}, Guilherme and {Carbajo-Hijarrubia}, Juan and {Carter}, David and {Casamiquela}, Laia and {Conrado}, Ana M. and {Corcho-Caballero}, Pablo and {Costantin}, Luca and {Deason}, Alis and {de Burgos}, Abel and {De Grandi}, Sabrina and {Di Matteo}, Paola and {Dom{\'\i}nguez-G{\'o}mez}, Jes{\'u}s and {Dorda}, Ricardo and {Drake}, Alyssa and {Dutta}, Rajeshwari and {Erkal}, Denis and {Feltzing}, Sofia and {Ferr{\'e}-Mateu}, Anna and {Feuillet}, Diane and {Figueras}, Francesca and {Fossati}, Matteo and {Franciosini}, Elena and {Frasca}, Antonio and {Fumagalli}, Michele and {Gallazzi}, Anna and {Garc{\'\i}a-Benito}, Rub{\'e}n and {Gentile Fusillo}, Nicola and {Gebran}, Marwan and {Gilbert}, James and {Gledhill}, T.~M. and {Gonz{\'a}lez Delgado}, Rosa M. and {Greimel}, Robert and {Guarcello}, Mario Giuseppe and {Guerra}, Jose and {Gullieuszik}, Marco and {Haines}, Christopher P. and {Hardcastle}, Martin J. and {Harris}, Amy and {Haywood}, Misha and {Helmi}, Amina and {Hernandez}, Nauzet and {Herrero}, Artemio and {Hughes}, Sarah and {Ir{\v{s}}i{\v{c}}}, Vid and {Jablonka}, Pascale and {Jarvis}, Matt J. and {Jordi}, Carme and {Kondapally}, Rohit and {Kordopatis}, Georges and {Krogager}, Jens-Kristian and {La Barbera}, Francesco and {Lam}, Man I. and {Larsen}, S{\o}ren S. and {Lemasle}, Bertrand and {Lewis}, Ian J. and {Lhom{\'e}}, Emilie and {Lind}, Karin and {Lodi}, Marcello and {Longobardi}, Alessia and {Lonoce}, Ilaria and {Magrini}, Laura and {Ma{\'\i}z Apell{\'a}niz}, Jes{\'u}s and {Marchal}, Olivier and {Marco}, Amparo and {Martin}, Nicolas F. and {Matsuno}, Tadafumi and {Maurogordato}, Sophie and {Merluzzi}, Paola and {Miralda-Escud{\'e}}, Jordi and {Molinari}, Emilio and {Monari}, Giacomo and {Morelli}, Lorenzo and {Mottram}, Christopher J. and {Naylor}, Tim and {Negueruela}, Ignacio and {O{\~n}orbe}, Jose and {Pancino}, Elena and {Peirani}, S{\'e}bastien and {Peletier}, Reynier F. and {Pozzetti}, Lucia and {Rainer}, Monica and {Ramos}, Pau and {Read}, Shaun C. and {Rossi}, Elena Maria and {R{\"o}ttgering}, Huub J.~A. and {Rubi{\~n}o-Mart{\'\i}n}, Jose Alberto and {Sabater}, Jose and {San Juan}, Jos{\'e} and {Sanna}, Nicoletta and {Schallig}, Ellen and {Schiavon}, Ricardo P. and {Schultheis}, Mathias and {Serra}, Paolo and {Shimwell}, Timothy W. and {Sim{\'o}n-D{\'\i}az}, Sergio and {Smith}, Russell J. and {Sordo}, Rosanna and {Sorini}, Daniele and {Soubiran}, Caroline and {Starkenburg}, Else and {Steele}, Iain A. and {Stott}, John and {Stuik}, Remko and {Tolstoy}, Eline and {Tortora}, Crescenzo and {Tsantaki}, Maria and {Van der Swaelmen}, Mathieu and {van Weeren}, Reinout J. and {Vergani}, Daniela},
        title = "{The wide-field, multiplexed, spectroscopic facility WEAVE: Survey design, overview, and simulated implementation}",
      journal = {\mnras},
         year = 2024,
        month = may,
       volume = {530},
       number = {3},
        pages = {2688-2730},
          doi = {10.1093/mnras/stad557},
archivePrefix = {arXiv},
       eprint = {2212.03981},
 primaryClass = {astro-ph.IM},
       adsurl = {https://ui.adsabs.harvard.edu/abs/2024MNRAS.530.2688J}
}

@ARTICLE{Christlieb2019,
       author = {{Christlieb}, N. and {Battistini}, C. and {Bonifacio}, P. and {Caffau}, E. and {Ludwig}, H.-G. and {Asplund}, M. and {Barklem}, P. and {Bergemann}, M. and {Church}, R. and {Feltzing}, S. and {Ford}, D. and {Grebel}, E.~K. and {Hansen}, C.~J. and {Helmi}, A. and {Kordopatis}, G. and {Kovalev}, M. and {Korn}, A. and {Lind}, K. and {Quirrenbach}, A. and {Rybizki}, J. and {Sk{\'u}lad{\'o}ttir}, {\'A}. and {Starkenburg}, E.},
        title = "{4MOST Consortium Survey 2: The Milky Way Halo High-Resolution Survey}",
      journal = {The Messenger},
         year = 2019,
        month = mar,
       volume = {175},
        pages = {26-29},
          doi = {10.18727/0722-6691/5121},
archivePrefix = {arXiv},
       eprint = {1903.02468},
 primaryClass = {astro-ph.GA},
       adsurl = {https://ui.adsabs.harvard.edu/abs/2019Msngr.175...26C}
}

@ARTICLE{Skuladottir2023,
       author = {{Sk{\'u}lad{\'o}ttir}, {\'A}. and {Puls}, A.~A. and {Amarsi}, A.~M. and {Battaglia}, G. and {Buder}, S. and {Campbell}, S. and {Cardona-Barrero}, S. and {Christlieb}, N. and {Feuillet}, D.~K. and {Gelli}, V. and {Hansen}, C.~J. and {Hill}, V. and {Ibata}, R. and {Jablonka}, P. and {Kacharov}, N. and {Karakas}, A. and {Koch-Hansen}, A.~J. and {Lind}, K. and {Lombardo}, L. and {Lucchesi}, R.~E.~R. and {Lugaro}, M. and {Martin}, N. and {Massari}, D. and {Nordlander}, T. and {Reichert}, M. and {Rossi}, M. and {Ruiter}, A.~J. and {Salvadori}, S. and {Seitenzahl}, I.~R. and {Tolstoy}, E. and {Xylakis-Dornbusch}, T. and {Youakim}, K.~C.},
        title = "{The 4MOST Survey of Dwarf Galaxies and their Stellar Streams (4DWARFS)}",
      journal = {The Messenger},
         year = 2023,
        month = mar,
       volume = {190},
        pages = {19-21},
          doi = {10.18727/0722-6691/5304},
       adsurl = {https://ui.adsabs.harvard.edu/abs/2023Msngr.190...19S}
}

@ARTICLE{Salvadori2012,
       author = {{Salvadori}, Stefania and {Ferrara}, Andrea},
        title = "{First stars in damped Ly{\ensuremath{\alpha}} systems}",
      journal = {\mnras},
         year = 2012,
        month = mar,
       volume = {421},
       number = {1},
        pages = {L29-L33},
          doi = {10.1111/j.1745-3933.2011.01200.x},
archivePrefix = {arXiv},
       eprint = {1111.6637},
 primaryClass = {astro-ph.CO},
       adsurl = {https://ui.adsabs.harvard.edu/abs/2012MNRAS.421L..29S}
}

@ARTICLE{vandenBosch2008,
       author = {{van den Bosch}, Frank C. and {Aquino}, Daniel and {Yang}, Xiaohu and {Mo}, H.~J. and {Pasquali}, Anna and {McIntosh}, Daniel H. and {Weinmann}, Simone M. and {Kang}, Xi},
        title = "{The importance of satellite quenching for the build-up of the red sequence of present-day galaxies}",
      journal = {\mnras},
         year = 2008,
        month = jun,
       volume = {387},
       number = {1},
        pages = {79-91},
          doi = {10.1111/j.1365-2966.2008.13230.x},
archivePrefix = {arXiv},
       eprint = {0710.3164},
 primaryClass = {astro-ph},
       adsurl = {https://ui.adsabs.harvard.edu/abs/2008MNRAS.387...79V}
}

@ARTICLE{Larson1980,
       author = {{Larson}, R.~B. and {Tinsley}, B.~M. and {Caldwell}, C.~N.},
        title = "{The evolution of disk galaxies and the origin of S0 galaxies}",
      journal = {\apj},
         year = 1980,
        month = may,
       volume = {237},
        pages = {692-707},
          doi = {10.1086/157917},
       adsurl = {https://ui.adsabs.harvard.edu/abs/1980ApJ...237..692L}
}

@ARTICLE{Kapferer2009,
       author = {{Kapferer}, W. and {Sluka}, C. and {Schindler}, S. and {Ferrari}, C. and {Ziegler}, B.},
        title = "{The effect of ram pressure on the star formation, mass distribution and morphology of galaxies}",
      journal = {\aap},
         year = 2009,
        month = may,
       volume = {499},
       number = {1},
        pages = {87-102},
          doi = {10.1051/0004-6361/200811551},
archivePrefix = {arXiv},
       eprint = {0903.3818},
 primaryClass = {astro-ph.CO},
       adsurl = {https://ui.adsabs.harvard.edu/abs/2009A&A...499...87K}
}

@ARTICLE{Bekki2014,
       author = {{Bekki}, Kenji},
        title = "{Galactic star formation enhanced and quenched by ram pressure in groups and clusters}",
      journal = {\mnras},
         year = 2014,
        month = feb,
       volume = {438},
       number = {1},
        pages = {444-462},
          doi = {10.1093/mnras/stt2216},
archivePrefix = {arXiv},
       eprint = {1311.3010},
 primaryClass = {astro-ph.CO},
       adsurl = {https://ui.adsabs.harvard.edu/abs/2014MNRAS.438..444B}
}

@ARTICLE{Koutsouridou2019,
       author = {{Koutsouridou}, I. and {Cattaneo}, A.},
        title = "{Bursting and quenching in satellite galaxies}",
      journal = {\mnras},
         year = 2019,
        month = dec,
       volume = {490},
       number = {4},
        pages = {5375-5389},
          doi = {10.1093/mnras/stz2916},
archivePrefix = {arXiv},
       eprint = {1911.01395},
 primaryClass = {astro-ph.GA},
       adsurl = {https://ui.adsabs.harvard.edu/abs/2019MNRAS.490.5375K}
}

@ARTICLE{Koutsouridou2022,
       author = {{Koutsouridou}, I. and {Cattaneo}, A.},
        title = "{Probing the link between quenching and morphological evolution}",
      journal = {\mnras},
         year = 2022,
        month = nov,
       volume = {516},
       number = {3},
        pages = {4194-4211},
          doi = {10.1093/mnras/stac2240},
archivePrefix = {arXiv},
       eprint = {2209.12883},
 primaryClass = {astro-ph.GA},
       adsurl = {https://ui.adsabs.harvard.edu/abs/2022MNRAS.516.4194K}
}

@ARTICLE{Cattaneo2006,
       author = {{Cattaneo}, A. and {Dekel}, A. and {Devriendt}, J. and {Guiderdoni}, B. and {Blaizot}, J.},
        title = "{Modelling the galaxy bimodality: shutdown above a critical halo mass}",
      journal = {\mnras},
         year = 2006,
        month = aug,
       volume = {370},
       number = {4},
        pages = {1651-1665},
          doi = {10.1111/j.1365-2966.2006.10608.x},
archivePrefix = {arXiv},
       eprint = {astro-ph/0601295},
 primaryClass = {astro-ph},
       adsurl = {https://ui.adsabs.harvard.edu/abs/2006MNRAS.370.1651C}
}

@ARTICLE{Birnboim2003,
       author = {{Birnboim}, Yuval and {Dekel}, Avishai},
        title = "{Virial shocks in galactic haloes?}",
      journal = {\mnras},
         year = 2003,
        month = oct,
       volume = {345},
       number = {1},
        pages = {349-364},
          doi = {10.1046/j.1365-8711.2003.06955.x},
archivePrefix = {arXiv},
       eprint = {astro-ph/0302161},
 primaryClass = {astro-ph},
       adsurl = {https://ui.adsabs.harvard.edu/abs/2003MNRAS.345..349B}
}

@ARTICLE{Dekel2006,
       author = {{Dekel}, Avishai and {Birnboim}, Yuval},
        title = "{Galaxy bimodality due to cold flows and shock heating}",
      journal = {\mnras},
         year = 2006,
        month = may,
       volume = {368},
       number = {1},
        pages = {2-20},
          doi = {10.1111/j.1365-2966.2006.10145.x},
archivePrefix = {arXiv},
       eprint = {astro-ph/0412300},
 primaryClass = {astro-ph},
       adsurl = {https://ui.adsabs.harvard.edu/abs/2006MNRAS.368....2D}
}

@ARTICLE{Koutsouridou2023,
       author = {{Koutsouridou}, I. and {Salvadori}, S. and {Sk{\'u}lad{\'o}ttir}, {\'A}. and {Rossi}, M. and {Vanni}, I. and {Pagnini}, G.},
        title = "{The energy distribution of the first supernovae}",
      journal = {\mnras},
         year = 2023,
        month = oct,
       volume = {525},
       number = {1},
        pages = {190-210},
          doi = {10.1093/mnras/stad2304},
archivePrefix = {arXiv},
       eprint = {2309.00045},
 primaryClass = {astro-ph.GA},
       adsurl = {https://ui.adsabs.harvard.edu/abs/2023MNRAS.525..190K}
}

@ARTICLE{Belokurov2018,
       author = {{Belokurov}, V. and {Erkal}, D. and {Evans}, N.~W. and {Koposov}, S.~E. and {Deason}, A.~J.},
        title = "{Co-formation of the disc and the stellar halo}",
      journal = {\mnras},
         year = 2018,
        month = jul,
       volume = {478},
       number = {1},
        pages = {611-619},
          doi = {10.1093/mnras/sty982},
archivePrefix = {arXiv},
       eprint = {1802.03414},
 primaryClass = {astro-ph.GA},
       adsurl = {https://ui.adsabs.harvard.edu/abs/2018MNRAS.478..611B}
}

@ARTICLE{Naidu2021,
       author = {{Naidu}, Rohan P. and {Conroy}, Charlie and {Bonaca}, Ana and {Zaritsky}, Dennis and {Weinberger}, Rainer and {Ting}, Yuan-Sen and {Caldwell}, Nelson and {Tacchella}, Sandro and {Han}, Jiwon Jesse and {Speagle}, Joshua S. and {Cargile}, Phillip A.},
        title = "{Reconstructing the Last Major Merger of the Milky Way with the H3 Survey}",
      journal = {\apj},
         year = 2021,
        month = dec,
       volume = {923},
       number = {1},
          eid = {92},
        pages = {92},
          doi = {10.3847/1538-4357/ac2d2d},
archivePrefix = {arXiv},
       eprint = {2103.03251},
 primaryClass = {astro-ph.GA},
       adsurl = {https://ui.adsabs.harvard.edu/abs/2021ApJ...923...92N}
}

@ARTICLE{Helmi2018,
       author = {{Helmi}, Amina and {Babusiaux}, Carine and {Koppelman}, Helmer H. and {Massari}, Davide and {Veljanoski}, Jovan and {Brown}, Anthony G.~A.},
        title = "{The merger that led to the formation of the Milky Way's inner stellar halo and thick disk}",
      journal = {\nat},
         year = 2018,
        month = oct,
       volume = {563},
       number = {7729},
        pages = {85-88},
          doi = {10.1038/s41586-018-0625-x},
archivePrefix = {arXiv},
       eprint = {1806.06038},
 primaryClass = {astro-ph.GA},
       adsurl = {https://ui.adsabs.harvard.edu/abs/2018Natur.563...85H}
}

@ARTICLE{Planck2014,
       author = {{Planck Collaboration} and {Ade}, P.~A.~R. and {Aghanim}, N. and {Armitage-Caplan}, C. and {Arnaud}, M. and {Ashdown}, M. and {Atrio-Barandela}, F. and {Aumont}, J. and {Baccigalupi}, C. and {Banday}, A.~J. and {Barreiro}, R.~B. and {Bartlett}, J.~G. and {Battaner}, E. and {Benabed}, K. and {Beno{\^\i}t}, A. and {Benoit-L{\'e}vy}, A. and {Bernard}, J.-P. and {Bersanelli}, M. and {Bielewicz}, P. and {Bobin}, J. and {Bock}, J.~J. and {Bonaldi}, A. and {Bond}, J.~R. and {Borrill}, J. and {Bouchet}, F.~R. and {Bridges}, M. and {Bucher}, M. and {Burigana}, C. and {Butler}, R.~C. and {Calabrese}, E. and {Cappellini}, B. and {Cardoso}, J.-F. and {Catalano}, A. and {Challinor}, A. and {Chamballu}, A. and {Chary}, R.-R. and {Chen}, X. and {Chiang}, H.~C. and {Chiang}, L.-Y. and {Christensen}, P.~R. and {Church}, S. and {Clements}, D.~L. and {Colombi}, S. and {Colombo}, L.~P.~L. and {Couchot}, F. and {Coulais}, A. and {Crill}, B.~P. and {Curto}, A. and {Cuttaia}, F. and {Danese}, L. and {Davies}, R.~D. and {Davis}, R.~J. and {de Bernardis}, P. and {de Rosa}, A. and {de Zotti}, G. and {Delabrouille}, J. and {Delouis}, J.-M. and {D{\'e}sert}, F.-X. and {Dickinson}, C. and {Diego}, J.~M. and {Dolag}, K. and {Dole}, H. and {Donzelli}, S. and {Dor{\'e}}, O. and {Douspis}, M. and {Dunkley}, J. and {Dupac}, X. and {Efstathiou}, G. and {Elsner}, F. and {En{\ss}lin}, T.~A. and {Eriksen}, H.~K. and {Finelli}, F. and {Forni}, O. and {Frailis}, M. and {Fraisse}, A.~A. and {Franceschi}, E. and {Gaier}, T.~C. and {Galeotta}, S. and {Galli}, S. and {Ganga}, K. and {Giard}, M. and {Giardino}, G. and {Giraud-H{\'e}raud}, Y. and {Gjerl{\o}w}, E. and {Gonz{\'a}lez-Nuevo}, J. and {G{\'o}rski}, K.~M. and {Gratton}, S. and {Gregorio}, A. and {Gruppuso}, A. and {Gudmundsson}, J.~E. and {Haissinski}, J. and {Hamann}, J. and {Hansen}, F.~K. and {Hanson}, D. and {Harrison}, D. and {Henrot-Versill{\'e}}, S. and {Hern{\'a}ndez-Monteagudo}, C. and {Herranz}, D. and {Hildebrandt}, S.~R. and {Hivon}, E. and {Hobson}, M. and {Holmes}, W.~A. and {Hornstrup}, A. and {Hou}, Z. and {Hovest}, W. and {Huffenberger}, K.~M. and {Jaffe}, A.~H. and {Jaffe}, T.~R. and {Jewell}, J. and {Jones}, W.~C. and {Juvela}, M. and {Keih{\"a}nen}, E. and {Keskitalo}, R. and {Kisner}, T.~S. and {Kneissl}, R. and {Knoche}, J. and {Knox}, L. and {Kunz}, M. and {Kurki-Suonio}, H. and {Lagache}, G. and {L{\"a}hteenm{\"a}ki}, A. and {Lamarre}, J.-M. and {Lasenby}, A. and {Lattanzi}, M. and {Laureijs}, R.~J. and {Lawrence}, C.~R. and {Leach}, S. and {Leahy}, J.~P. and {Leonardi}, R. and {Le{\'o}n-Tavares}, J. and {Lesgourgues}, J. and {Lewis}, A. and {Liguori}, M. and {Lilje}, P.~B. and {Linden-V{\o}rnle}, M. and {L{\'o}pez-Caniego}, M. and {Lubin}, P.~M. and {Mac{\'\i}as-P{\'e}rez}, J.~F. and {Maffei}, B. and {Maino}, D. and {Mandolesi}, N. and {Maris}, M. and {Marshall}, D.~J. and {Martin}, P.~G. and {Mart{\'\i}nez-Gonz{\'a}lez}, E. and {Masi}, S. and {Massardi}, M. and {Matarrese}, S. and {Matthai}, F. and {Mazzotta}, P. and {Meinhold}, P.~R. and {Melchiorri}, A. and {Melin}, J.-B. and {Mendes}, L. and {Menegoni}, E. and {Mennella}, A. and {Migliaccio}, M. and {Millea}, M. and {Mitra}, S. and {Miville-Desch{\^e}nes}, M.-A. and {Moneti}, A. and {Montier}, L. and {Morgante}, G. and {Mortlock}, D. and {Moss}, A. and {Munshi}, D. and {Murphy}, J.~A. and {Naselsky}, P. and {Nati}, F. and {Natoli}, P. and {Netterfield}, C.~B. and {N{\o}rgaard-Nielsen}, H.~U. and {Noviello}, F. and {Novikov}, D. and {Novikov}, I. and {O'Dwyer}, I.~J. and {Osborne}, S. and {Oxborrow}, C.~A. and {Paci}, F. and {Pagano}, L. and {Pajot}, F. and {Paladini}, R. and {Paoletti}, D. and {Partridge}, B. and {Pasian}, F. and {Patanchon}, G. and {Pearson}, D. and {Pearson}, T.~J. and {Peiris}, H.~V. and {Perdereau}, O. and {Perotto}, L. and {Perrotta}, F. and {Pettorino}, V. and {Piacentini}, F. and {Piat}, M. and {Pierpaoli}, E. and {Pietrobon}, D. and {Plaszczynski}, S. and {Platania}, P. and {Pointecouteau}, E.},
        title = "{Planck 2013 results. XVI. Cosmological parameters}",
      journal = {\aap},
         year = 2014,
        month = nov,
       volume = {571},
          eid = {A16},
        pages = {A16},
          doi = {10.1051/0004-6361/201321591},
archivePrefix = {arXiv},
       eprint = {1303.5076},
 primaryClass = {astro-ph.CO},
       adsurl = {https://ui.adsabs.harvard.edu/abs/2014A&A...571A..16P}
}

@ARTICLE{Behroozi2013,
       author = {{Behroozi}, Peter S. and {Wechsler}, Risa H. and {Wu}, Hao-Yi},
        title = "{The ROCKSTAR Phase-space Temporal Halo Finder and the Velocity Offsets of Cluster Cores}",
      journal = {\apj},
         year = 2013,
        month = jan,
       volume = {762},
       number = {2},
          eid = {109},
        pages = {109},
          doi = {10.1088/0004-637X/762/2/109},
archivePrefix = {arXiv},
       eprint = {1110.4372},
 primaryClass = {astro-ph.CO},
       adsurl = {https://ui.adsabs.harvard.edu/abs/2013ApJ...762..109B}
}

@ARTICLE{Bryan1998,
       author = {{Bryan}, Greg L. and {Norman}, Michael L.},
        title = "{Statistical Properties of X-Ray Clusters: Analytic and Numerical Comparisons}",
      journal = {\apj},
         year = 1998,
        month = mar,
       volume = {495},
       number = {1},
        pages = {80-99},
          doi = {10.1086/305262},
archivePrefix = {arXiv},
       eprint = {astro-ph/9710107},
 primaryClass = {astro-ph},
       adsurl = {https://ui.adsabs.harvard.edu/abs/1998ApJ...495...80B}
}

@ARTICLE{Jeon2014,
       author = {{Jeon}, Myoungwon and {Pawlik}, Andreas H. and {Bromm}, Volker and {Milosavljevi{\'c}}, Milo{\v{s}}},
        title = "{Recovery from Population III supernova explosions and the onset of second-generation star formation}",
      journal = {\mnras},
         year = 2014,
        month = nov,
       volume = {444},
       number = {4},
        pages = {3288-3300},
          doi = {10.1093/mnras/stu1980},
archivePrefix = {arXiv},
       eprint = {1407.0034},
 primaryClass = {astro-ph.GA},
       adsurl = {https://ui.adsabs.harvard.edu/abs/2014MNRAS.444.3288J}
}

@ARTICLE{Mannucci2008,
       author = {{Mannucci}, F. and {Maoz}, D. and {Sharon}, K. and {Botticella}, M.~T. and {Della Valle}, M. and {Gal-Yam}, A. and {Panagia}, N.},
        title = "{The supernova rate in local galaxy clusters}",
      journal = {\mnras},
         year = 2008,
        month = jan,
       volume = {383},
       number = {3},
        pages = {1121-1130},
          doi = {10.1111/j.1365-2966.2007.12603.x},
archivePrefix = {arXiv},
       eprint = {0710.1094},
 primaryClass = {astro-ph},
       adsurl = {https://ui.adsabs.harvard.edu/abs/2008MNRAS.383.1121M}
}

@ARTICLE{Iwamoto99,
       author = {{Iwamoto}, Koichi and {Brachwitz}, Franziska and {Nomoto}, Ken'ICHI and {Kishimoto}, Nobuhiro and {Umeda}, Hideyuki and {Hix}, W. Raphael and {Thielemann}, Friedrich-Karl},
        title = "{Nucleosynthesis in Chandrasekhar Mass Models for Type IA Supernovae and Constraints on Progenitor Systems and Burning-Front Propagation}",
      journal = {\apjs},
         year = 1999,
        month = dec,
       volume = {125},
       number = {2},
        pages = {439-462},
          doi = {10.1086/313278},
archivePrefix = {arXiv},
       eprint = {astro-ph/0002337},
 primaryClass = {astro-ph},
       adsurl = {https://ui.adsabs.harvard.edu/abs/1999ApJS..125..439I}
}

@ARTICLE{Ritter2012,
       author = {{Ritter}, Jeremy S. and {Safranek-Shrader}, Chalence and {Gnat}, Orly and {Milosavljevi{\'c}}, Milo{\v{s}} and {Bromm}, Volker},
        title = "{Confined Population III Enrichment and the Prospects for Prompt Second-generation Star Formation}",
      journal = {\apj},
         year = 2012,
        month = dec,
       volume = {761},
       number = {1},
          eid = {56},
        pages = {56},
          doi = {10.1088/0004-637X/761/1/56},
archivePrefix = {arXiv},
       eprint = {1203.2957},
 primaryClass = {astro-ph.CO},
       adsurl = {https://ui.adsabs.harvard.edu/abs/2012ApJ...761...56R}
}

@ARTICLE{Bonifacio2021,
       author = {{Bonifacio}, P. and {Monaco}, L. and {Salvadori}, S. and {Caffau}, E. and {Spite}, M. and {Sbordone}, L. and {Spite}, F. and {Ludwig}, H. -G. and {Di Matteo}, P. and {Haywood}, M. and {Fran{\c{c}}ois}, P. and {Koch-Hansen}, A.~J. and {Christlieb}, N. and {Zaggia}, S.},
        title = "{TOPoS. VI. The metal-weak tail of the metallicity distribution functions of the Milky Way and the Gaia-Sausage-Enceladus structure}",
      journal = {aap},
         year = 2021,
        month = jul,
       volume = {651},
          eid = {A79},
        pages = {A79},
          doi = {10.1051/0004-6361/202140816},
archivePrefix = {arXiv},
       eprint = {2105.08360},
 primaryClass = {astro-ph.GA},
       adsurl = {https://ui.adsabs.harvard.edu/abs/2021A&A...651A..79B}
}

@article{SS19,
  title={Probing the existence of very massive first stars},
  author={Salvadori, S and Bonifacio, P and Caffau, E and Korotin, S and Andreevsky, S and Spite, M and Sk{\'u}lad{\'o}ttir, {\'A}},
  journal={Monthly Notices of the Royal Astronomical Society},
  volume={487},
  number={3},
  pages={4261--4284},
  year={2019},
  publisher={Oxford University Press}
}

@article{Hartwig15,
	Author = {Hartwig, Tilman and Bromm, Volker and Klessen, Ralf S and Glover, Simon CO},
	Journal = {Monthly Notices of the Royal Astronomical Society},
	Number = {4},
	Pages = {3892--3908},
	Publisher = {Oxford University Press},
	Title = {Constraining the primordial initial mass function with stellar archaeology},
	Volume = {447},
	Year = {2015}}

@ARTICLE{Schneider2003,
       author = {{Schneider}, R. and {Ferrara}, A. and {Salvaterra}, R. and {Omukai}, K. and
         {Bromm}, V.},
        title = "{Low-mass relics of early star formation}",
      journal = {\nat},
         year = 2003,
        month = apr,
       volume = {422},
       number = {6934},
        pages = {869-871},
          doi = {10.1038/nature01579},
archivePrefix = {arXiv},
       eprint = {astro-ph/0304254},
 primaryClass = {astro-ph},
       adsurl = {https://ui.adsabs.harvard.edu/abs/2003Natur.422..869S}
}

@ARTICLE{Griffen2016,
       author = {{Griffen}, Brendan F. and {Ji}, Alexander P. and {Dooley}, Gregory A. and {G{\'o}mez}, Facundo A. and {Vogelsberger}, Mark and {O'Shea}, Brian W. and {Frebel}, Anna},
        title = "{The Caterpillar Project: A Large Suite of Milky Way Sized Halos}",
      journal = {\apj},
         year = 2016,
        month = feb,
       volume = {818},
       number = {1},
          eid = {10},
        pages = {10},
          doi = {10.3847/0004-637X/818/1/10},
archivePrefix = {arXiv},
       eprint = {1509.01255},
 primaryClass = {astro-ph.GA},
       adsurl = {https://ui.adsabs.harvard.edu/abs/2016ApJ...818...10G}
}
\bibliographystyle{aasjournal}



\appendix

\section{Observational properties of the MW}
\label{sec: MWobs}

Here, we provide the literature references for the observational estimates of the MW properties at $z=0$ listed in Table~\ref{table: global_properties}.

Measurements of the total stellar mass of the MW span $M_* = (5$--$6.4)\times10^{10}\,{\rm M_\odot}$ \citep[e.g.,][]{McMillan2011,Bovy2013,Licquia2015,McMillan2017,Bland-Hawthorn2016,Cautun2020}.
Estimates for the mass of the ISM, 
$M_{\rm gas} \simeq (0.7$--$1.0)\times10^{10}\,{\rm M_\odot}$ \citep{Feriere2001, Bovy2013, Misiriotis2006, Bovy2013, Nakanishi2016}, imply gas-to-stellar mass ratios $M_{\rm gas}/M_* \simeq 0.1$--$0.2$.
Measurements of the total SFR lie in the range ${\rm SFR}\simeq 1$--$3\,{\rm M_\odot\,yr^{-1}}$
\citep{Misiriotis2006,Chomiuk2011,Licquia2015,Bland-Hawthorn2016, Elia2022}.

Observed mass-loading factors are typically lower limits, as they probe only specific gas phases (e.g., warm ionized or molecular).
For the MW, \citet{Fox2019} estimate a disk-wide mass-loading factor $\eta_{\rm MW}\sim0.1\pm0.06$ from high-velocity clouds, noting that contributions from low- and intermediate-velocity clouds and hot gas would increase this value. Measurements in external galaxies find comparable or higher values:
$\eta\simeq0.38$ for star-forming galaxies with
$M_*\sim5\times10^{10}\,{\rm M_\odot}$ 
\citep[see their Eq~16]{Chisholm2017}, median $\eta\simeq0.25$ for star-forming and intermediate objects with
$M_*>10^9\,{\rm M_\odot}$ \citep{Pino2019}, and
$\eta\simeq0.1$--$0.2$ for normal galaxies with $9.0< {\rm log}(M_*/M_\odot)<11.7$ at $0.6<z<2.7$ \citep{Schreiber2019}.

Observational estimates of the SNIa rate in late-type star-forming galaxies include: 
$0.14^{+0.045}_{-0.035}\:$SNuM (SN per century per $10^{10}\:{\rm M_\odot}$ of stellar mass) from \cite{Mannucci2008}, $0.11{\pm 0.02}\:$SNuM from \cite{Li2011}, and $0.11^{+0.04}_{-0.03}\:$SNuM for galaxies with stellar masses $4-8 \times 10^{10}\:{\rm M_\odot}$ from \cite{Grauz2013}. For ccSNe, observational studies estimate rates of $3.85^{+0.85}_{-0.7}\:$SN/century \citep{Mannucci2008}, $2.3{\pm 0.48}\:$SN/century \citep{Li2011}, $3.2^{+7.3}_{-2.6}\:$SN/century \citep{Adams2013}, and $1.6{\pm 0.48}\:$SN/century \citep{Rozwadowska2021}.

\section{The impact of Type Ia SNe}
\label{sec: SNIa}

\begin{figure}
\begin{center}
\includegraphics[width=0.5\hsize]{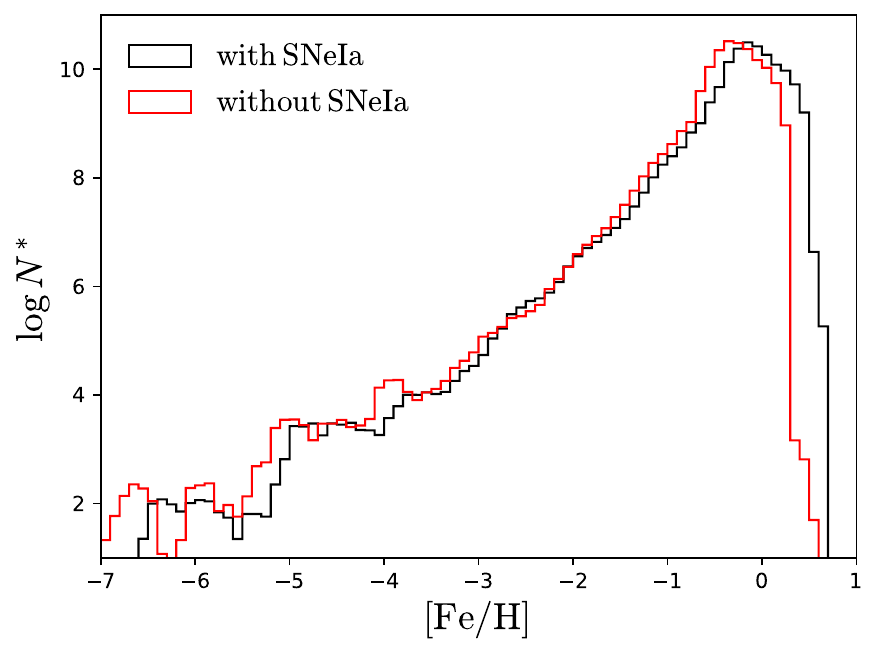}
\caption{Metallicity distribution function of a Caterpillar MW-analogue at $z=0$, with (black line) and without (red line) accounting for Type Ia SNe.}
\label{fig: MDF-noIa}
\end{center}
\end{figure}

Fig.~\ref{fig: MDF-noIa} shows the impact of SNe~Ia to the MW MDF at $z=0$. We find that their contribution becomes significant at $\rm[Fe/H]>-0.7$ and especially at $\rm[Fe/H]>0$, where the model without SNe~Ia predicts $63\%$ fewer stars.

\end{document}